\documentclass[journal]{IEEEtran}
\usepackage[T1]{fontenc}
\usepackage[utf8]{inputenc}
\usepackage{amsmath,amssymb,amsfonts}
\usepackage{graphicx}
\usepackage{tikz}
\usetikzlibrary{arrows.meta, positioning, shapes.geometric}
\usepackage{cite}
\usepackage{url}
\usepackage{hyperref}
\usepackage{booktabs}
\usepackage{array}
\usepackage{multirow}
\usepackage{pgfplots}
\usepackage{pgfplotstable}
\pgfplotsset{compat=1.18}
\usepackage{color}
\usepackage{xcolor}
\usepackage{algorithmic}
\usepackage{array}
\usetikzlibrary{calc}
\usepackage[caption=false,font=normalsize,labelfont=sf,textfont=sf]{subfig}
\usepackage{textcomp}
\usepackage{stfloats}
\usepackage{url}
\usepackage{verbatim}
\usepackage{graphicx}
\def\BibTeX{{\rm B\kern-.05em{\sc i\kern-.025em b}\kern-.08em
    T\kern-.1667em\lower.7ex {E}\kern-.125emX}}
\usepackage{balance}

\begin{document}

\title{Lyapunov Exponents in Power Systems: A Survey of Theory and Applications in Stability, Control, Forecasting, and Protection}

\author{Mohammad Almomani, \IEEEmembership{graduate student member, IEEE},
        and Venkataramana Ajjarapu, \IEEEmembership{Fellow, IEEE}


\IEEEauthorblockA{\textit{Department of Electrical \& Computer Engineering, Iowa State University, Ames, IA, USA} \\
Emails: \{mmomani, vajjarap\}@iastate.edu}}

\maketitle  

\small 

\begin{abstract}
Lyapunov exponents (LEs) are fundamental metrics from nonlinear dynamical systems theory that quantify the average exponential rate of divergence or convergence of nearby trajectories in phase space. Over the past three decades, LEs have emerged as powerful tools across a wide range of power system problems, including transient and voltage stability assessment, chaos detection and mitigation, short-term load and renewable energy forecasting, and fault diagnosis. This paper presents a comprehensive review of these applications, covering more than 165 publications spanning from the early demonstrations of chaos in simplified machine models (1990s) to the most recent data-driven and machine-learning-augmented approaches applied to inverter-based resource (IBR)-dominated grids (2024--2026). The review is organized into four principal application domains: (i) stability analysis and real-time monitoring, (ii) chaos characterization and control, (iii) forecasting and predictability analysis, and (iv) protection, fault diagnosis, and condition monitoring. For each domain, we trace the evolution of methods, compare model-based and data-driven LE estimation strategies, identify key benchmarks and validation gaps, and outline the
most pressing open research problems. 
\end{abstract}


\small
\section{Introduction}
\label{sec:intro}

\subsection{Motivation and Context}
\label{subsec:motivation} 

Modern power systems are undergoing the most profound structural transformation since the introduction of alternating current. The rapid integration of inverter-based resources (IBRs)-wind turbines, photovoltaic (PV) systems, and battery energy storage -is displacing conventional synchronous generators (SGs) and fundamentally altering the dynamic behavior of the grid. Declining synchronous inertia, reduced inherent damping, and increased stochastic variability from renewable generation all amplify the nonlinear characteristics of the network \cite{kazma_stability_2025, peng_measuring_2025, ramos-guerrero_measuring_2025}. Classical small-signal stability analysis, based on linearization and eigenvalue computation around an operating point, was designed for a world of large, inertia-rich synchronous machines. It provides only local information and cannot capture the large-disturbance, global nonlinear behavior that increasingly governs system security \cite{wadduwage_power_2013, li_chaotic_2009}.

At the same time, the wide deployment of phasor measurement units (PMUs) and wide-area measurement systems (WAMS) has created an unprecedented volume of high-resolution, time-synchronized measurement data. This data creates an opportunity to move beyond model-dependent stability assessment toward real-time, measurement-driven methods that can respond to the grid's actual dynamic state rather than a precomputed model of it \cite{dasgupta_real-time_2013, wei_model-free_2018, huang_wide-area_2018}.

Lyapunov exponents (LEs) sit at the intersection of these two trends. As a tool from nonlinear dynamical systems theory, LEs quantify the average exponential rate at which two trajectories that are initially close diverge in the system's phase space. A positive largest Lyapunov exponent (LLE) signals exponential divergence and is the mathematical signature of chaos and instability; a negative LLE signals exponential convergence and confirms asymptotic stability. Because LEs are intrinsically nonlinear metrics-they make no assumption of small perturbations or proximity to an equilibrium-they are naturally suited to the analysis of stressed, post-fault, and large-disturbance behavior. Furthermore, LEs can be estimated directly from time-series data using phase-space reconstruction, enabling a model-free implementation compatible with PMU-based monitoring architectures \cite{dasgupta_pmu-based_2015, rashidi_lyapunov_2016, huang_wide-area_2018}.

The result is a literature that has grown substantially over three decades, branching from early theoretical demonstrations of chaos in power system models \cite{lai_illustrating_1993, srivastava_chaotic_1993, lo_chaotic_1998} to highly practical PMU-based real-time stability monitors \cite{dasgupta_real-time_2013, safavizadeh_framework_2019, pinzon_pmu-based_2019}, renewable generation forecasters \cite{safari_novel_2018, mao_wind_2018, yufei_photovoltaic_2016}, and equipment fault diagnostic systems \cite{qin_partial_2018, yang_chaotic_2020, yang-yang_new_2016}. Despite this breadth, no comprehensive review has yet organized these contributions under a unified framework, compared model-based and data-driven approaches across all application domains, or addressed the most recent shift toward IBR-dominated grids. This paper fills that gap.

\subsection{The Lyapunov Exponent vs. the Lyapunov Function: A Critical Distinction}
\label{subsec:distinction}

A persistent source of confusion in the power systems literature is the use of the term Lyapunov method to refer to two entirely distinct tools, both named after the Russian mathematician Aleksandr Lyapunov but serving very different purposes.

\textbf{Lyapunov functions} are scalar-valued, energy-like functions $V(\mathbf{x}) > 0$ whose time derivative along system trajectories satisfies $\dot{V}(\mathbf{x}) \leq 0$. They are used to \emph{prove} the stability of an equilibrium point without computing trajectories explicitly. In power systems, they form the foundation of direct stability methods and transient energy functions (TEF), and they underpin a large family of Lyapunov-based control design techniques, including backstepping, passivity-based control, and input-to-state stability (ISS) analysis \cite{kazma_stability_2025}. This is a rich and well-established literature, but it is \emph{not} the subject of this review.

\textbf{Lyapunov exponents}, by contrast, are numerical quantities computed from a system's trajectories, or directly from time-series data. The $i$-th Lyapunov exponent $\lambda_i$ measures the time-averaged exponential growth or decay rate of the $i$-th principal perturbation direction in phase space. Geometrically, an infinitesimal ball of nearby initial conditions evolves into a perturbation ellipsoid along the system trajectory; the Lyapunov exponents quantify the average stretching or contraction rates of the ellipsoid's principal axes
. Formally,
\begin{equation}
\lambda_i = \lim_{t \to \infty} \frac{1}{t}   \ln \frac{\|\delta\mathbf{x}_i(t)\|}{\|\delta\mathbf{x}_i(0)\|},
\label{eq:le_def}
\end{equation}
where $\delta\mathbf{x}_i(t)$ is the $i$-th orthonormal perturbation vector evolved under the variational, or linearized, flow. The ordered set $\{\lambda_1 \geq \lambda_2 \geq \cdots \geq \lambda_n\}$ is called the Lyapunov spectrum. Unlike Lyapunov functions, which are used to analytically prove stability, Lyapunov exponents are trajectory-based indicators used to \emph{measure} the degree of divergence, convergence, chaos, or predictability in an observed or simulated dynamic response \cite{wadduwage_power_2013, hayes_viable_2018}.


This review is exclusively concerned with Lyapunov \emph{exponents} in the sense of \eqref{eq:le_def}. Therefore, all references to Lyapunov methods in the following sections refer to LE-based analysis, estimation, and application, not to Lyapunov function-based control design.

\subsection{Historical Evolution}
\label{subsec:history} 

\begin{figure*}[!ht]
\centering
\resizebox{0.98\textwidth}{!}{%
\begin{tikzpicture}[
    timeline/.style={-{Latex[length=3mm]}, thick},
    phase/.style={
        rectangle,
        rounded corners,
        draw=black,
        thick,
        align=center,
        text width=4.0cm,
        minimum height=1.15cm,
        inner sep=6pt,
        font=\small\bfseries
    },
    year/.style={font=\bfseries\small}
]

\def\xA{1.2}
\def\xB{5.9}
\def\xC{10.6}
\def\xD{15.3}

\draw[timeline] (0.2,0) -- (17.0,0);

\foreach \x/\label in {
    \xA/{Early 1990s},
    \xB/{2000s--Early 2010s},
    \xC/{Mid-2010s},
    \xD/{2020s--Present}
}{
    \draw[thick] (\x,0.25) -- (\x,-0.16);
    \node[year, below=0.30cm] at (\x,0) {\label};
}

\node[phase, above=0.25cm] (p1) at (\xA,0) {
    Phase 1:
    Theoretical foundations
};

\node[phase, above=0.25cm] (p2) at (\xB,0) {
    Phase 2: Model-based stability analysis
};

\node[phase, above=0.25cm] (p3) at (\xC,0) {
    Phase 3: PMU/WAMS-based real-time methods
};

\node[phase, above=0.25cm] (p4) at (\xD,0) {
    Phase 4: Machine learning integration and IBR grids
};

\foreach \x/\p in {
    \xA/p1,
    \xB/p2,
    \xC/p3,
    \xD/p4
}{
}

\end{tikzpicture}%
}
\caption{Timeline of Lyapunov exponent applications in power systems.}
\label{fig:le_timeline}
\end{figure*}
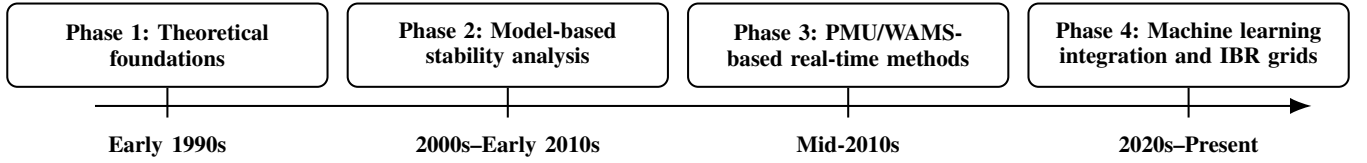

The application of LE theory to power systems can be traced through four broadly identifiable phases. The evolution of LE applications in power systems is summarized in Fig.~\ref{fig:le_timeline}.


\textbf{Phase 1 - Theoretical foundations (early 1990s):} The earliest power system applications of LEs were primarily concerned with confirming that chaotic oscillations were theoretically possible in simplified, single-machine infinite-bus (SMIB) and two-machine models. Studies by Lai
\textit{et al.} \cite{lai_illustrating_1993}, Srivastava and Srivastava \cite{srivastava_chaotic_1993}, and Lo and Zhu \cite{lo_chaotic_1998} used bifurcation diagrams and LLE computation to demonstrate period-doubling cascades and strange attractors in power system ODE models. This phase established the theoretical legitimacy of chaos analysis in power engineering, though the models were too simple for direct operational use.

\textbf{Phase 2 - Model-based stability analysis (2000s--early 2010s):} Researchers moved from confirming chaos to using the LLE as a quantitative stability indicator. Key contributions include the demonstration that positive LLE correlates with transient instability following a fault \cite{liu_detection_1994}, and the systematic study of LLE as a transient stability index on benchmark test systems \cite{wadduwage_power_2013, wadduwage_investigation_2013}. Parallel work on forecasting applied phase-space reconstruction and LLE-based chaos verification to load and wind power time series \cite{choi_daily_1996, mori_short-term_1996, michanos_short-term_2003}.

\textbf{Phase 3 - PMU/WAMS-based real-time methods (mid-2010s):} The availability of high-rate PMU data enabled a critical shift from model-based to data-driven, model-free LE estimation. Dasgupta \textit{et al.} pioneered real-time rotor angle stability and short-term voltage stability monitoring using model-free LLE estimation from PMU measurements \cite{dasgupta_real-time_2013, dasgupta_pmu-based_2015}, an approach subsequently extended to short-term voltage stability \cite{dasgupta_pmu-based_2012, ge_improved_2018, safavizadeh_framework_2019}, coherency detection \cite{khaitan_vantage_2013}, and PMU placement optimization \cite{rashidi_lyapunov_2016}. This phase produced the majority of the stability-monitoring literature reviewed here.

\textbf{Phase 4 - Machine learning integration and IBR grids (2020s--present):} The most recent phase is characterized by two concurrent developments. First, LEs are increasingly combined with machine learning classifiers-decision trees, convolutional neural networks (CNNs), graph neural networks, and physics-informed LSTMs-to create hybrid stability assessors that retain physical interpretability while improving classification accuracy \cite{umbereen_investigating_2024, yao_physics-informed_2025, sayyeda_maximum_2024}. Second, and more fundamentally, the grid is transitioning to IBR dominance, and several recent papers have begun to address how LE-based methods must adapt to the radically different dynamics of converter-controlled resources, where inertia is virtual or absent and synchronization is software-defined \cite{peng_measuring_2026, kazma_stability_2025, kazma_generalizable_2025, wang_transient_2024}.

\subsection{Scope of This Review}
\label{subsec:scope}

\begin{figure*}[!ht]
\centering
\resizebox{\textwidth}{!}{%
\begin{tikzpicture}[
    >=Latex,
    box/.style={
        rectangle,
        rounded corners,
        draw=black,
        thick,
        align=center,
        inner sep=6pt,
        font=\footnotesize
    },
    titlebox/.style={
        box,
        fill=gray!10,
        minimum width=3.0cm,
        minimum height=0.5cm,
        font=\small\bfseries
    },
    branchbox/.style={
        box,
        fill=blue!6,
        minimum width=3.0cm,
        minimum height=0.5cm,
        font=\footnotesize\bfseries
    },
    domainbox/.style={
        box,
        fill=green!6,
        text width=5.2cm,
        minimum height=2.0cm,
        align=left,
        font=\footnotesize
    },
    outbox/.style={
        box,
        fill=red!5,
        text width=4cm,
        minimum height=4.2cm,
        align=left,
        font=\footnotesize
    },
    arrow/.style={thick,->}
]

\node[titlebox] (scope) at (0,0) {Scope of This Review};

\node[branchbox] (included) at (-5.0,-1.4) {Included in Review};
\node[branchbox] (excluded) at (4.5,-1.4) {Excluded from Review};

\draw[arrow] (scope) -- ++(0,-0.6) -| (included);
\draw[arrow] (scope) -- ++(0,-0.6) -|(excluded);

\node[box, fill=blue!4, text width=6.3cm, minimum height=1.1cm] (corpus) at (-5.0,-2.7) {
Applications of \textbf{Lyapunov exponents} in power systems.
\textbf{165+ publications, 1991--2026}
};

\draw[arrow] (included) -- (corpus);

\node[domainbox] (d1) at (-7.0,-4.9) {
\textbf{1. Stability assessment.} 
Section~\ref{sec:stability}\\[1mm]
-- Transient stability\\
-- Short-term voltage stability\\
-- Rotor-angle dynamics\\
-- Coherency detection
};

\node[domainbox] (d2) at (-1.0,-4.9) {
\textbf{2. Chaos analysis and control. }
Section~\ref{sec:chaos}\\[1mm]
-- Chaos detection\\
-- Bifurcation behavior\\
-- LLE-based confirmation\\
-- Chaos suppression control
};

\node[domainbox] (d3) at (-7.0,-7.4) {
\textbf{3. Forecasting and predictability}\\
Section~\ref{sec:forecasting}\\[1mm]
-- Load, wind, PV, and price forecasting\\
-- Chaotic phase-space methods\\
-- LLE-based predictability horizon
};

\node[domainbox] (d4) at (-1.0,-7.4) {
\textbf{4. Protection and condition monitoring}\\
Section~\ref{sec:protection}\\[1mm]
-- Islanding detection\\
-- Fault and partial discharge diagnosis\\
-- Vibration and equipment health
};

\draw[arrow] (corpus.south) -- ++(0,-0.23) -| (d1.north);
\draw[arrow] (corpus.south) -- ++(0,-0.23) -| (d2.north);
\draw[arrow] (d1.south) -- (d3.north);
\draw[arrow] (d2.south) -- (d4.north);

\node[outbox] (outscope) at (4.5,-5.8) {
\textbf{Not covered}\\[1mm]
-- Lyapunov function-based direct stability methods\\
-- Transient energy functions\\
-- Lyapunov-based control synthesis without LE computation\\
-- Pure bifurcation analysis without LE computation\\
-- Converter-only LE studies not tied to power-system behavior\\
-- ML stability methods without an LE component
};

\draw[arrow] (excluded) -- (outscope);

\end{tikzpicture}%
}
\caption{Scope of this review. The paper focuses on Lyapunov exponent applications in power systems and organizes the literature into four main application domains, while excluding related topics that do not explicitly involve LE computation.}
\label{fig:scope_review}
\end{figure*}
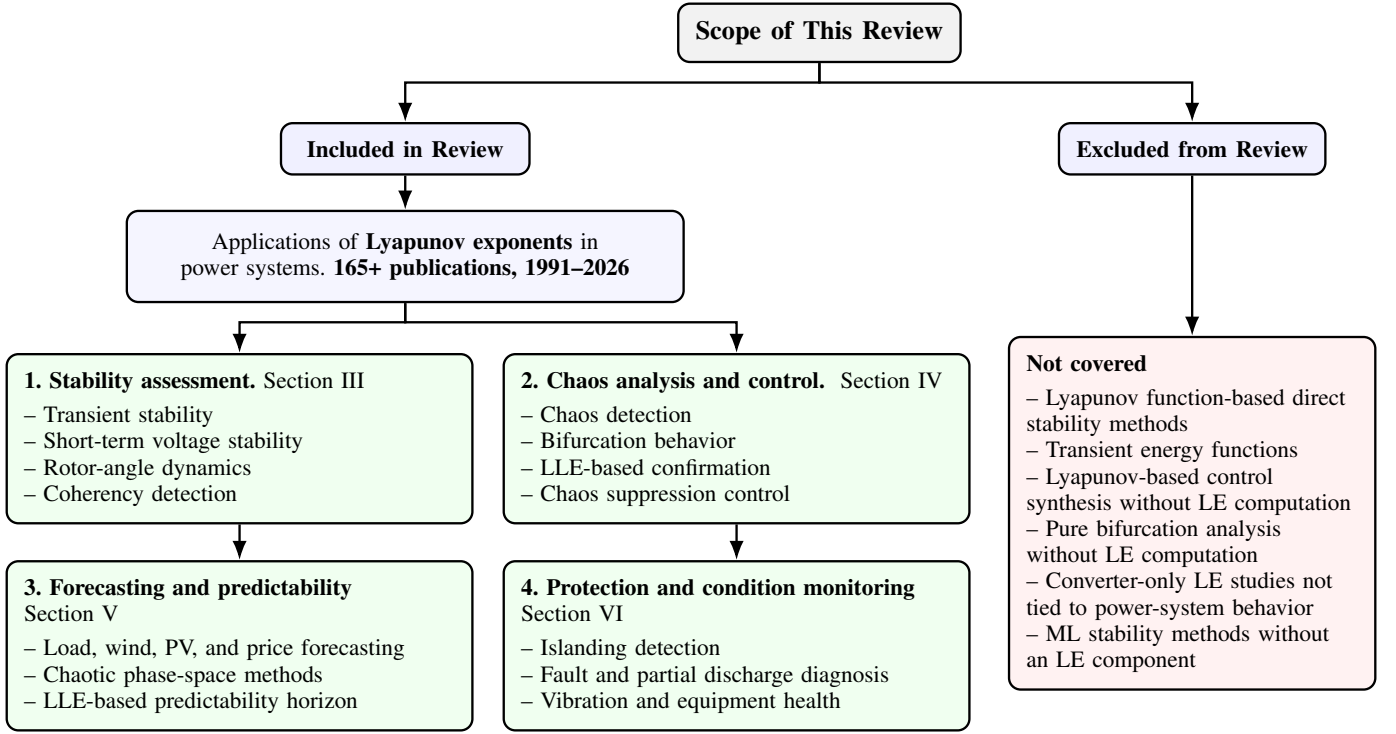

This review focuses on applications of Lyapunov exponents (LEs), in the strict sense of \eqref{eq:le_def}, to power system problems. The reviewed literature was systematically assembled from more than 165 publications spanning 1991--2026. As summarized in Fig.~\ref{fig:scope_review}, the corpus is organized around four major application domains: stability assessment and real-time monitoring, chaos analysis and control, forecasting and predictability analysis, and protection, fault diagnosis, and condition monitoring.

The scope is intentionally restricted to studies in which LE computation plays a direct methodological role. Therefore, related but distinct areas are excluded, including Lyapunov function-based direct stability methods, transient energy functions, Lyapunov-based control synthesis without LE computation, purely bifurcation-based studies without LE analysis, converter-level LE studies not directly connected to power system behavior, and machine-learning stability methods that do not incorporate an LE component. This boundary is used to maintain a clear distinction between LE-based analysis and broader Lyapunov-theoretic or nonlinear-dynamics approaches in power engineering.

\subsection{Four Application Domains at a Glance}
\label{subsec:domains}

To orient the reader before the detailed sections, we briefly characterize each domain.

\textbf{Stability assessment} is by far the largest application domain in the surveyed literature, encompassing approximately 85 publications. The unifying idea is that the LLE computed from the post-fault system trajectory, whether from a model or directly from PMU measurements, provides a quantitative, physically grounded stability indicator. Applications range from critical clearing time (CCT) computation \cite{wadduwage_power_2013, banerjee_lyapunov_2014} and structural stability mapping \cite{wadduwage_power_2013} to real-time monitoring from streaming PMU data \cite{dasgupta_real-time_2013, pinzon_pmu-based_2019,
almomani_novel_2025}.

\textbf{Chaos analysis and control} (approximately 35 publications) addresses the observation, well established since the 1990s, that power system models can exhibit chaotic oscillations under certain loading conditions, parameter combinations, or contingencies. LEs serve as the primary diagnostic: a positive LLE confirms chaos, while tracking how LLE varies with a control parameter reveals bifurcation boundaries. The control literature in this domain focuses on restoring negative LLE through FACTS devices \cite{srivastava_elimination_1998}, power system stabilizers \cite{li_chaotic_2009}, fractional-order controllers \cite{min_fractional-order_2017, nangrani_fractional_2018}, and synergetic control \cite{al-hussein_chaos_2021}.

\textbf{Forecasting and predictability} (42 publications) is the second-largest domain. The common workflow is: (1) verify that the time series of interest (load, wind speed, PV output, electricity price) is chaotic by computing a positive LLE; (2) reconstruct the attractor using Takens embedding; and (3) apply a local-linear or machine-learning predictor in the reconstructed phase space. The predictability horizon itself can be estimated from the LLE via $T_{\max} = \lambda_1^{-1} \ln (\Delta/\delta_0)$, where $\Delta$ and $\delta_0$ are the acceptable and initial prediction errors, respectively \cite{mohamed_variability_2007}. This provides a rigorous upper bound on how far ahead any predictor can remain accurate-a meta-forecasting result independent of the model used.

\textbf{Protection and fault diagnosis} (14 publications) is the most recent application domain. These papers exploit the observation that fault conditions, equipment degradation, or topology changes alter the chaotic properties of a measurable signal (voltage waveform, vibration signal, partial discharge pulse train). The LLE and related chaos metrics (correlation dimension, approximate entropy) are extracted as features for classification. Applications include microgrid islanding detection \cite{yang-yang_new_2016, amini_hybrid_2025}, partial discharge type identification in transformers \cite{qin_partial_2018}, and vibration-based fault diagnosis in high-voltage circuit breakers and turbines \cite{yang_vibration_2019, yang_chaotic_2020}.

\subsection{Cross-Cutting Themes}
\label{subsec:themes}

Several analytical distinctions recur across all four domains and are explicitly addressed in every section of this review.

\textbf{Model-based vs.\ data-driven LE estimation.} Model-based approaches compute LEs from the variational equations of an analytical power system model \cite{wadduwage_power_2013, kazma_stability_2025}. They provide theoretical rigor and can be used for offline stability mapping, but require a validated model and are computationally intensive for large systems. Data-driven (model-free) approaches estimate LEs directly from measured time-series data using phase-space reconstruction \cite{hayes_viable_2018, wei_model-free_2018}. They are compatible with PMU-based monitoring but are sensitive to noise, window length, and embedding parameters.

\textbf{Largest LE vs.\ full Lyapunov spectrum.} Most power system applications use only the LLE ($\lambda_1$) because its sign determines stability and instability. The full LE spectrum $\{\lambda_1, \lambda_2, \ldots, \lambda_n\}$ provides additional information-the Kaplan--Yorke dimension, the metric entropy, and information about the attractor's structure-but is computationally more demanding and has been exploited in only a small fraction of the reviewed papers \cite{kazma_stability_2025, aghtar_lyapunov_2023}.

\textbf{Conventional SG grid vs.\ IBR-dominated grid.} Nearly all papers before 2020 analyze grids dominated by synchronous generators, for which the swing equation provides a well-understood nonlinear model. IBR dynamics are governed by inner control loops operating on microsecond timescales, producing qualitatively different phase-space behavior. Adapting LE methods to IBR-dominated grids is identified throughout this review as one of the most important open problems \cite{peng_measuring_2025, peng_measuring_2026, kazma_generalizable_2025}.

\subsection{Contributions of This Review}
\label{subsec:contributions}
Relative to existing surveys of power system stability and nonlinear dynamics, this review makes the following original contributions:

\begin{enumerate}
  \item \textbf{Unified framework across four domains.} To the best of our knowledge, this is the first review to systematically cover stability assessment, chaos control, forecasting, and condition monitoring under a single LE-based framework, revealing cross-domain methodological parallels that are not visible when each application is surveyed in isolation.

  \item \textbf{Explicit model-based vs.\ data-driven comparison.} We explicitly compare LE estimation pipelines across both paradigms in each application domain, identifying the trade-offs in accuracy, computational cost, noise sensitivity, and real-time deployability.

  \item \textbf{Coverage of modern grid contexts.} The review includes the most recent literature on IBR-dominated grids, renewable integration uncertainty \cite{kazma_stability_2025}, and hybrid ML-LE assessment methods \cite{ umbereen_investigating_2024}, contexts that existing surveys do not address.

  \item \textbf{Critical gap analysis.} Each section concludes with an explicit identification of open problems, methodological limitations, and under-explored directions, providing a research agenda for the community.

\end{enumerate}

\subsection{Paper Organization}
\label{subsec:organization}

The remainder of this paper is organized as follows. Section \ref{sec:foundations} reviews the mathematical foundations of Lyapunov exponent theory and the principal estimation algorithms used in the power systems literature, including phase-space reconstruction, the Wolf, Kantz, and Rosenstein methods, and the discrete-QR approach for model-based spectrum computation. Section \ref{sec:stability} covers stability assessment and real-time monitoring, tracing the evolution from offline model-based CCT analysis to PMU-based real-time voltage and angle stability monitors. Section \ref{sec:chaos} addresses chaos analysis, bifurcation characterization, and LE-guided chaos control. Section \ref{sec:forecasting} reviews the forecasting and predictability literature. Section \ref{sec:protection} covers protection, fault diagnosis, and condition monitoring. Section \ref{sec:emerging} discusses emerging topics and cross-cutting themes, including IBR grids, ML-LE hybrids, and Koopman operator connections. Section \ref{sec:conclusion} presents conclusions and directions for future research.

%
%

%
\section{Foundations of Chaos Theory and Lyapunov Exponents}
\label{sec:foundations}

This section establishes the mathematical foundations required for all subsequent sections. We begin with power system dynamic models that exhibit nonlinear and chaotic behavior (Section \ref{subsec:ps_models}), introduce chaos theory concepts in the power-system context(Section \ref{subsec:chaos_concepts}), provide a rigorous definition of the Lyapunov exponent spectrum (Section \ref{subsec:le_def}), describe the principal model-based estimation algorithms (Section \ref{subsec:model_based}), present the phase-space reconstruction framework that enables data-driven estimation (Section \ref{subsec:psr}), describe the main data-driven estimation algorithms (Section \ref{subsec:data_driven}), discuss the predictability horizon derived from the LLE (Section \ref{subsec:predictability}), and close with a comparative summary and practical guidelines (Section \ref{subsec:algo_comparison}).

\subsection{Power System Dynamic Models}
\label{subsec:ps_models}

Nonlinear and chaotic phenomena in power systems arise from the interplay of electromagnetic, mechanical, and control dynamics. The mathematical models used across the reviewed literature fall into three broad categories.

\subsubsection{The Swing Equation and SMIB Model}

The single-machine infinite-bus (SMIB) model is the most widely used test bed for nonlinear and chaotic analysis \cite{qin_random_2014}. The machine rotor dynamics are governed by the swing equation:
\begin{equation}
M\ddot{\delta} + D\dot{\delta} + P_{\max}\sin\delta = P_m,
\label{eq:swing}
\end{equation}
where $\delta$ is the rotor angle, $\omega = \dot{\delta}$ is the speed deviation, $M$ is the inertia constant, $D$ is the damping coefficient, $P_{\max}$ is the maximum electrical power, and $P_m$ is the mechanical input. Written as a first-order system with state vector $\mathbf{x} = [\delta,\, \omega]^T$:
\begin{equation}
\dot{\mathbf{x}} = \mathbf{f}(\mathbf{x}) = \begin{bmatrix} \omega \\   \dfrac{1}{M}\bigl(P_m - P_{\max}\sin\delta - D\omega\bigr)
  \end{bmatrix}.
\label{eq:smib_ode}
\end{equation}
Under periodic external forcing (e.g., oscillating $P_m = \bar{P}_m + A\sin(\Omega t)$), \eqref{eq:swing} becomes a driven nonlinear oscillator capable of exhibiting period-doubling bifurcations and chaos for certain parameter combinations \cite{qin_random_2014}. The addition of random parameter perturbations, such as stochastic damping $D = \bar{D} + \rho\xi(t)$ where $\xi(t)$ is a noise process, can further induce or amplify chaos \cite{qin_random_2014}.

\subsubsection{Three-Bus Test System with Induction Motor Load}

In addition to the classical SMIB model, several studies have employed a three-bus test system to investigate nonlinear and chaotic dynamics in power systems \cite{lai_illustrating_1993, srivastava_chaotic_1993}. This configuration extends the SMIB framework by incorporating a dynamic load model, thereby capturing more realistic interactions between generation and load dynamics.

The system consists of a synchronous generator connected to an infinite bus through a transmission network, with a load bus interfacing both. The load is modeled as a composite load that includes a simplified induction motor in parallel with a constant power (P--Q) load and a constant impedance component.

The synchronous generator dynamics are still described by the swing equation in \eqref{eq:swing}. However, unlike the classical SMIB case, the electrical power output depends on the network interconnection and load dynamics, introducing additional nonlinear coupling. The active and reactive power demand at the load bus are expressed as:
\begin{align}
P_d &= P_0 + P_1 + K_{p\omega}\dot{\delta} + K_{pv}\bigl(V + T\dot{V}\bigr), \\ Q_d &= Q_0 + Q_1 + K_{q\omega}\dot{\delta} + K_{qv}V + K_{qv2}V^2,
\end{align}
where $V$ is the load bus voltage magnitude, and the coefficients $K_{p\omega}, K_{pv}, K_{q\omega}, K_{qv}, K_{qv2}$ characterize the sensitivity of the load to frequency and voltage variations. The term $T\dot{V}$ represents the dynamic voltage response of the load.

Several papers have extended the analysis to include voltage-sensitive and frequency-sensitive load models, on-load tap changer (OLTC) dynamics, and static var system (SVS) control loops \cite{srivastava_chaotic_1993}. These additions enrich the state space and reveal more complex routes to instability, including quasiperiodic attractors and boundary crises, which are absent in the basic SMIB model. Fractional-order generalizations of these models, based on Caputo or conformable derivatives, have also been studied, and their LE analysis is discussed in Section \ref{sec:chaos} \cite{yan_dynamics_2021, chen_analysis_2024}.

\subsubsection{Multi-Machine and DAE Models}

For multi-machine systems, the state equations are extended to include all generator rotor dynamics, excitation systems, governor models, and network algebraic constraints. The general differential-algebraic equation (DAE) form is:
\begin{align}
\dot{\mathbf{x}}_d &= \mathbf{f}(\mathbf{x}_d,\,                       \mathbf{x}_a,\, \mathbf{u}), \label{eq:dae_diff}\\
\mathbf{0} &= \mathbf{g}(\mathbf{x}_d,\, \mathbf{x}_a), \label{eq:dae_alg}
\end{align}
where $\mathbf{x}_d \in \mathbb{R}^{n_d}$ are the differential states (rotor angles, speeds, flux linkages, controller states), $\mathbf{x}_a \in \mathbb{R}^{n_a}$ are the algebraic states (bus voltages and angles satisfying the power flow equations), and $\mathbf{u}$ represents control inputs or disturbances. This DAE formulation is used in the most rigorous LE-based analyses \cite{wadduwage_power_2013, kazma_stability_2025,lo_chaotic_1998, hayes_viable_2018} and provides the basis for model-based LE computation discussed in Section \ref{subsec:model_based}. When the algebraic constraints are locally solvable (i.e., the Jacobian $\partial \mathbf{g}/\partial \mathbf{x}_a$ is invertible), the DAE can be reduced to an ODE via the implicit function theorem, yielding an equivalent system:
\begin{equation}
\dot{\mathbf{x}}_d = \mathbf{F}(\mathbf{x}_d, \mathbf{u}),
\label{eq:reduced_ode}
\end{equation}
on which standard variational analysis and LE computation can be performed \cite{kazma_stability_2025}. \textcolor{black}{However, LE-based characterization of ill-conditioned DAE regimes, particularly near loadability-induced voltage-collapse points where $\partial \mathbf{g}/\partial \mathbf{x}_a$ becomes singular or nearly singular, remains largely unexplored.}

\subsection{Chaos Theory Fundamentals in the Power System
Context}
\label{subsec:chaos_concepts}

\subsubsection{Definition and Signatures of Chaos}

A deterministic dynamical system governed by \eqref{eq:smib_ode} or \eqref{eq:reduced_ode} is said to exhibit \emph{chaos} if its trajectories are (i) bounded, (ii) aperiodic, and (iii) sensitively dependent on initial conditions-that is, two trajectories starting arbitrarily close together diverge exponentially over time. The sensitivity to initial conditions is quantified by the Lyapunov exponents (Section \ref{subsec:le_def}); a positive largest LE is the primary mathematical signature of chaos \cite{li_chaotic_2009, nangrani_trends_2020}.

A canonical early demonstration on the IEEE 14-bus system is given in \cite{li_chaotic_2009}, where standard transient-stability differential-algebraic simulation tools were used to reveal Hopf bifurcation, period-doubling cascades, and a strange attractor-showing that complex nonlinear phenomena arise in realistic models without requiring simplified assumptions.

\subsubsection{Routes to Chaos}

Three main routes to chaos are documented in power systems \cite{srivastava_chaotic_1993, yibei_research_2011}:

\begin{enumerate}
  \item \textbf{Period-doubling (Feigenbaum) route:} As a parameter (e.g., load level, forcing amplitude) is varied, the system undergoes a cascade of bifurcations in which the period of oscillation doubles repeatedly: $T \to 2T \to 4T \to 8T \to \cdots$, converging at a finite parameter value to a chaotic orbit. This is the most commonly observed route in power system models \cite{lo_chaotic_1998, li_chaotic_2009}.

  \item \textbf{Quasiperiodic route:} A sequence of Hopf bifurcations generates oscillations at incommensurate frequencies. The resulting quasiperiodic trajectory lies on a torus, which subsequently breaks down into chaos through a torus-collision or torus-doubling mechanism.

  \item \textbf{Intermittency route:} Alternating intervals of nearly periodic behavior and chaotic bursts, with the average burst frequency increasing toward a bifurcation point.
\end{enumerate}

The \emph{Lyapunov exponent spectrum} distinguishes all these regimes quantitatively (Table \ref{tab:attractor_types}), a point emphasized in \cite{srivastava_chaotic_1993}.

\begin{table}[!t]
  \caption{Attractor Classification via the Lyapunov Spectrum for an $n$-Dimensional Continuous-Time System (ordered $\lambda_1 \geq \lambda_2 \geq \cdots \geq \lambda_n$)}
  \label{tab:attractor_types}
  \centering
  \begin{tabular}{ll}
    \toprule
    \textbf{Attractor type} & \textbf{LE signature}\\
    \midrule
    Stable fixed point  & All $\lambda_i < 0$\\
    periodic motion         & $\lambda_1 = 0$, rest $< 0$ \\
    Quasiperiodic motion & $\lambda_{1,2,3} = 0$, rest $< 0$\\
    Chaos               & $\lambda_1 > 0, \lambda_2= 0$, $\lambda_n < 0$ \\
    Hyperchaos (Superchaos)         & $\lambda_1 > \lambda_2 > 0, \lambda_3 = 0$, rest $< 0$\\
    \bottomrule
  \end{tabular}
\end{table}

\subsubsection{Chaos vs.\ Voltage Collapse}

An important distinction, relevant to Section \ref{sec:stability}, is between chaos and voltage collapse. Chaotic oscillations are \emph{bounded}: the trajectory wanders on a strange attractor but does not diverge to infinity. Voltage collapse, by contrast, is an \emph{unbounded} divergence of the trajectory away from the feasibility region. In some parameter regimes, the two phenomena are connected: a chaotic attractor can undergo a boundary crisis, in which it collides with an unstable manifold and suddenly disappears, after which the trajectory collapses \cite{srivastava_chaotic_1993}. This crisis mechanism provides a nonlinear-dynamics explanation of certain voltage-collapse events and motivates the use of LEs as early-warning indicators.

\subsection{Definition of the Lyapunov Exponent Spectrum}
\label{subsec:le_def}

\subsubsection{Variational Equation and Perturbation Ellipsoid}

Consider the autonomous ODE \eqref{eq:reduced_ode} with a reference trajectory $\mathbf{x}^*(t)$ starting from $\mathbf{x}^*(0) = \mathbf{x}_0$. A perturbed trajectory $\mathbf{x}^*(t) + \delta\mathbf{x}(t)$ satisfies, to first order, the \emph{variational equation}:
 
\begin{equation}
\delta\dot{\mathbf{x}}(t) = \mathbf{J}\bigl(\mathbf{x}^*(t)\bigr)\,\delta\mathbf{x}(t),
\label{eq:variational}
\end{equation}
 
where $\mathbf{J}(\mathbf{x}) = \partial\mathbf{f}/\partial\mathbf{x}$ is the Jacobian of the vector field. The solution is expressed through the \emph{state-transition matrix} (or fundamental solution matrix) $\boldsymbol{\Phi}(t,0)$:
 
\begin{equation}
\delta\mathbf{x}(t) = \boldsymbol{\Phi}(t,0)\,\delta\mathbf{x}(0), \qquad \dot{\boldsymbol{\Phi}}(t,0) = \mathbf{J}(\mathbf{x}^*(t))\, \boldsymbol{\Phi}(t,0).
\label{eq:stm}
\end{equation}

\begin{figure}[!t]
\centering
\resizebox{\linewidth}{!}{
\begin{tikzpicture}[
    scale=0.95,
    axis/.style={-{Latex[length=2.2mm]}, thick},
    manifold/.style={gray!55, thin},
    traj/.style={blue!70!black, very thick},
    pert/.style={gray!70, thin},
    vec/.style={-{Latex[length=2mm]}, thick, black},
    lab/.style={font=\small},
    smalllab/.style={font=\scriptsize},
    propbox/.style={
        rectangle,
        rounded corners,
        draw=black,
        thick,
        fill=gray!8,
        align=left,
        text width=4.0cm,
        inner sep=6pt,
        font=\footnotesize
    }
]

\coordinate (O) at (0,0);
\draw[axis] (O) -- (0,2.25) node[above] {$x_2$};
\draw[axis] (O) -- (-1.15,-0.85) node[left] {$x_1$};
\draw[axis] (O) -- (1.45,-0.38) node[right] {$x_3$};

\coordinate (c0) at (0.95,0.22);
\draw[gray!70, thick] (c0) ellipse [x radius=0.34, y radius=0.34];
\draw[gray!50, dashed] ($(c0)+(-0.34,0)$) -- ($(c0)+(0.34,0)$);
\draw[gray!50, dashed] ($(c0)+(0,-0.34)$) -- ($(c0)+(0,0.34)$);

\coordinate (ct) at (6.55,1.30);
\draw[gray!65, thick] (ct) ellipse [x radius=0.42, y radius=1.15];
\draw[gray!50, dashed] ($(ct)+(0,-1.15)$) -- ($(ct)+(0,1.15)$);
\draw[gray!50, dashed] ($(ct)+(-0.42,0)$) -- ($(ct)+(0.42,0)$);

\draw[traj] (c0) .. controls (2.0,0.10) and (2.9,0.35) .. (3.85,0.62)
             .. controls (4.8,0.88) and (5.75,1.02) .. (ct);

\draw[traj,-{Latex[length=2mm]}] (2.85,0.37) -- (3.0,0.4);
\draw[traj,-{Latex[length=2mm]}] (5.35,0.99) -- (5.55,1.03);

\coordinate (p0) at (0.98,0.55);
\coordinate (pt) at (6.55,2.50);
\draw[pert] (p0) .. controls (1.9,0.55) and (2.8,0.90) .. (3.8,1.25)
            .. controls (4.8,1.55) and (5.8,1.75) .. (pt);

\draw[pert,-{Latex[length=1.8mm]}] (5.0,1.60) -- (5.1,1.62);

\draw[vec] (c0) -- (p0) node[midway,left,smalllab] {$\delta x_0$};
\draw[vec] (ct) -- (pt) node[midway,right,smalllab] {$\delta x(t)$};

\node[lab, blue!70!black] at (4.65,0.5) {$\phi(x_0,t)$};
\node[smalllab, gray!30!black] at (3.85,1.85) {$\phi(x_0+\delta x_0,t)$};
\node[smalllab] at (0.85,0.05) {$x_0$};

\node[font=\large] at (4.2,-1.10)
{$\|\delta x(t)\| \approx e^{\lambda_i t}\,\|\delta x(0)\|$};

\node[propbox] (props) at (9.8,0.9) {%
\textbf{Key properties of Lyapunov exponents}\\[1mm]
\textbf{Stability / chaos:}\\
If all $\lambda_i<0$, perturbations decay; if $\lambda_1>0$, nearby trajectories separate exponentially.\\[1mm]
\textbf{Dissipation:}\\
$\sum_{i=1}^{n}\lambda_i<0$ implies net phase-space volume contraction.\\[1mm]
\textbf{Fractal dimension:}\\
$d_{KY}=j+\dfrac{\sum_{i=1}^{j}\lambda_i}{|\lambda_{j+1}|}$.\\[1mm]
\textbf{Invariance:}\\
LEs are unchanged under smooth coordinate transformations.
};

\draw[-{Latex[length=2mm]}, thick, gray!70] (7.0,1.0) -- (7.5,1.0);

\end{tikzpicture}}
\caption{Schematic illustration of divergence of nearby trajectories and the geometric interpretation of Lyapunov exponents. A small initial perturbation sphere evolves into an ellipsoid under the linearized flow. The singular stretching rates determine the Lyapunov spectrum.}
\label{fig:lyapunov_exponent_schematic}
\end{figure}
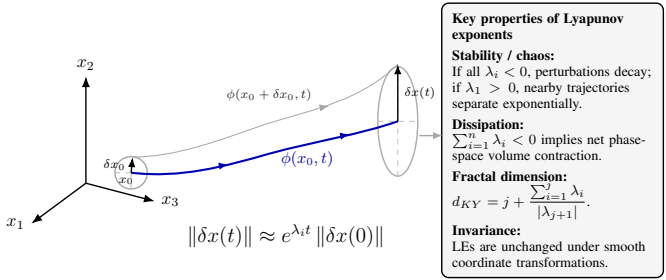

An initial $n$-dimensional unit sphere of perturbations $\delta\mathbf{x}(0)$ is mapped by $\boldsymbol{\Phi}(t,0)$ into an ellipsoid whose semi-axes grow (or shrink) exponentially at distinct rates, as illustrated in Fig.~\ref{fig:lyapunov_exponent_schematic}. The $i$-th Lyapunov exponent $\lambda_i$ is defined as the asymptotic growth rate of the $i$-th principal semi-axis \cite{wadduwage_power_2013,
kazma_stability_2025}:
 
\begin{equation}
\lambda_i = \lim_{t\to\infty}   \frac{1}{t}\ln\sigma_i\bigl(\boldsymbol{\Phi}(t,0)\bigr),   \quad i = 1, 2, \ldots, n,
\label{eq:le_sigma}
\end{equation}
where $\sigma_i(\cdot)$ denotes the $i$-th singular value of the transition matrix, ordered so that $\lambda_1 \geq \lambda_2 \geq \cdots \geq \lambda_n$.

\subsubsection{Multiplicative Ergodic Theorem}

The existence and trajectory-independence (almost sure, with respect to an ergodic invariant measure) of the limits in \eqref{eq:le_sigma} is guaranteed by the \emph{Oseledets Multiplicative Ergodic Theorem} \cite{kazma_stability_2025}. This theorem ensures that for ergodic dynamical systems, the LE spectrum $\{\lambda_i\}$ is an intrinsic property of the attractor, not of the initial condition or the specific realization, provided the system is sufficiently ergodic.

\subsubsection{Key Properties}

Several properties of LEs are directly exploited in the power systems applications reviewed in this paper:

\begin{enumerate}
  \item \textbf{Stability criterion:} For a $C^1$-differentiable flow, all $\lambda_i < 0$ implies asymptotic Lyapunov stability of the equilibrium \cite{wadduwage_power_2013}. Conversely, $\lambda_1 > 0$ implies chaos.

  \item \textbf{Liouville's theorem:} The sum of all Lyapunov exponents equals the time-averaged divergence of the vector field along the trajectory: $\sum_{i=1}^n \lambda_i = \langle \nabla \cdot \mathbf{f} \rangle_t$. For dissipative power system models, $\sum \lambda_i < 0$ \cite{kazma_stability_2025}.

  \item \textbf{Kaplan--Yorke dimension:} The fractal dimension of the attractor can be estimated from the LE spectrum as $d_{KY} = j + \sum_{i=1}^{j}\lambda_i/|\lambda_{j+1}|$, where $j$ is the largest index such that $\sum_{i=1}^{j}\lambda_i \geq 0$.

  \item \textbf{Invariance:} LEs are invariant under smooth coordinate transformations. The stability region and structural stability analysis in        \cite{wadduwage_power_2013} exploit this invariance to map stability boundaries across different fault scenarios and operating points.
\end{enumerate}

\subsubsection{Finite-Time Lyapunov Exponents}
\label{subsec:ftle}

The asymptotic definition \eqref{eq:le_sigma} is not directly applicable in power system monitoring, where decisions must be made within a few seconds after a disturbance. The \emph{finite-time Lyapunov exponent} (FTLE) is defined over a window of length $T$:
\begin{equation}
\Lambda_i(t_0, T) = \frac{1}{T}   \ln\sigma_i\bigl(\boldsymbol{\Phi}(t_0{+}T,\, t_0)\bigr).
\label{eq:ftle}
\end{equation}
The FTLE $\Lambda_1(t_0,T)$ converges to $\lambda_1$ as $T\to\infty$ under ergodicity but can differ substantially for short windows, introducing a bias-variance trade-off: shorter windows improve detection speed but increase sensitivity to noise and transient oscillation content. The influence of window length on FTLE reliability has been studied in \cite{guo_lyapunov_2014, said_real-time_2024, niu_transient_2024}, and adaptive window selection is an active research direction (Section \ref{sec:stability}).

\subsection{Model-Based LE Computation}
\label{subsec:model_based}

When a validated mathematical model is available, LEs can be computed directly from the variational equation \eqref{eq:variational} and the state-transition matrix \eqref{eq:stm}.

\subsubsection{Gram-Schmidt Re-Orthonormalization (GSR)}

The most widely used model-based algorithm is the \emph{Gram-Schmidt re-orthonormalization} (GSR) method, attributed to Benettin \textit{et al.} and used in \cite{wadduwage_power_2013, yan_pmu-based_2012, bosetti_transient_2018}. The key difficulty in computing the full LE spectrum from \eqref{eq:stm} directly is that all columns of $\boldsymbol{\Phi}(t,0)$ tend to align with the direction of fastest growth ($\lambda_1$) as $t \to \infty$, causing severe ill-conditioning (collapse of the perturbation ellipsoid onto its largest axis).

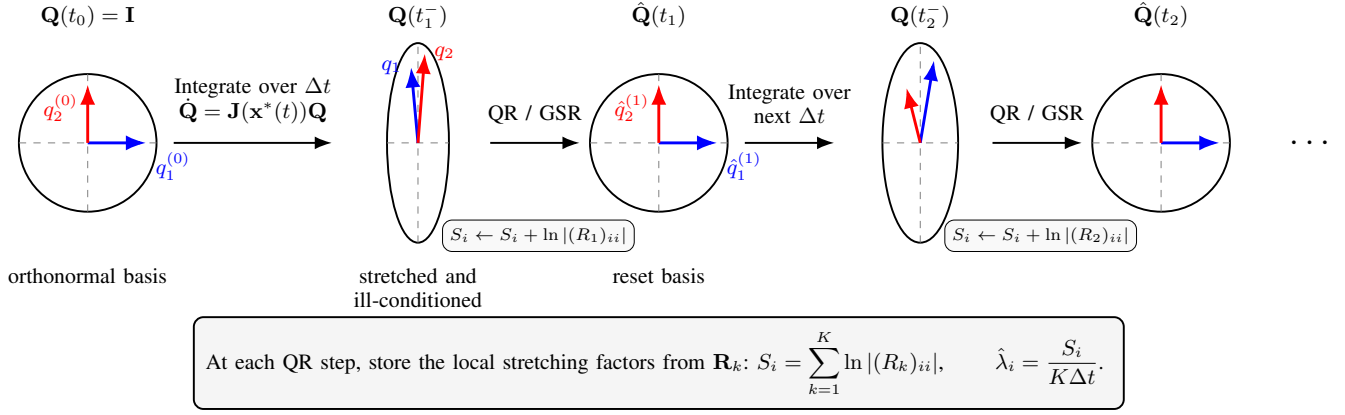
\begin{figure*}[!t]
\centering
\resizebox{0.98\textwidth}{!}{%
\begin{tikzpicture}[
    >=Latex,
    font=\small,
    flow/.style={-{Latex[length=2.5mm]}, thick},
    smallnote/.style={
        rectangle,
        rounded corners,
        draw=black,
        thin,
        align=center,
        fill=gray!5,
        inner sep=3pt,
        font=\scriptsize
    },
    note/.style={
        rectangle,
        rounded corners,
        draw=black,
        thick,
        align=center,
        fill=gray!8,
        inner sep=5pt
    }
]

\coordinate (A) at (-1,0);
\coordinate (B) at (3.8,0);
\coordinate (C) at (7.3,0);
\coordinate (D) at (11.1,0);
\coordinate (E) at (14.6,0);

\draw[thick] (A) circle (1.0);
\draw[dashed, gray] ($(A)+(-1.0,0)$) -- ($(A)+(1.0,0)$);
\draw[dashed, gray] ($(A)+(0,-1.0)$) -- ($(A)+(0,1.0)$);

\draw[->, very thick, blue] (A) -- ($(A)+(0.85,0)$) node[below right] {$q_1^{(0)}$};
\draw[->, very thick, red]  (A) -- ($(A)+(0,0.85)$) node[below left] {$q_2^{(0)}$};

\node[above=1.55cm] at (A) {$\mathbf{Q}(t_0)=\mathbf{I}$};
\node[below=1.70cm] at (A) {orthonormal basis};

\draw[flow] ($(A)+(1.25,0)$) -- ($(B)+(-1.25,0)$)
node[midway, above=0.15cm, align=center]
{Integrate over $\Delta t$\\
$\dot{\mathbf{Q}}=\mathbf{J}(\mathbf{x}^*(t))\mathbf{Q}$};

\draw[thick] (B) ellipse [x radius=0.45, y radius=1.45];
\draw[dashed, gray] ($(B)+(0,-1.45)$) -- ($(B)+(0,1.45)$);
\draw[dashed, gray] ($(B)+(-0.45,0)$) -- ($(B)+(0.45,0)$);

\draw[->, very thick, blue] (B) -- ($(B)+(- 0.1,1.1)$) node[ left] {$q_1$};
\draw[->, very thick, red]  (B) -- ($(B)+(0.1,1.3)$) node[ right] {$q_2$};

\node[above=1.55 cm] at (B) {$\mathbf{Q}(t_1^-)$};
\node[below=1.70cm, align=center] at (B) {stretched and\\ill-conditioned};

\draw[flow] ($(B)+(1.05,0)$) -- ($(C)+(-1.15,0)$)
node[midway, above=0.15cm] {QR / GSR};

\node[smallnote] at (5.55,-1.35)
{$S_i \leftarrow S_i+\ln |(R_1)_{ii}|$};

\draw[thick] (C) circle (1.0);
\draw[dashed, gray] ($(C)+(-1.0,0)$) -- ($(C)+(1.0,0)$);
\draw[dashed, gray] ($(C)+(0,-1.0)$) -- ($(C)+(0,1.0)$);

\draw[->, very thick, blue] (C) -- ($(C)+(0.85,0)$) node[below right] {$\hat{q}_1^{(1)}$};
\draw[->, very thick, red]  (C) -- ($(C)+(0,0.85)$) node[below left] {$\hat{q}_2^{(1)}$};

\node[above=1.55 cm] at (C) {$\hat{\mathbf{Q}}(t_1)$};
\node[below=1.70cm] at (C) {reset basis};

\draw[flow] ($(C)+(1.25,0)$) -- ($(D)+(-1.25,0)$)
node[midway, above=0.15cm, align=center]
{Integrate over\\next $\Delta t$};

\draw[thick] (D) ellipse [x radius=0.55, y radius=1.45];
\draw[dashed, gray] ($(D)+(0,-1.45)$) -- ($(D)+(0,1.45)$);
\draw[dashed, gray] ($(D)+(-0.55,0)$) -- ($(D)+(0.55,0)$);

\draw[->, very thick, blue] (D) -- ($(D)+(0.20,1.20)$);
\draw[->, very thick, red]  (D) -- ($(D)+(-0.2,0.8)$);

\node[above=1.55cm] at (D) {$\mathbf{Q}(t_2^-)$};

\draw[flow] ($(D)+(1.05,0)$) -- ($(E)+(-1.15,0)$)
node[midway, above=0.15cm] {QR / GSR};

\node[smallnote] at (12.85,-1.35)
{$S_i \leftarrow S_i+\ln |(R_2)_{ii}|$};

\draw[thick] (E) circle (1.0);
\draw[dashed, gray] ($(E)+(-1.0,0)$) -- ($(E)+(1.0,0)$);
\draw[dashed, gray] ($(E)+(0,-1.0)$) -- ($(E)+(0,1.0)$);

\draw[->, very thick, blue] (E) -- ($(E)+(0.85,0)$);
\draw[->, very thick, red]  (E) -- ($(E)+(0,0.85)$);

\node[above=1.55cm] at (E) {$\hat{\mathbf{Q}}(t_2)$};

\node[font=\Large] at (16.8,0) {$\cdots$};

\node[note, minimum width=8.5cm] at (7.3,-3.2)
{
At each QR step, store the local stretching factors from $\mathbf{R}_k$: 
$\displaystyle S_i=\sum_{k=1}^{K}\ln |(R_k)_{ii}|$,
\qquad
$\displaystyle \hat{\lambda}_i=\frac{S_i}{K\Delta t}$.
};

\end{tikzpicture}%
}
\caption{Schematic illustration of Gram--Schmidt re-orthonormalization for Lyapunov exponent computation. An orthonormal perturbation basis is propagated over each interval, deforms into an ellipsoid, and is then replaced by a new orthonormal basis through QR decomposition. The logarithms of the diagonal entries of $\mathbf{R}_k$ are accumulated at each re-orthonormalization step to estimate the Lyapunov exponents.}
\label{fig:gsr_circle_ellipse}
\end{figure*}

The GSR algorithm resolves this by periodically re-orthonormalizing the columns of a matrix of perturbation vectors $\mathbf{Q}(t) \in \mathbb{R}^{n \times n}$ as explianed in Figure \ref{fig:gsr_circle_ellipse}:
\begin{enumerate}
  \item Initialize $\mathbf{Q}(0) = \mathbf{I}_n$.
  \item Integrate the $n$ variational equations        simultaneously:        $\dot{\mathbf{Q}}(t) = \mathbf{J}(\mathbf{x}^*(t))\,        \mathbf{Q}(t)$, over one renormalization interval        $[t_k,\, t_{k+1}]$.
  \item Perform QR decomposition:        $\mathbf{Q}(t_{k+1}) = \hat{\mathbf{Q}}_{k+1}        \mathbf{R}_{k+1}$, where $\hat{\mathbf{Q}}_{k+1}$        has orthonormal columns and $\mathbf{R}_{k+1}$ is        upper triangular.
  \item Accumulate the logarithms of the diagonal entries:        $S_i = S_i + \ln |(\mathbf{R}_{k+1})_{ii}|$.
  \item Set $\mathbf{Q}(t_{k+1}) \leftarrow        \hat{\mathbf{Q}}_{k+1}$ and repeat.
\end{enumerate}
After $K$ intervals, the $i$-th LE estimate is:
$\hat{\lambda}_i = S_i / (K \Delta t)$.

\subsubsection{Discrete-QR Method for DAE Systems}

For DAE power system models \eqref{eq:dae_diff}--\eqref{eq:dae_alg}, Hayes and Milano \cite{hayes_viable_2018} proposed an efficient discrete-QR method that reuses the Jacobian matrices already factorized during implicit integration. Using the trapezoidal or backward-Euler scheme with step $\Delta t$, the transition matrix is approximated as a product of inverse-Jacobian factors:
\begin{equation}
\boldsymbol{\Phi}(k\Delta t,\, 0) \approx \prod_{i=1}^{k} \bigl[\boldsymbol{\Gamma} - \Delta t\,   \mathbf{J}(\mathbf{x}^*(t_{k-i+1}))\bigr]^{-1} \boldsymbol,
\label{eq:discrete_QR}
\end{equation}
where $\boldsymbol{\Gamma}$ is the left-hand-side matrix of the semi-implicit DAE. Since the Jacobian $\mathbf{J}$ is already factorized (LU or sparse direct) during the DAE solve, computing $\boldsymbol{\Phi}$ requires only additional triangular back-substitution steps at essentially zero additional factorization cost. This approach was demonstrated on systems up to 1,479 buses \cite{hayes_viable_2018}, significantly advancing the scalability of model-based LE computation.

The Kazma--Taha framework \cite{kazma_stability_2025} further extends this approach by introducing a parameterized \emph{state-deformation tensor} $\tilde{\boldsymbol{\Xi}}(\boldsymbol{\gamma})$ that selects state variables relevant to a given bus's stability:
\begin{equation}
\tilde{\boldsymbol{\Xi}}(\boldsymbol{\gamma}) = \sum_j \gamma_j \sum_{i=1}^{N} (\boldsymbol{\varphi}_i^0)^T \boldsymbol{\varphi}_i^0,
\label{eq:deformation_tensor}
\end{equation}
where $\boldsymbol{\varphi}_i^0$ are columns of the discretized transition matrix and $\gamma_j \in \{0,1\}$ selects state $j$ for inclusion. This enables per-bus stability assessment using a single scalar metric $\ln\det(\tilde{\boldsymbol{\Xi}})$, which is provably equivalent (Theorem 1 in \cite{kazma_stability_2025}) to the sum of the bus-level LE spectrum.

\subsection{Phase-Space Reconstruction for Data-Driven Estimation}
\label{subsec:psr}

When a system model is unavailable or uncertain, LEs can be estimated from observed scalar time-series data $\{x(k)\}$ using the theory of \emph{phase-space reconstruction}.

\subsubsection{Takens Embedding Theorem}

The Takens Embedding Theorem states that a single scalar measurement $x(t) = h(\mathbf{x}(t))$ can reconstruct the dynamics of a d-dimensional system through delay embedding. The delay-embedding map
\begin{equation}
\mathbf{s}(t) = \bigl[x(t),\, x(t{+}\tau),\,   x(t{+}2\tau),\, \ldots,\, x(t{+}(m{-}1)\tau)\bigr]^T
\label{eq:embedding}
\end{equation}
with embedding dimension $m \geq 2d + 1$ and time delay $\tau > 0$ produces a diffeomorphic image of the original attractor in $\mathbb{R}^m$. Consequently, the LE spectrum of the reconstructed dynamics approximates the true LE spectrum.

In power system applications, the scalar observable is typically a PMU measurement: rotor angle difference $\delta_{ij}(t)$ \cite{dasgupta_pmu-based_2012, wei_model-free_2018}, bus voltage magnitude or angle \cite{ge_improved_2018, pinzon_pmu-based_2019}, or a load or renewable power time series \cite{mohamed_variability_2007, yufei_photovoltaic_2016}.
 
 \subsubsection{Selection of Embedding Parameters}

Correct selection of the time delay $\tau$ and embedding dimension $m$ is critical. Two standard criteria are used throughout the reviewed literature:

\textbf{Time delay $\tau$:} The delay should be chosen so that $x(t)$ and $x(t+\tau)$ are sufficiently independent to span the attractor, yet close enough in time to reflect dynamic continuity. Two common methods are: (i) the \emph{first zero} of the autocorrelation function (ACF) defined as $g(n)=\frac{\sum_k x_k x_{k+n}}{\sum_k |x_k|^2}$, and (ii) the \emph{first minimum} of the average mutual information (AMI) function $I(\tau) = \sum_{x,x'} p(x,x') \ln[p(x,x')/(p(x)p(x'))]$, which identifies nonlinear independence\cite{pinzon_chaos_2018}. The AMI criterion is generally preferred for nonlinear systems.

\textbf{Embedding dimension $m$:} The standard method is \emph{false nearest neighbors} (FNN): for each point in the $m$-dimensional reconstruction, its nearest neighbor is identified; if their distance increases substantially when the dimension is raised to $m+1$, the neighbor is false (an artifact of projection) \cite{aghbalou_long_2018}. This is quantified using $D = \frac{\left|x(t+k\tau) - x^{(r)}(t+k\tau)\right|} {\sqrt{\sum_{k=1}^{d-1} \left[x(t+k\tau) - x^{(r)}(t+k\tau)\right]^2}} > \text{tol}$, where $x(t)$ is the scalar time series, $\tau$ is the time delay, $k$ indexes the embedding components, $d$ is the embedding dimension, $x^{(r)}(\cdot)$ denotes the nearest neighbor trajectory, and $\text{tol}$ is a predefined threshold. The minimum $m$ at which the fraction of false neighbors falls below a threshold (typically $5\%$--$10\%$) is selected. An alternative is the \emph{C-C method} \cite{yufei_photovoltaic_2016}, which determines $\tau$ and $m$ simultaneously from statistics of the correlation integral, offering computational efficiency for short data records.

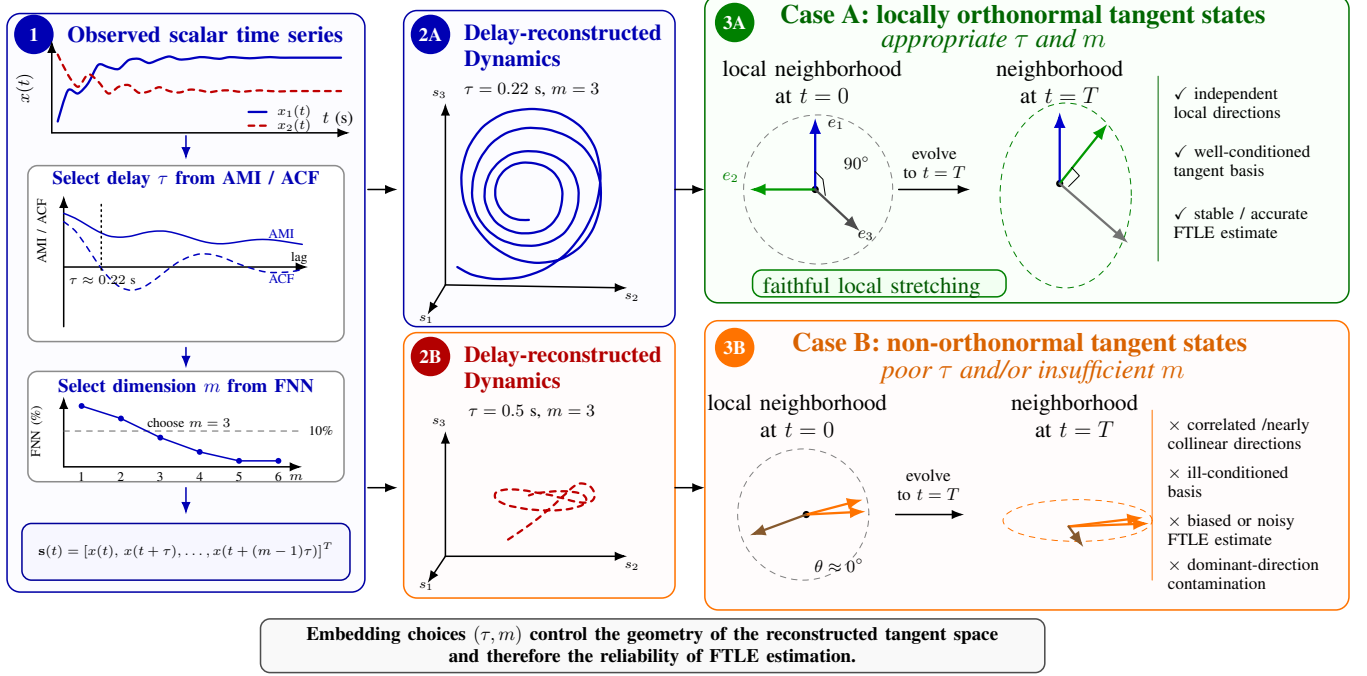
\begin{figure*}[!t]
\centering
\resizebox{0.98\textwidth}{!}{%
\begin{tikzpicture}[
    >=Latex,
    font=\small,
    scale=1.1,
    line join=round,
    line cap=round,
    arr/.style={-Latex,thick},
    panelblue/.style={draw=blue!65!black,rounded corners=7pt,thick,fill=blue!1},
    panelgreen/.style={draw=green!45!black,rounded corners=7pt,thick,fill=green!2},
    panelred/.style={draw=orange!90!red,rounded corners=7pt,thick,fill=red!1},
    subbox/.style={draw=black!45,rounded corners=5pt,thick,fill=white},
    note/.style={draw,rounded corners=5pt,thick,align=center,inner sep=4pt},
    badge/.style={circle,draw,thick,minimum size=7mm,inner sep=0pt,font=\bfseries,text=white},
    tiny/.style={font=\scriptsize}
]

\node[font=\bfseries\Large] at (10.7,11.2)
{Phase-Space Reconstruction and Its Impact on FTLE Estimation};

\draw[panelblue] (0.0,0.9) rectangle (6.0,10.6);
\node[badge,fill=blue!70!black,draw=blue!70!black] at (0.45,10.25) {1};
\node[anchor=west,text=blue!70!black,font=\bfseries\large] at (0.9,10.25)
{Observed scalar time series};

\draw[arr] (0.75,8.60) -- (0.75,10.15);
\draw[arr] (0.75,8.60) -- (5.75,8.60);
\node[rotate=90] at (0.30,9.35) {$x(t)$};
\node at (5.55,8.83) {$t$ (s)};

\draw[blue!75!black,very thick] plot[smooth] coordinates { (0.85,8.79) (1.02,9.31) (1.19,9.26) (1.36,9.43) (1.53,9.74) (1.70,9.70) (1.87,9.67) (2.04,9.82) (2.21,9.84) (2.38,9.77) (2.55,9.83) (2.72,9.88) (2.89,9.83) (3.06,9.84) (3.23,9.87) (3.39,9.86) (3.56,9.84) (3.73,9.87) (3.90,9.86) (4.07,9.85) (4.24,9.86) (4.41,9.87) (4.58,9.86) (4.75,9.86) (4.92,9.86) (5.09,9.86) (5.26,9.86) (5.43,9.86) (5.60,9.86) };

\draw[red!75!black,very thick,dashed] plot[smooth] coordinates {
(0.85,9.912) (1.02,9.576) (1.19,9.371) (1.36,9.582) (1.53,9.459) (1.70,9.259) (1.87,9.371) (2.04,9.400) (2.21,9.271) (2.38,9.294) (2.55,9.365) (2.72,9.300) (2.89,9.276) (3.06,9.324) (3.23,9.312) (3.39,9.282) (3.56,9.306) (3.73,9.312) (3.90,9.288) (4.07,9.300) (4.24,9.312) (4.41,9.300) (4.58,9.294) (4.75,9.300) (4.92,9.300) (5.09,9.300) (5.26,9.300) (5.43,9.300) (5.60,9.300) };
\draw[blue!75!black,very thick] (4.05,8.95) -- (4.35,8.95);
\node[tiny,anchor=west] at (4.42,8.95) {$x_1(t)$};
\draw[red!75!black,very thick,dashed] (4.05,8.72) -- (4.35,8.72);
\node[tiny,anchor=west] at (4.42,8.72) {$x_2(t)$};


\draw[arr,blue!70!black] (3.0,8.55) -- (3.0,8.15);

\draw[subbox] (0.35,5.10) rectangle (5.65,8.05);
\node[text=blue!70!black,font=\bfseries] at (3.0,7.82)
{Select delay $\tau$ from AMI / ACF};

\draw[arr] (0.95,5.35) -- (0.95,7.55);
\draw[arr] (0.95,6.35) -- (5.05,6.35);
\node[tiny,rotate=90] at (0.58,6.95) {AMI / ACF};
\node[tiny] at (4.9,6.5) {lag};

\draw[blue!75!black,thick] plot[smooth] coordinates {
(0.95,7.25) (1.14,7.17) (1.33,7.06) (1.52,6.95) (1.71,6.87)
(1.90,6.85) (2.09,6.87) (2.28,6.92) (2.47,6.95) (2.66,6.95)
(2.85,6.91) (3.05,6.85) (3.24,6.79) (3.43,6.75) (3.62,6.74)
(3.81,6.76) (4.00,6.79) (4.19,6.81) (4.38,6.81) (4.57,6.79)
(4.76,6.76) (4.95,6.73)
};

\draw[blue!75!black,thick,dashed] plot[smooth] coordinates {
(0.95,7.11) (1.14,6.96) (1.33,6.70) (1.52,6.41) (1.71,6.17)
(1.90,6.01) (2.09,5.96) (2.28,6.01) (2.47,6.13) (2.66,6.28)
(2.85,6.42) (3.05,6.52) (3.24,6.57) (3.43,6.56) (3.62,6.51)
(3.81,6.43) (4.00,6.36) (4.19,6.30) (4.38,6.26) (4.57,6.26)
(4.76,6.27) (4.95,6.31)
};

\draw[dotted,thick] (1.58,6.30) -- (1.58,7.45);
\node[tiny] at (1.60,6.13) {$\tau\approx0.22$ s};
\node[tiny,blue!75!black] at (4.60,6.92) {AMI};
\node[tiny,blue!75!black] at (4.60,6.16) {ACF};

\draw[arr,blue!70!black] (3.0,5.00) -- (3.0,4.65);

\draw[subbox] (0.35,2.75) rectangle (5.65,4.60);
\node[text=blue!70!black,font=\bfseries] at (3.0,4.37)
{Select dimension $m$ from FNN};

\draw[arr] (0.95,3.00) -- (0.95,4.18);
\draw[arr] (0.95,3.00) -- (4.95,3.00);
\node[tiny,rotate=90] at (0.50,3.60) {FNN (\%)};
\node[tiny] at (4.85,2.82) {$m$};

\draw[black!60,dashed] (0.95,3.60) -- (4.95,3.60);
\node[tiny,anchor=west] at (4.95,3.60) {10\%};

\draw[blue!75!black,thick] plot coordinates {
(1.25,4.02) (1.91,3.81) (2.57,3.49) (3.23,3.25) (3.89,3.10) (4.55,3.10)
};

\foreach \p in {(1.25,4.02),(1.91,3.81),(2.57,3.49),(3.23,3.25),(3.89,3.10),(4.55,3.10)}
{
    \fill[blue!75!black] \p circle (1.4pt);
}

\foreach \x/\m in {1.25/1,1.91/2,2.57/3,3.23/4,3.89/5,4.55/6}
{
    \node[tiny] at (\x,2.83) {\m};
}

\node[tiny] at (3.05,3.75) {choose $m=3$};

\draw[arr,blue!70!black] (3.0,2.65) -- (3.0,2.20);

\draw[subbox,fill=blue!3,draw=blue!60!black] (0.25,1.10) rectangle (5.75,2.05);
\node[align=center,tiny] at (3.0,1.60)
{$\mathbf{s}(t)=[x(t),\,x(t+\tau),\ldots,x(t+(m-1)\tau)]^T$};

\draw[arr] (6.05,7.65) -- (6.55,7.65);

\draw[arr] (6.05,2.65) -- (6.55,2.65);

\draw[panelblue] (6.65,5.35) rectangle (11.2,10.65);
\draw[panelred] (6.65,0.8) rectangle (11.2,5.25);

\node[badge,fill=blue!70!black,draw=blue!70!black] at (7.10,10.25) {2A};
\node[badge,fill=red!70!black,draw=red!70!black] at (7.10,4.85) {2B};

\node[anchor=west,text=blue!70!black,font=\bfseries\large] at (7.55,10.25) {Delay-reconstructed};
\node[anchor=west,text=blue!70!black,font=\bfseries\large] at (7.55,9.82) {Dynamics};

\node[anchor=west,text=red!70!black,font=\bfseries\large] at (7.55,4.85) {Delay-reconstructed};
\node[anchor=west,text=red!70!black,font=\bfseries\large] at (7.55,4.42) {Dynamics};

\node[font=\small] at (8.80,9.32) {$\tau=0.22$ s, $m=3$};
\node[font=\small] at (8.80,3.92) {$\tau=0.5$ s, $m=3$};

\draw[arr] (7.35,6.05) -- (7.05,5.55);
\draw[arr] (7.35,6.05) -- (10.40,6.00);
\draw[arr] (7.35,6.05) -- (7.35,9.10);
\node[tiny] at (7.03,5.48) {$s_1$};
\node[tiny] at (10.48,5.84) {$s_2$};
\node[tiny] at (7.25,9.25) {$s_3$};

\draw[arr] (7.35,1.5) -- (7.05,1.05);
\draw[arr] (7.35,1.5) -- (10.40,1.5);
\draw[arr] (7.35,1.5) -- (7.35,3.6);
\node[tiny] at (7.03,1.0) {$s_1$};
\node[tiny] at (10.48,1.34) {$s_2$};
\node[tiny] at (7.25,3.75) {$s_3$};

\draw[blue!75!black,very thick] plot[smooth] coordinates {
(7.55,6.35)
(7.80,6.22) (8.18,6.15) (8.62,6.18) (9.02,6.32) (9.38,6.58)
(9.68,6.96) (9.88,7.40) (9.95,7.88) (9.86,8.34) (9.62,8.70)
(9.25,8.93) (8.80,9.00) (8.35,8.93) (7.98,8.70) (7.73,8.34)
(7.62,7.90) (7.66,7.45) (7.84,7.05) (8.12,6.72) (8.48,6.50)
(8.88,6.42) (9.28,6.48) (9.60,6.68) (9.82,6.98) (9.94,7.35)
(9.95,7.73) (9.85,8.08) (9.66,8.36) (9.38,8.55) (9.05,8.66)
(8.70,8.68) (8.38,8.60) (8.12,8.42) (7.93,8.16) (7.82,7.85)
(7.80,7.52) (7.88,7.22) (8.05,6.98) (8.28,6.81) (8.56,6.72)
(8.86,6.71) (9.12,6.78) (9.34,6.92) (9.49,7.14) (9.57,7.39)
(9.58,7.65) (9.51,7.89) (9.37,8.08) (9.17,8.21) (8.93,8.28)
(8.68,8.29) (8.45,8.24) (8.25,8.11) (8.10,7.93) (8.02,7.71)
(8.01,7.48) (8.08,7.27) (8.20,7.10) (8.38,6.98) (8.58,6.92)
(8.79,6.92) (8.99,6.97) (9.14,7.08) (9.25,7.24) (9.30,7.43)
(9.31,7.62) (9.26,7.80) (9.16,7.94) (9.01,8.04) (8.83,8.09)
(8.65,8.10) (8.48,8.06) (8.34,7.97) (8.24,7.84) (8.19,7.68)
(8.19,7.52) (8.24,7.38) (8.33,7.26) (8.45,7.18) (8.60,7.14)
(8.75,7.14)
};
\draw[red!75!black,very thick,dashed] plot[smooth] coordinates {
(8.40,1.78) (8.58,1.90) (8.75,2.02) (8.91,2.14) (9.05,2.26)
(9.18,2.37) (9.28,2.47) (9.37,2.55) (9.44,2.62) (9.50,2.67)
(9.55,2.70) (9.60,2.72) (9.65,2.72) (9.70,2.71) (9.75,2.69)
(9.79,2.66) (9.82,2.62) (9.84,2.58) (9.85,2.54) (9.85,2.50)
(9.84,2.47) (9.82,2.44) (9.79,2.42) (9.75,2.41) (9.70,2.40)
(9.64,2.40) (9.58,2.41) (9.51,2.42) (9.43,2.44) (9.34,2.47)
(9.24,2.49) (9.14,2.52) (9.03,2.54) (8.92,2.56) (8.81,2.58)
(8.71,2.60) (8.61,2.61) (8.52,2.62) (8.44,2.62) (8.37,2.62)
(8.31,2.61) (8.26,2.60) (8.22,2.58) (8.19,2.56) (8.18,2.53)
(8.18,2.50) (8.19,2.47) (8.22,2.44) (8.27,2.41) (8.33,2.38)
(8.41,2.35) (8.50,2.33) (8.61,2.31) (8.72,2.29) (8.84,2.28)
(8.96,2.27) (9.08,2.27) (9.20,2.27) (9.31,2.28) (9.41,2.29)
(9.50,2.31) (9.57,2.33) (9.63,2.35) (9.67,2.38) (9.70,2.41)
(9.71,2.44) (9.71,2.47) (9.69,2.50) (9.66,2.53) (9.61,2.55)
(9.55,2.57) (9.47,2.58) (9.38,2.59) (9.28,2.59) (9.18,2.59)
(9.08,2.58) (8.98,2.57) (8.89,2.56) (8.81,2.54) (8.75,2.52)
};




\draw[arr] (11.2,7.65) -- (11.7,7.65);

\draw[arr] (11.2,2.65) -- (11.7,2.65);

\draw[panelgreen] (11.7,5.75) rectangle (22.5,10.9);

\node[badge, draw=green!50!black, text=white, fill=green!50!black] at (12.20,10.45) {3A};
\node[anchor=west, text=green!40!black, font=\bfseries\Large] at (12.85,10.55)
{Case A: locally orthonormal tangent states};
\node[anchor=west, text=green!40!black, font=\itshape\Large] at (14.55,10.15)
{appropriate $\tau$ and $m$};

\node[align=center, font=\large] at (13.5,9.45) {local neighborhood\\ at $t=0$};
\draw[dashed, gray] (13.6,7.65) circle (1.25);
\fill (13.55,7.65) circle (0.06);

\draw[->, very thick, blue!80!black] (13.55,7.65) -- (13.55,8.85);
\draw[->, very thick, green!60!black] (13.55,7.65) -- (12.45,7.65);
\draw[->, very thick, black!70] (13.55,7.65) -- (14.30,6.95);
\node[anchor=west] at (13.65,8.75) {$e_1$};
\node[anchor=east, text=green!50!black] at (12.38,7.85) {$e_2$};
\node[anchor=west] at (14.15,6.88) {$e_3$};
\draw (13.72,7.55) -- (13.68,7.82) -- (13.55,7.95);
\node at (14.25,8.10) {$90^\circ$};

\draw[arr] (14.95,7.65) -- (16.15,7.65);
\node[align=center] at (15.55,8.10) {evolve\\ to $t=T$};

\node[align=center, font=\large] at (17.65,9.45) {neighborhood\\ at $t=T$};
\draw[dashed, draw=green!50!black] (17.75,7.55) ellipse (1.10 and 1.55);
\fill (17.65,7.75) circle (0.06);
\draw[->, very thick, blue!80!black] (17.65,7.75) -- (17.65,8.95);
\draw[->, very thick, green!60!black] (17.65,7.75) -- (18.45,8.75);
\draw[->, very thick, black!55] (17.65,7.75) -- (18.80,6.75);
\draw (17.8,7.7) -- (17.98,7.88) -- (17.85,8.02);

\draw[green!40!black] (19.30,6.45) -- (19.30,9.55);
\node[anchor=west, align=left] at (19.45,9.10) {$\checkmark$ independent\\ local directions};
\node[anchor=west, align=left] at (19.45,8.10) {$\checkmark$ well-conditioned\\ tangent basis};
\node[anchor=west, align=left] at (19.45,7.10) {$\checkmark$ stable / accurate\\ FTLE estimate};

\draw[note, draw=green!50!black, fill=green!5] (12.5,5.85) rectangle (16.75,6.30);
\node[text=green!35!black, font=\large] at (14.5,6) {faithful local stretching};


\draw[panelred] (11.7,0.6) rectangle (22.5,5.45);

\node[badge, draw=orange!90!red, text=white, fill=orange!90!red] at (12.20,5.00) {3B};
\node[anchor=west, text=orange!90!red!85!black, font=\bfseries\Large] at (13.05,5.05)
{Case B: non-orthonormal tangent states};
\node[anchor=west, text=orange!90!red!85!black, font=\itshape\Large] at (14.55,4.60)
{poor $\tau$ and/or insufficient $m$};

\node[align=center, font=\large] at (13.25,3.85) {local neighborhood\\ at $t=0$};
\draw[dashed, gray] (13.35,2.20) circle (1.10);
\fill (13.40,2.20) circle (0.06);

\draw[->, very thick, orange!90!red] (13.40,2.20) -- (14.35,2.45);
\draw[->, very thick, orange!90!red] (13.40,2.20) -- (14.40,2.25);
\draw[->, very thick, brown!70!black] (13.40,2.20) -- (12.45,1.85);
\node at (13.95,1.35) {$\theta \approx 0^\circ$};

\draw[arr] (14.90,2.20) -- (16.05,2.20);
\node[align=center] at (15.45,2.70) {evolve\\ to $t=T$};

\node[align=center, font=\large] at (17.90,3.85) {neighborhood\\ at $t=T$};
\draw[dashed, draw=orange!90!red] (17.95,2.10) ellipse (1.25 and 0.35);
\draw[->, very thick, orange!90!red] (17.8,2) -- (19.15,2.05);
\draw[->, very thick, orange!90!red] (17.8,2) -- (19.05,2.15);
\draw[->, very thick, brown!70!black] (17.8,2) -- (18.05,1.65);

\draw[orange!90!red] (19.20,1.10) -- (19.20,3.95);
\node[anchor=west, align=left] at (19.35,3.55) {$\times$ correlated /nearly \\ collinear directions};
\node[anchor=west, align=left] at (19.35,2.75) {$\times$ ill-conditioned\\ basis};
\node[anchor=west, align=left] at (19.35,1.95) {$\times$ biased or noisy\\ FTLE estimate};
\node[anchor=west, align=left] at (19.35,1.20) {$\times$ dominant-direction\\ contamination};


\draw[draw=black!70,rounded corners=5pt,thick,fill=gray!5]
(4.25,-0.45) rectangle (17.4,0.45);

\node[align=center,font=\bfseries] at (10.82,0.00)
{Embedding choices $(\tau,m)$ control the geometry of the reconstructed tangent space\\
and therefore the reliability of FTLE estimation.};

\end{tikzpicture}%
}
\caption{Representative schematic of phase-space reconstruction for data-driven Lyapunov analysis. A scalar time series is delay-embedded using Takens reconstruction, where the time delay $\tau$ and embedding dimension $m$ are selected using AMI/ACF and FNN criteria. The reconstructed geometry determines whether local tangent directions are well-conditioned and nearly orthonormal, which improves FTLE estimation, or nearly collinear and ill-conditioned, which leads to biased or noisy FTLE estimates.}
\label{fig:voltage_psr_ftle}
\end{figure*}

\textcolor{red}{Figure~\ref{fig:voltage_psr_ftle} presents a representative schematic of phase-space reconstruction for data-driven Lyapunov exponent estimation. A scalar time series is first embedded using Takens delay coordinates, where the time delay $\tau$ is selected from AMI/ACF criteria and the embedding dimension $m$ is selected using the FNN criterion. The figure emphasizes that these embedding parameters directly affect the geometry of the reconstructed state space and, consequently, the quality of the estimated FTLE.}

\textcolor{red}{When $\tau$ and $m$ are appropriately selected, the reconstructed trajectory unfolds the underlying dynamics and produces locally independent tangent directions. In this case, the local perturbation basis remains well-conditioned and approximately orthonormal, so the observed stretching of nearby trajectories provides a reliable FTLE estimate. In contrast, poor selection of $\tau$ or insufficient embedding dimension can cause projection overlap, false neighbors, and nearly collinear tangent directions. This produces an ill-conditioned local tangent basis, where the estimated stretching may be dominated by a single direction or contaminated by noise, leading to biased or unreliable FTLE estimates. Thus, the figure highlights that accurate data-driven LE estimation depends not only on the FTLE algorithm itself, but also on the quality of the phase-space reconstruction.}

\subsection{Data-Driven LE Estimation Algorithms}
\label{subsec:data_driven}

Three algorithms dominate the data-driven LE estimation literature in power systems. All three estimate the LLE $\lambda_1$ from a reconstructed phase-space trajectory.

\subsubsection{Wolf Method}

The \emph{Wolf method} \cite{wang_analysis_2019} (Wolf \textit{et al.}, 1985) directly tracks the evolution of a fiducial trajectory pair. Given the reconstructed trajectory $\{\mathbf{s}(k)\}$, the algorithm:
\begin{enumerate}
  \item Locates the nearest neighbor $\mathbf{s}(k_0)$        to the reference point $\mathbf{s}(0)$ with        initial separation $L_0 = \|\mathbf{s}(k_0)        - \mathbf{s}(0)\|$.
  \item Evolves both trajectories until their separation        exceeds a threshold $\epsilon$, recording the        final separation $L_0'$.
  \item Replaces the evolved neighbor with the nearest available point that preserves the orientation of the separation vector and repeats.
  \item Accumulates $\lambda_1 \approx        \frac{1}{t_N}\sum_{k}\ln(L_k'/L_k)$.
\end{enumerate}
The Wolf method gives a positive definite estimate and converges reliably for long, low-noise records. Its main limitation is sensitivity to the replacement criterion and poor performance on short or noisy data.

\subsubsection{Rosenstein Method}

The \emph{Rosenstein method} (Rosenstein \textit{et al.}, 1993) is designed for short, noisy data and is the preferred algorithm in many power system PMU-based implementations \cite{banerjee_lyapunov_2014,wang_short-term_2009}. It estimates $\lambda_1$ from the average divergence of all nearest neighbor pairs simultaneously:
\begin{equation}
d_j(t) = C_j\,e^{\lambda_1 t},
\label{eq:rosenstein_divergence}
\end{equation}
where $d_j(t) = \|\mathbf{s}(j+t) - \mathbf{s}(\hat{j}+t)\|$ is the distance between the $j$-th point and its nearest neighbor $\hat{j}$ (excluding temporally close points). The LLE is estimated as the slope of:
\begin{equation}
\bar{y}(t) = \frac{1}{N}\sum_{j=1}^{N} \ln d_j(t) \approx \ln C + \lambda_1 t,
\label{eq:rosenstein_slope}
\end{equation}
obtained by linear regression over the initial linear region of $\bar{y}(t)$.

\subsubsection{Kantz Method}

The \emph{Kantz method} (Kantz, 1994) is similar to the Rosenstein method but uses a neighborhood of radius $r$ rather than a single nearest neighbor, averaging over all points within the ball $B(\mathbf{s}(t_0), r)$:
\begin{equation}
S(t) = \frac{1}{|\mathcal{T}|} \sum_{t_0\in\mathcal{T}} \ln\!\left(\frac{1}{|U_\epsilon(t_0)|} \sum_{\mathbf{s}_j\in U_\epsilon(t_0)} \|\mathbf{s}(t_0+t) - \mathbf{s}(j+t)\|\right),
\label{eq:kantz}
\end{equation}
where $U_\epsilon(t_0)$ is the $\epsilon$-neighborhood. The LLE is the slope of $S(t)$ in the linear scaling region. Kantz's method is more robust to noise than the Wolf method and is used in several power system load forecasting and predictability studies
\cite{mohamed_variability_2007}.

\subsubsection{Small-Data Method}

For very short data records (common in real-time PMU windows), the \emph{small-data method} provides an alternative by exploiting all pairs within a tight neighborhood and averaging their short-time divergence rates \cite{yufei_photovoltaic_2016, chen_identification_2009}. It is equivalent to the Rosenstein method with aggressive neighborhood selection and is particularly suited to embedded microprocessor implementations.

\subsubsection{Online / Recursive Estimation}

For real-time power system applications, batch estimation algorithms are too slow. Several works have proposed recursive implementations that update the LLE estimate as each new PMU measurement arrives \cite{wei_model-free_2018, ge_improved_2018, pinzon_pmu-based_2019}. The core idea is to maintain a sliding window of the reconstructed trajectory and update the nearest-neighbor divergence estimate incrementally, avoiding full recomputation at each time step. The adaptive time-window strategy of \cite{wei_model-free_2018} further adjusts the window length dynamically based on the estimated oscillation frequency, improving detection speed in the presence of inter-area swings.

\subsection{Predictability Horizon from the LLE}
\label{subsec:predictability}

A fundamental application of the LLE in the forecasting domain (Section \ref{sec:forecasting}) is the \emph{predictability horizon}: the maximum time interval over which a chaotic trajectory can be reliably predicted. If two initial conditions differ by $\delta_0$ (measurement uncertainty), their separation grows as $\delta(t) \approx \delta_0\,e^{\lambda_1 t}$. Defining the maximum acceptable prediction error as $\Delta$, the maximum prediction horizon is:
\begin{equation} T_{\max} = \frac{1}{\lambda_1} \ln\!\left(\frac{\Delta}{\delta_0}\right).
\label{eq:predict_horizon}
\end{equation}
This formula, due originally to Lorenz, was applied to power system load forecasting by \cite{mohamed_variability_2007} to bound the maximum reliable prediction horizon of the Tunisian daily peak-load series. The key implication for power engineers is that \eqref{eq:predict_horizon} provides a model-independent, empirically computable upper bound on any forecasting method's accuracy-a limit set by the physics of the system's nonlinear dynamics rather than by model choice.

\subsection{Comparative Summary and Practical Guidelines}
\label{subsec:algo_comparison}

 Table \ref{tab:algo_summary} summarizes the principal LE estimation algorithms discussed in this section, comparing their data requirements, computational complexity, noise robustness, and typical power system applications.

\begin{table*}[!t]
  \renewcommand{\arraystretch}{1.35}
  \caption{Summary of Principal Lyapunov Exponent Estimation
  Methods Used in Power System Applications}
  \label{tab:algo_summary}
  \centering
  \begin{tabular}{p{2.4cm}p{2.0cm}p{2.0cm}p{1.5cm}p{2.0cm}p{3.6cm}}
    \toprule
    \textbf{Method} & \textbf{Model} &    \textbf{Data needed} & \textbf{Noise} &    \textbf{Output} & \textbf{Typical application} \\
    & \textbf{required?} & & \textbf{robustness} & & \\
    \midrule
    GSR \cite{wadduwage_power_2013,yan_pmu-based_2012,bosetti_transient_2018}      & Yes & System equations      & N/A (model) & Full spectrum      & Offline TSA, stability mapping \\
    Discrete-QR \cite{hayes_viable_2018}      & Yes & DAE + Jacobian      & N/A (model) & LLE (or spectrum)      & Large-scale offline LE computation \\
    Wolf \cite{wang_analysis_2019}      & No & Long clean record      & Low & LLE only      & Chaos confirmation, forecasting \\
    Rosenstein \cite{banerjee_lyapunov_2014,wang_short-term_2009}      & No & Short noisy record      & Medium & LLE only      & PMU-based monitoring \\
    Kantz \cite{mohamed_variability_2007}      & No & Moderate, moderate noise      & Medium--high & LLE only      & Load/wind forecasting,        predictability \\
    Small-data \cite{yufei_photovoltaic_2016}      & No & Very short record      & Medium & LLE only      & Ultra-short-term PV/wind prediction \\
    Online recursive \cite{wei_model-free_2018}      & No & Streaming PMU data      & Medium & FTLE online      & Real-time stability monitoring \\
    Stochastic LE \cite{verdejo_application_2015}      & Yes & Linearized stochastic model      & N/A (model) & tochastic LLE      & Uncertainty-aware stability margin \\
    \bottomrule
  \end{tabular}
\end{table*}

Several practical guidelines emerge from the comparative analysis:

\begin{enumerate}

  \item \textbf{Use model-based GSR or discrete-QR for rigorous offline analysis.} When a validated power system model is available, model-based methods give the most accurate LE spectrum. The discrete-QR approach \cite{hayes_viable_2018} is preferred for DAE systems at scale.

  \item \textbf{Use the Rosenstein or Kantz method for PMU-based estimation.} These methods are most robust to the noise and short window lengths encountered in online monitoring contexts.
  \item \textbf{Select embedding parameters rigorously.}        Using the ACF to set $\tau$ is acceptable for linear time series; for nonlinear power data, the        AMI minimum should be used. The FNN criterion should be applied to the $m$ selection; ad hoc choices of $m = 2$ or $m = 3$ without verification can lead to unreliable estimates.

  \item \textbf{Interpret FTLEs in context.}        A positive FTLE in the first 100--200, ms post-fault does not always indicate instability;        transient oscillations can temporarily raise the windowed divergence rate. This contamination problem is addressed by several adaptive-threshold methods reviewed in Section \ref{sec:stability}.

  \item \textbf{LLE sign is necessary but not sufficient for chaos confirmation.} As demonstrated in        \cite{lai_illustrating_1993}, complex phase portraits near voltage collapse boundaries are not always chaotic; a positive LLE must be verified with consistent Poincaré sections and power spectra.

\end{enumerate}

%
%

%
\section{Transient and Voltage Stability Assessment}
\label{sec:stability}

Stability assessment is by far the largest application domain in the surveyed literature, with approximately 85 publications spanning from the first PMU-based chaos detection paper in 1994 \cite{liu_detection_1994} to the most recent work on IBR-dominated grids in 2025--2026 \cite{peng_measuring_2025, peng_measuring_2026}. The unifying principle across all papers is that the sign of the largest Lyapunov exponent (LLE) computed from the post-fault or post-disturbance trajectory provides a physically grounded, model-independent stability indicator: a sustained negative LLE implies exponential convergence toward the post-fault equilibrium (stability), while a positive LLE implies trajectory divergence (instability).

This section is organized as follows. Section \ref{subsec:model_tsa} covers model-based approaches for transient stability analysis and critical clearing time determination. Section \ref{subsec:pmu_rotor} presents PMU-based real-time rotor-angle stability monitoring. Section \ref{subsec:pmu_voltage} addresses PMU-based short-term voltage stability monitoring. Section \ref{subsec:coherency} covers coherency identification and controlled islanding. Section \ref{subsec:contingency} discusses contingency ranking and security assessment. Section \ref{subsec:stochastic} reviews stochastic and uncertainty-aware stability analysis. Section \ref{subsec:ml_hybrid} addresses hybrid machine-learning and LE methods. Section \ref{subsec:ibr_stability} covers the emerging literature on IBR-dominated grids. Section \ref{subsec:stability_gaps} identifies open problems.

\subsection{Model-Based Transient Stability Analysis and Critical Clearing Time}
\label{subsec:model_tsa}

\subsubsection{LLE as a Transient Stability Certificate}

The theoretical basis for using LEs in transient stability analysis (TSA) was established in the model-based literature, which computes LEs directly from variational equations of the power system model rather than from measured data.

The seminal work by Wadduwage \textit{et al.} \cite{wadduwage_power_2013} demonstrated the use of the full LE spectrum for TSA on the IEEE 3-machine 9-bus and 16-machine 69-bus benchmark systems. The LE spectrum is computed from the post-fault swing-equation model using Gram-Schmidt re-orthonormalization (Section \ref{subsec:model_based}). The key results are:
\begin{enumerate}
  \item \textbf{Stability criterion:} A negative LLE        ($\lambda_1 < 0$) in the post-fault system implies exponential convergence to the stable equilibrium,        while a positive LLE ($\lambda_1 > 0$) implies instability.
  \item \textbf{CCT determination:} The critical clearing time (CCT) is identified as the fault duration at which $\lambda_1$ transitions from negative to positive. This approach agrees closely with time-domain simulation (TDS) results but provides a nonlinear-dynamics interpretation independent of the specific energy function.
  \item \textbf{LE invariance:} LEs are invariant with respect to the initial conditions within the stability region, a property exploited to map the full stability region in parameter space without repeated simulation from many initial conditions.
  \item \textbf{Structural stability:} LEs can quantify structural stability-how the stability boundary shifts under changes in generator output levels or load, providing a richer picture than binary pass/fail CCT comparison.
\end{enumerate}

A companion investigation \cite{wadduwage_investigation_2013} validated the applicability of this LE-based framework against Extended Equal Area Criterion (EEAC) and Potential Energy Boundary Surface (PEBS) methods. A critical practical finding was that convergence of the asymptotic LE requires long integration times (on the order of 500\,s), which renders the direct approach unsuitable for real-time use. This observation motivated the subsequent shift toward finite-time LEs and PMU-based estimation, discussed in Sections \ref{subsec:pmu_rotor} and \ref{subsec:pmu_voltage}.

The framework was extended by Bosetti and Khan \cite{bosetti_transient_2018} to include \emph{covariant Lyapunov vectors} (CLVs), which provide not only the scalar LE values but also the directions of the expanding and contracting manifolds of the post-fault attractor. This enables characterization of local stability structure near the stability boundary, including the detection of hyperbolicity and basin-boundary geometry in multi-machine oscillating systems.

\subsubsection{Scalable Model-Based Computation}

A major scalability advance was achieved by Hayes and Milano \cite{hayes_viable_2018}, who developed the discrete-QR method (described in Section \ref{subsec:model_based}) that reuses Jacobian factors from implicit numerical integration. This reduced the additional computational cost of LLE computation to essentially zero beyond that of the DAE integration itself, enabling the first LLE computation on a 1,479-bus dynamic model of the all-island Irish transmission system. The method was also validated on stochastic DAE models with Ornstein-Uhlenbeck load disturbances, demonstrating that LLE remains negative for non-chaotic stochastic orbits and correctly identifies instability when it occurs.

The Kazma--Taha framework \cite{kazma_stability_2025, kazma_generalizable_2025} further extends model-based LE analysis to renewable resource allocation under uncertainty. By defining a parameterized state-deformation tensor whose $\log\det$ is equivalent to the sum of all bus-level LEs (Section \ref{subsec:model_based}), this work enables joint voltage, frequency, and rotor-angle stability assessment in a single framework, and formulates optimal renewable siting as a submodular optimization problem with a provable 99\% near-optimality guarantee via greedy algorithms. Validation covers 9-, 39-, and 200-bus standard test systems.

A practical MATLAB-based analysis tool,
PSLE\_TOOL\cite{kamble_psle_tool_2015}, implements
model-based LE computation for multiple generator state
variables simultaneously (rotor angle, speed, internal
voltage, field current), enabling operators to visualize
LE evolution and identify individual generator
contributions to system instability on the IEEE 39-bus
system.

\subsubsection{Stochastic Stability via the Almost-Sure Lyapunov Exponent}

Verdejo \textit{et al.} \cite{verdejo_stability_2012, verdejo_application_2015} developed a rigorous framework for the \emph{almost-sure Lyapunov exponent} of power systems under persistent Markovian disturbances. The model linearizes the nonlinear DAE around an operating point, yielding:

\begin{equation}
\Delta\dot{\mathbf{x}} = \mathbf{A}(p,\,\xi_t)\, \Delta\mathbf{x},
\label{eq:stoch_linear}
\end{equation}

where $\xi_t$ is an Ornstein--Uhlenbeck stochastic process representing random generator current or load voltage fluctuations. The almost-sure MLE $\lambda$ satisfies: $\lambda < 0$ iff the zero equilibrium of \eqref{eq:stoch_linear} is almost-surely exponentially stable; $\lambda \geq 0$ implies instability. Three numerical methods (trajectory averaging, sphere projection, and unit-sphere dynamics) were compared on 4-, 10-, 16-machine test systems, establishing that trajectory averaging (Method 1) is the most computationally efficient for large systems. Crucially, this framework shows that the maximum tolerable stochastic disturbance intensity before stability loss can be computed as a function of $\lambda$, providing an uncertainty-aware stability margin not available from deterministic analysis.

\subsection{PMU-Based Real-Time Rotor-Angle Stability Monitoring}
\label{subsec:pmu_rotor}

\subsubsection{Pioneer Work: Windowed LLE from Phasor Measurements}

The shift from model-based offline analysis to measurement-based real-time monitoring was pioneered by Liu \textit{et al.}\cite{liu_detection_1994}, who proposed estimating the largest windowed Lyapunov exponent directly from real-time phasor measurements of rotor angle and frequency obtained from PMUs. The swing-equation parameters were identified via least-squares from the PMU data, and the variational equation was integrated over a sliding window of approximately 1,s. A positive windowed exponent was interpreted as evidence of a transiently chaotic power swing, i.e., a swing that is sensitively dependent on initial conditions and may lead to instability even though it does not immediately lose synchronism.

This early work established three ideas that persist throughout the subsequent literature: (i) the LLE can be estimated from PMU time series without a full system model; (ii) a finite window of 1--2,s is sufficient for a practical stability decision; and (iii) the positive-LLE criterion correctly identifies contingencies that will eventually become unstable before any overtly visible divergence.

\subsubsection{Model-Aided PMU-Based Monitoring}

Yan \textit{et al.}\cite{yan_pmu-based_2011} developed an online MLE computation scheme that combines PMU measurements with a limited structural model to estimate unobserved states via trapezoidal integration and Newton-Raphson correction. The MLE is computed using Gram-Schmidt re-orthonormalization applied to the estimated system trajectory. This approach was validated on a WECC-like 200-bus model, demonstrating correct stability prediction including multi-swing instability. The key finding is that incorporating even a partial structural model (generator inertias and network topology) substantially improves MLE accuracy compared to fully model-free estimation, particularly for slow-divergence contingencies.

\subsubsection{Fully Model-Free Rotor-Angle Monitoring}

Dasgupta \textit{et al.}\cite{dasgupta_pmu-based_2015} introduced a purely model-free framework that requires only arithmetic and logarithmic operations on PMU rotor-angle difference time series. The finite-time LLE is estimated as:
\begin{equation}
\lambda^R(k\Delta t) = \frac{1}{Nk\Delta t}
\sum_{m=1}^{N} \ln\!\left(
\frac{|\theta_i^R((k{+}m)\Delta t) -
       \theta_i^R((k{+}m{-}1)\Delta t)|}
     {|\theta_i^R(m\Delta t) -
       \theta_i^R((m{-}1)\Delta t)|}
\right),
\label{eq:dasgupta_mle}
\end{equation}
where $\theta_i^R(t) = \theta_i(t) - \theta_\mathrm{ref}(t)$ is the relative rotor angle of generator $i$ with respect to a reference. A pairwise version identifies coherent generator groups online: generators $i$ and $j$ are coherent if $\lambda^R(k\Delta t) < 0$, and incoherent (swinging apart) if $\lambda^R(k\Delta t) > 0$.

Wang \textit{et al.}\cite{wang_pmu-based_2016} extended this framework by incorporating frequency-dependent load models and excitation-system voltage dynamics, comparing four modeling configurations: load-reduced (LR), load-reduced with exciter (LRE), structure-preserving (SP), and structure-preserving with exciter (SPE). The SP and SPE configurations significantly improved trajectory estimation and MLE accuracy for slow-divergence and multi-swing scenarios, demonstrating that the quality of the underlying state estimation directly impacts MLE reliability.

\subsubsection{Critical Generator Pair and Adaptive Windows}

Huang \textit{et al.}\cite{huang_wide-area_2018} proposed a fast TSA method that combines LLE computation with a\emph{critical generator pair} (CGP) identification algorithm. Rather than tracking all generator-pair angles, the method first identifies the pair with the largest angular separation as the CGP, then computes LLE only on the CGP trajectory. The stability decision combines the LLE zero-crossing time with the sign of the relative rotor speed $\Delta\omega_\mathrm{ZCP}$ at the crossing:
\begin{equation}
\text{Stable iff } \lambda_1 \to \text{negative, or }
\bigl(\lambda_1 \text{ crosses 0 and }
\Delta\omega_\mathrm{ZCP} < 0\bigr).
\label{eq:huang_rule}
\end{equation}
This criterion exploits the physical fact that a returning swing (converging back to equilibrium) has negative relative rotor speed at the LLE zero-crossing, while a diverging swing has positive $\Delta\omega$. The approach was validated on 39- and 10-bus systems.

Guo \textit{et al.}\cite{guo_lyapunov_2014} specifically studied the effect of window-size selection on MLE
reliability. They proposed an adaptive window
$W = 1/f_\mathrm{dom}$ proportional to the inverse of
the dominant oscillation frequency obtained by online FFT
of the rotor-angle signal. A consistency check comparing
two MLE estimates from slightly shifted windows further
reduces false predictions near the stability boundary.
Banerjee \textit{et al.}\cite{banerjee_lyapunov_2014}
and Wadduwage \textit{et al.}\cite{wadduwage_hybrid_2015}
combined LLE estimation with trajectory-separation heuristics
to improve detection speed, demonstrating reliable TSA on
IEEE 9-bus and 68-bus systems.

Wei \textit{et al.}\cite{wei_model-free_2018} proposed
a recursive nearest-neighbor MLE estimator with an
adaptive time-window strategy specifically designed for
streaming PMU data. Rather than using only the sign of the
final LLE value, their method monitors the \emph{slope
trend} of the LLE curve: a consistently rising trend
toward positive values is classified as unstable even
before the LLE crosses zero, enabling earlier detection.
This was validated on the IEEE 39-bus system with various
fault types.

\subsubsection{Predictive MLE and Trajectory Extrapolation}

Zong \textit{et al.}\cite{zong_transient_2024} proposed
the \emph{Predictive Maximum Lyapunov Exponent} (PMLE),
which replaces measured trajectory divergence with
predicted trajectory divergence. The relative rotor speed
is predicted using a switching regression model
(exponential for monotone trajectories,
Fourier-series for oscillating ones), and the relative
rotor angle is extrapolated using a self-memory nonlinear
model. The PMLE is then computed from the predicted
trajectories, providing an earlier stability decision
than any retrospective MLE estimate. Stability is
classified via a geometric criterion in the
$(PMLE,\, \mathrm{d}PMLE/\mathrm{d}t)$ state plane.
Validation on the IEEE 39-bus and a 2,131-bus system
demonstrated average detection times approximately 40\%
shorter than competing model-free MLE methods.

Niu \textit{et al.}\cite{niu_transient_2024} shifted the
observable from generator angles to critical
\emph{transmission branch} active power responses,
deriving a relationship between branch power oscillation
dynamics and the underlying rotor-angle divergence. MLE
is then estimated from the branch power time series,
making the method applicable even when direct rotor-angle
PMU measurements are unavailable or incomplete.

\subsubsection{Modified LLE Formulations for
Practical Deployment}

Said and Matsuo \cite{said_real-time_2024} proposed the
\emph{Instantaneous Lyapunov Exponent} (ILE) framework,
which eliminates the sliding-window and phase-space
reconstruction steps entirely. The ILE is a local,
instantaneous decay/growth rate computed from a single
trajectory sample, complemented by four
Malthusian Parameter Estimators (MPE) that interpret
stability via the local exponential growth rate of a
linearized first-order differential equation.
This approach achieves near-instantaneous stability
indication from streaming PMU data, though at the cost
of reduced noise robustness compared to windowed methods.

\subsection{PMU-Based Short-Term Voltage Stability
Monitoring}
\label{subsec:pmu_voltage}

\subsubsection{Foundational PMU-Voltage-LE Framework}

Short-term voltage stability (STVS) monitoring using
PMU-based LLE was introduced by Dasgupta
\textit{et al.}\cite{dasgupta_pmu-based_2012}, who
applied the same model-free finite-time LLE framework
to PMU voltage magnitude vectors rather than rotor-angle
differences. The system-level LLE is:
\begin{equation}
\Lambda(k\Delta t) = \frac{1}{Nk\Delta t}
\sum_{m=1}^{N} \ln\!\left(
\frac{\|\mathbf{V}_{(k+m)\Delta t} -
      \mathbf{V}_{(k+m-1)\Delta t}\|}
     {\|\mathbf{V}_{m\Delta t} -
      \mathbf{V}_{(m-1)\Delta t}\|}
\right),
\label{eq:dasgupta_stvs}
\end{equation}
where $\mathbf{V}_t \in \mathbb{R}^{n_\mathrm{bus}}$ is
the PMU voltage magnitude vector. An individual-bus LLE
uses the scalar $v_i(t)$ in place of the vector
$\mathbf{V}_t$, enabling bus-level instability ranking.
A key theoretical result in \cite{dasgupta_real-time_2013}
proves that if any individual bus LLE becomes positive,
the system-level LLE must also become positive
asymptotically, providing a formal justification for
bus-level monitoring.

Validation on the IEEE 162-bus system showed that the
system LLE remains negative for stable post-fault
trajectories and becomes positive during voltage collapse
\cite{dasgupta_real-time_2013}. A real-time cyber-physical
implementation using the chain RTDS $\to$ SEL-421 PMU
$\to$ OpenPDC $\to$ MATLAB was demonstrated on a 9-bus
system \cite{reddy_pmu_2014}, confirming that the full
PMU-PDC-computational chain can support LLE-based STVS
monitoring at practical measurement rates.

\subsubsection{Oscillation Contamination and the Phase-Correction Problem}

The most significant practical challenge in PMU-voltage LLE
estimation is \emph{oscillation contamination}: when
post-fault voltage trajectories exhibit damped oscillations
(which are normal and stable), the derivative-based LLE
estimator \eqref{eq:dasgupta_stvs} can spuriously show
positive values because the oscillation increases the
numerator of the logarithm during each cycle. This problem
was systematically studied by Ge \textit{et al.}
\cite{ge_improved_2018}, who proposed fitting the
post-contingency voltage as a sum of damped oscillatory
modes:
\begin{equation}
v_i(t) = v_{i,\infty} +
\sum_{k=1}^{K} A_k\,e^{\sigma_k t}\cos(\omega_k t + \phi_k),
\label{eq:damped_mode}
\end{equation}
and using the \emph{mean trend} $v_{i,\infty}$ (the
DC-component of the voltage trajectory) instead of the
raw voltage in \eqref{eq:dasgupta_stvs}. The resulting
\emph{phase-corrected LLE} eliminates spurious oscillatory
contamination and correctly classifies both stable
oscillatory recovery and true voltage collapse.

Building on this approach, Safavizadeh
\textit{et al.}\cite{safavizadeh_framework_2019}
developed a comprehensive real-time STVS framework using
PMU streams, with explicit treatment of fault detection
logic, communication delay, averaging of recent PMU
packets for noise robustness, GUI, and alarm generation.
The framework achieved STVS identification approximately
5\,s before actual emergency conditions in simulation
studies, providing a practical time budget for
remedial action.

\subsubsection{Streaming LLE Variants and Stability Margin
Quantification}

Gubitz \textit{et al.}\cite{gubitz_online_2018}
introduced three streaming LLE variants designed for
causal (non-future-looking) computation:
\begin{itemize}
  \item \emph{ALE} (Average LLE): average slope between
        successive LLE extrema.
  \item \emph{TALE} (Time-weighted ALE): time-weighted
        version with emphasis on recent behavior.
  \item \emph{M-TALE} (Modified TALE): TALE normalized
        by the voltage overshoot magnitude.
\end{itemize}
These variants not only detect instability (binary
classification) but also \emph{quantify} the stability
margin: M-TALE provides a continuous severity indicator
correlated with how far the trajectory is from the
stability boundary. This is a critical advance over
pure sign-based detection, as it enables graded alerts
and proportional remedial action. The connection between
LLE magnitude and Battery Energy Storage Systems (BESS) activation was also demonstrated:
a persistent positive LLE above a threshold
triggers BESS dispatch, improving the voltage recovery
profile.

\subsubsection{Modified Reference Benchmarks and
Threshold Selection}

Thilakarathne and Wadduwage \cite{thilakarathne_modified_2018}
identified that the standard PMU-based LLE formula can
produce false alarms because the denominator of the
logarithmic ratio is anchored to the initial measurement
segment, which may contain noise-induced fluctuations
unrelated to stability. They proposed replacing the
initial-condition denominator with a reference
based on a predefined instability benchmark, yielding
a modified exponent that is referenced to a known unstable
trajectory shape rather than to an arbitrary initial
fluctuation. This substantially reduces false-positive
alarms for manageable transients on the IEEE 9-bus system.

Khaitan \textit{et al.}\cite{khaitan_thrust_2017}
proposed the THRUST (Transient Hybrid Robust Unification
of Stability Testing) framework, which computes a
\emph{System Stability Index} (SSI) combining voltage
and relative rotor-angle LLE values into a unified
indicator. This was validated on the IEEE 39-bus and a
13,029-bus system, making it one of the largest-scale
LLE-based stability assessments in the literature.
The key contribution of THRUST is breadth: by jointly
covering rotor-angle and voltage instability, it avoids
the common limitation of methods that address only one
stability mode.

\subsubsection{Chaos-Theoretic Embedding for STVS}

Pinzon \textit{et al.}\cite{pinzon_chaos_2018,
pinzon_pmu-based_2019, pinzon_real-time_2019} explicitly
incorporated chaos-theoretic phase-space embedding
(Section \ref{subsec:psr}) into the STVS monitoring
pipeline. Rather than applying LLE directly to raw
voltage samples, they optimize embedding parameters
$(\tau, m, \epsilon)$ using Monte Carlo simulation over
many operating conditions and contingencies, and then
apply Particle Swarm Optimization (PSO) to minimize
classification error. This offline calibration step
ensures that the online LLE computation uses a
consistently reliable parameter set. Comparison between
Asamata and Peng et al. \cite{aslanian_analyzing_2020}
showed that the basic LLE formula without careful
parameter tuning is unreliable for windows shorter
than 2\,s, confirming the necessity of the offline
optimization step.

\subsubsection{Dual-Signal and Multi-Observable
Extensions}

Peng \textit{et al.}\cite{jin_transient_2024}
proposed computing two parallel LLE indicators from the
same PMU data: $M_1$ from voltage magnitude time series
and $M_2$ from voltage phase-angle time series. When $M_1$
exhibits oscillatory contamination, the decision is
delegated to $M_2$. This dual-signal strategy avoids the
need for explicit mode-fitting (as in
\cite{ge_improved_2018}) and improves robustness for
MMC-HVDC system contingencies.

A related approach by Almomani \textit{et al.} \cite{almomani_novel_2025} proposes novel data-driven indices for early STVS quantification, extending the LLE framework toward continuous severity scoring rather than binary detection, and validating the approach on standard IEEE test systems. The proposed method distinguishes between slow voltage recovery and oscillatory behavior of the voltage waveform by employing empirical mode decomposition (EMD) to separate the signal into fast and slow time-scale components. Lyapunov exponents (LE) are then calculated for each component individually.

\subsection{Generator Coherency Identification and
Controlled Islanding}
\label{subsec:coherency}

Generator coherency-the grouping of machines that swing
together after a disturbance-is essential for dynamic
equivalencing, controlled islanding, and wide-area
protection. LLE-based coherency identification exploits
the fact that two generators are coherent if and only if
their relative rotor angle $\Delta\delta_{ij}(t) =
\delta_i(t) - \delta_j(t)$ has a negative LLE (the angle
difference converges), and incoherent if its LLE is
positive (the angle diverges).

\subsubsection{VANTAGE Framework}

Khaitan and McCalley \cite{khaitan_vantage_2013} proposed
the VANTAGE (Viability ANalysis of Transient groups via
phasors) framework, which groups generators iteratively:
a reference generator is selected, all other generators
are assigned to it if their pairwise LLE with the
reference is negative, and the algorithm terminates when
all generator pairs within each group have negative LLEs.
This approach requires no predefined number of groups
(unlike $k$-means clustering), is compatible with either
model-based or model-free LLE estimation, and has average
complexity $\mathcal{O}(NR)$ where $N$ is the number of
generators and $R$ the number of coherent groups.

\subsubsection{Fast Coherency with Auxiliary Conditions}

Huang \textit{et al.}\cite{huang_wams-based_2020}
extended the LLE-based coherency framework using relative
angular speed deviations as auxiliary conditions to
accelerate detection for non-oscillatory cases. Their
method defines five swing trajectory types and an
additional pair of conditions:

\begin{itemize}
  \item \textbf{Criterion 1:} If the LLE--$t$ curve of the relative rotor angle of a generator pair (GP) crosses the zero line, the sign of the angular velocity at the zero-crossing point ($\omega_{\text{ZCP}}$) determines coherency. Specifically, $\omega_{\text{ZCP}} < 0$ indicates coherent generators, while $\omega_{\text{ZCP}} > 0$ indicates noncoherent generators. The corresponding time $t_{\text{ZCP}}$ defines the coherency identification instant.
  
  \item \textbf{Criterion 2:} If the LLE--$t$ curve does not cross the zero line, the final value of the LLE is used for assessment. A negative final LLE value indicates coherent generators, whereas a positive final LLE value indicates noncoherent generators.
\end{itemize}

These auxiliary conditions detect coherency without
waiting for the LLE to converge, significantly reducing
the identification time compared to VANTAGE on the same
test cases.

\subsubsection{Coherency-Based Islanding}

Rashidi and Farjah \cite{rashidi_based_2016} integrated
LLE-based coherency identification into a complete
transient instability prediction and mitigation
framework. The framework: (i) uses LLE computed from
post-fault voltage magnitudes as features for a Support
Vector Machine (SVM) classifier to predict instability;
(ii) identifies coherent generator groups via pairwise
LLE; (iii) determines the optimal islanding boundary
minimizing generation-load imbalance between islands
while maintaining connectivity; and (iv) is robust to
network topology changes and measurement errors.
Validation on the IEEE 39-bus system demonstrated that
the method can identify correct islanding plans within
the post-fault observation window. A comprehensive
taxonomy of coherency detection methods is provided in
\cite{koochi_taxonomy_2019}, where LLE-based methods
are classified among the measurement-based approaches.

\subsection{Contingency Ranking and Security Assessment}
\label{subsec:contingency}

Beyond binary stability detection, the LLE can be used
as a \emph{severity} metric to rank contingencies by
their dynamic risk.

Nangrani and Bhat \cite{nangrani_power_2015,
nangrani_identification_2015} proposed combining the
classical MW Performance Index (MWPI)-which captures
static overload severity-with a \emph{Chaotic
Performance Index} (CPI) derived from the LLE of discrete
power-flow time series recorded under incremental load
variations. The combined index ranks contingencies that
are static-load-safe but dynamically divergent, which
the MWPI alone cannot detect. The method identifies the
most vulnerable transmission line as the one whose
disturbance produces the highest LLE across all bus-power
sequences.

Aghtar \textit{et al.}\cite{aghtar_lyapunov_2023}
proposed a hybrid static-dynamic contingency ranking
framework using the \emph{full Lyapunov spectrum} (not
only the LLE) as the dynamic ranking metric. The
framework first screens critical lines using static
voltage stability indices (L-index, FVSI, loading index),
then computes the average LE $\lambda_a = \sum_i
\lambda_i $ from the reconstructed phase space of
the selected lines' power time series. A more negative
$\lambda_a$ indicates stronger stability; a less negative
or positive $\lambda_a$ indicates proximity to or
crossing of the instability boundary. This was validated
on the IEEE 36-bus and IEEE 118-bus systems.

At the largest scale, R\"{u}eger \textit{et al.}
\cite{rueger_lyapunov_2019} applied an LLE-based severity
metric to the ENTSO-E continental dynamic model with
approximately 23,000 buses and 6,000 generators,
developing an automated workflow in PowerFactory and
Python. The LLE is computed from frequency oscillation
envelopes after disturbances and combined with
frequency amplitude deviation and rate-of-change-of-frequency
(RoCoF) into a multi-criteria global performance
indicator for disturbance ranking.

Ravikumar and Khaparde \cite{ravikumar_taxonomy_2018}
placed LLE-based indicators within a broader taxonomy of
five classes of PMU-derived catastrophic stability
indicators (time-domain, frequency-domain, energy-function,
Lyapunov-based, and AI-based), providing a useful
comparative framework for practitioners selecting
monitoring indices for wide-area protection systems.

\subsection{Stochastic Stability and Uncertainty
Propagation}
\label{subsec:stochastic}

Classical LLE-based TSA and STVS methods treat disturbances
as deterministic events. The growing share of stochastic
renewable generation requires an extended framework that
quantifies how uncertainty propagates through the nonlinear
grid dynamics and affects stability margins.

Namachchivaya \textit{et al.}\cite{arnold_stochastic_1991}
provided the mathematical foundations for the
almost-sure Lyapunov exponent of stochastic dynamical
systems. Verdejo \textit{et al.}\cite{verdejo_stability_2012,
verdejo_application_2015} applied this theory to power
systems, computing the stochastic MLE of the linearized
system under stochastic disturbances on test networks from 10 to
68 machines. The key operational insight is that the
stochastic MLE is a monotonically increasing function of
the disturbance intensity $\rho$: there exists a critical
$\rho^*$ at which $\lambda$ transitions from negative to
zero, defining the maximum tolerable noise intensity for
almost-sure stability.

The Kazma--Taha framework \cite{kazma_stability_2025}
extended this to the full nonlinear DAE setting,
introducing a state-deformation tensor that tracks
how uncertain renewable injections at different buses
propagate as perturbation growth rates through the
full nonlinear flow. Bus stability indices are computed
from the parameterized LE spectrum, and the optimal
bus selection for renewable hosting is solved as a
submodular optimization. This framework is the most
mathematically complete treatment of uncertainty
propagation via LEs in the power-systems literature.

\subsection{Optimal PMU Placement for Stability
Monitoring}
\label{subsec:pmu_placement}

Effective deployment of PMU-based LLE monitoring requires
identifying which buses must be instrumented to maximize
the stability-monitoring information content of the
measurement system.

Rashidi and Farjah\cite{rashidi_lyapunov_2016} proposed
an LLE-based PMU placement formulation that augments the
classical observability-coverage constraint with a
stability-significance objective. Buses whose individual
LLE values most accurately track the system-level LLE
are identified as \emph{critical buses} for stability
monitoring. The PMU placement problem is then formulated
to maximize measurement redundancy at these critical buses
without increasing the minimum PMU count required for
full observability. The approach was validated on the
IEEE 39-bus system under multiple operating conditions.

A complementary approach by Rashidi \textit{et al.}
\cite{rashidi_based_2016} demonstrated that even with
a limited number of optimally placed PMUs, LLE-based
instability prediction using SVM features achieves
accuracy comparable to full-observability methods,
confirming the practical value of stability-aware
PMU placement.

Yoon \textit{et al.}\cite{yoon_optimal_2020} extended
the LLE-based placement idea to the problem of
distributed generator (DG) siting in distribution
systems. Rather than placing sensors, this work places
generation: the LLE computed from measured voltage-frequency
time series identifies buses with the highest divergence
rate (most dynamically vulnerable), and DGs are sited
at these buses to reduce the positive LLE magnitude.
The approach avoids full dynamic modeling and provides
a data-driven, dynamics-aware DG placement criterion.

\subsection{Hybrid Machine Learning and LLE Methods}
\label{subsec:ml_hybrid}

The most recent phase of the stability-assessment
literature (2020-present) is characterized by the
fusion of LLE-based nonlinear dynamics indicators with
machine-learning classifiers, motivated by two
complementary goals: improving classification speed
relative to standalone LLE methods, and improving
physical interpretability relative to pure deep-learning
approaches.

\subsubsection{LLE as Feature for ML Classifiers}

Rashidi \textit{et al.}\cite{rashidi_based_2016} were
among the first to use LLE values as features for SVM
classification. Post-fault LLEs computed from voltage
magnitudes at critical PMU buses are fed as an
input vector to an SVM, which classifies the post-fault
state as stable or unstable. The method is robust to
network topology changes and measurement noise, and
generalizes to unseen contingency types within the
same operating conditions.

Oyama \textit{et al.}\cite{oyama_online_2020} proposed
a two-stage hybrid framework: Decision Trees (DTs) rapidly
classify first-swing-dominated contingencies from
pre-computed offline training data, and the remaining
ambiguous or multi-swing cases are evaluated by online
LLE computation. This explicit role-sharing exploits the
speed of DT inference for simple cases while reserving
the physics-grounded LLE evaluation for complex
near-boundary cases where data-driven classifiers are
less reliable.

Umbereen \textit{et al.}\cite{umbereen_investigating_2024}
compared Maximum Lyapunov Exponent (MLE) and Convolutional Neural Network (CNN) approaches for transient stability assessment (TSA) on the Nordic32 system. MLE demonstrates higher accuracy and reliability, particularly in unstable and fault scenarios, achieving up to 100\% accuracy for unstable cases and 98\% overall. In contrast, CNN enables faster decision-making but shows reduced accuracy under fault conditions, with significantly lower precision and recall. While CNN is effective when sufficient historical data are available, MLE is more suitable for data-scarce environments and provides slightly longer early warning times, indicating more robust instability detection.

The
comparison revealed that CNN provides faster
classification for simple first-swing cases, while MLE
gives earlier and more reliable warning for multi-swing
instability near the stability boundary.
The authors proposed a \emph{hybrid architecture}: CNN
for initial fast detection, MLE for verification and
severity scoring.

Dabou \textit{et al.}\cite{dabou_time_2021}
constructed a high-dimensional stability index (HDSI)
from 500+ interpretable features extracted from PMU
rotor-speed and rate-of-change-of-transient energy function (RoCoTE)
signals via five signal-processing engines (including
Lyapunov-based divergence metrics). These features
were used to train SVM, Random Forest, AdaBoost, and
DT classifiers on the IEEE 39-bus and 68-bus systems,
with LLE-derived features among the most discriminative.

\subsubsection{Koopman Operator Methods}

Two recent papers replaced phase-space reconstruction
with Koopman-operator-based dynamics extraction as a
preprocessing step before LLE computation
\cite{zhang_koopman_2024, gao_short-term_2025}.
Zhang \textit{et al.}\cite{zhang_koopman_2024} applied
Dynamic Mode Decomposition (DMD) to extract the dominant
Koopman eigenmode from a short post-fault voltage window
($\sim$0.5\,s), reconstructed the trend from this
dominant mode, and computed LLE on the reconstructed
trend rather than on raw oscillatory measurements.
This approach reduces the LLE-decision time from
1.24--2.66\,s (conventional MLE) to approximately
0.5\,s post-fault clearing on the Nordic32 system.

Gao \textit{et al.}\cite{gao_short-term_2025} combined
Hankel-matrix-enhanced EDMD (HEDMD) with predictive MLE
to achieve fast STVS prediction even from very short
measurement windows, validated on Nordic32 and a CEPRI
test system.



%
\subsection{IBR-Dominated and Hybrid Grids}
\label{subsec:ibr_stability}

The transition to inverter-based resources (IBRs)
introduces qualitatively different dynamics compared
to synchronous-generator-dominated grids: inertia is
virtual or absent, synchronization is software-defined,
and control-loop timescales span from microseconds to
seconds. This raises fundamental questions about whether
LE-based methods developed for SG grids remain valid
and how they must be adapted.

\subsubsection{DFIG and Wind Power Systems}

Several papers have studied the chaotic characteristics
of Doubly-Fed Induction Generator (DFIG) wind turbines
using the full LE spectrum
\cite{cao_chaotic_2025}.
Cao and Song \cite{cao_chaotic_2025} built a
nine-dimensional nonlinear DFIG model and computed the
complete LE spectrum as a function of converter control
parameters (proportional gains $k_{p2}$, $k_{i2}$) and
wind speed. Chaos occurs when any exponent becomes
positive, and the paper derives the chaos-parameter
boundary in the $(k_{p2}, k_{i2})$ plane-directly
useful for controller design to avoid chaotic regimes.

Shao \textit{et al.}\cite{shao_stability_2024}
studied combined wind-solar integration effects, showing
that renewable output variability can push the system
from periodic to chaotic voltage oscillation as renewable
reactive power injection crosses a critical threshold.
The LLE quantifies this threshold and provides a
safety margin for renewable dispatch curtailment.

\subsubsection{AC/DC Hybrid Systems with Commutation Failure}


Similarly, Wang \textit{et al.}\cite{wang_transient_2024}
developed a Switching System MLE (SSMLE) method for
hybrid grids containing grid-following (GFL) and
grid-forming (GFM) inverters, where voltage dynamics
involve mode switching between normal, LVRT, and
limiter-active regimes. Model-based MLE computation
with QR decomposition is applied to each subsystem's
differential equations, with switching compensation
applied at mode boundaries.

\subsubsection{Generalized Voltage Damping for IBR Grids}

The most theoretically complete treatment of STVS in
IBR-dominated grids is the two-part series by Peng
\textit{et al.}\cite{peng_measuring_2025,
peng_measuring_2026}. These papers introduce a
\emph{generalized voltage damping} (GVD) framework
that abandons raw-voltage-based LLE in favor of LLE
applied to \emph{voltage-related transient energy}
(VTE). The system-wise GVD (sGVD) and port-wise GVD
(pGVD) are defined as:
\begin{equation}
K_V^{(s)} = -\mathrm{MLE}(\mathrm{VTE}^{(s)}),
\quad
K_V^{(i)} = -\mathrm{MLE}(\mathrm{VTE}^{(i)}),
\label{eq:gvd}
\end{equation}
where $\mathrm{VTE}^{(s)}$ and $\mathrm{VTE}^{(i)}$ are
the system-wise and port-wise voltage transient energies,
respectively. The average of pGVDs across all ports
equals the sGVD, enabling device-level accountability:
each IBR's pGVD quantifies its individual contribution
to or deterioration of system STVS. This framework is
device-agnostic (applicable to any port with measurable
voltage and power) and provides a natural bridge between
the physics of energy dissipation and the empirical LLE
measurement.

\subsection{Wind and Renewable Impact on Stability}
\label{subsec:renewable_stability}

The impact of renewable energy integration on LLE-based
stability margins has been studied from both model-based
and data-driven perspectives.

Amiri and Selvaraj \cite{amiri_lyapunov_2019, selvaraj_investigation_2021}
applied model-based MLE to systematically quantify how
wind penetration degrades the stability margin on the
IEEE 39-bus system. By scanning initial conditions
in the state space, they mapped the stability region
boundary and showed that replacing synchronous generators
with FSIG wind turbines at certain buses contracts the
stability region and increases the MLE under the same
contingency. The implication is that MLE-based stability
region mapping can directly inform wind farm siting
decisions.


Kazma and Taha \cite{kazma_stability_2025} addressed
this more rigorously: their log-det stability metric
identifies which buses can optimally absorb stochastic
renewable injections while minimally propagating
instability across the network. Their renewable
allocation algorithm was validated on IEEE 9-, 39-,
and 200-bus systems.

\subsection{Open Problems in Stability Assessment}
\label{subsec:stability_gaps}

Despite the substantial progress reviewed above, several
critical gaps and open problems remain.

\textbf{(1) Oscillation contamination in STVS.}
The phase-correction approaches of
\cite{ge_improved_2018} and the Koopman-preprocessing
of \cite{zhang_koopman_2024, gao_short-term_2025} reduce
but do not fully resolve the problem of LLE contamination
by damped oscillations. No systematic, parameter-free
solution exists, particularly for mixed-mode oscillations.

\textbf{(2) FTLE window selection.}
The choice of window length $T$ for finite-time LLE
estimation critically affects both detection speed and
reliability. Shorter windows detect instability faster
but are more noise-sensitive; longer windows are more
reliable but delay the alarm. No universally accepted
adaptive criterion exists, though \cite{guo_lyapunov_2014}
and \cite{zong_transient_2024} represent partial solutions.

\textbf{(3) LLE under IBR dynamics.}
The theoretical basis for LLE-based stability assessment
was developed for inertia-dominated SG systems. IBR
control loops operate on microsecond timescales and
introduce non-smooth switching dynamics (LVRT, current
limiters, mode transitions) that are fundamentally
incompatible with classical smooth Lyapunov theory.
The GVD/pGVD framework \cite{peng_measuring_2025,
peng_measuring_2026} and switching MLE
\cite{ wang_transient_2024} represent
important advances, but a complete theoretical framework
for LLE-based TSA in fully IBR-dominated systems is
still lacking.


\textbf{(4) Scalability to large real-world systems.}
Most PMU-based LLE methods have been validated on
benchmark test systems of up to a few hundred buses.
Real-time deployment in systems with thousands of
PMUs, high measurement noise, communication latency,
missing data, and time-varying topology remains largely
undemonstrated. The THRUST validation on a 13,029-bus
system \cite{khaitan_thrust_2017} is a notable exception
but represents an offline study rather than a live
deployment.

\textbf{(5) Stability margin quantification.}
Current LLE-based methods provide binary stable/unstable
decisions or, at best, a qualitative severity ordering.
A rigorous, quantitative stability margin
(analogous to the energy margin in direct methods or the
damping ratio in modal analysis) derived from the LLE
and validated against physical instability thresholds
remains an open problem.

Table \ref{tab:stability_summary} provides a structured
summary of the key papers reviewed in this section,
organized by subproblem, approach, test system, and
identified limitations.

\begin{table*}[!t]
\renewcommand{\arraystretch}{1.3}
\caption{Summary of Key Lyapunov Exponent-Based Stability
Assessment Papers}
\label{tab:stability_summary}
\centering
\small
\begin{tabular}{p{3.2cm}p{2.2cm}p{2.8cm}p{2.6cm}p{4.0cm}}
\toprule
\textbf{Reference} &
\textbf{Stability type} &
\textbf{Approach} &
\textbf{Test system} &
\textbf{Key contribution / limitation} \\
\midrule
Wadduwage 2013 \cite{wadduwage_power_2013}
  & Transient (RA)
  & Model-based, GSR
  & IEEE 9-, 69-bus
  & CCT \& stability region; offline only \\
Hayes 2018 \cite{hayes_viable_2018}
  & Transient (RA)
  & Model-based, discrete-QR
  & Up to 1479-bus
  & Scalable DAE-LLE; no real-time \\
Kazma 2025 \cite{kazma_stability_2025}
  & Multi-mode (joint)
  & Model-based, logdet tensor
  & 9-, 39-, 200-bus
  & Renewable allocation; IBR pending \\
Liu 1994 \cite{liu_detection_1994}
  & Transient (RA)
  & PMU windowed LLE
  & 3-machine
  & First PMU-based chaos detection \\
Yan 2011 \cite{yan_pmu-based_2011}
  & Transient (RA)
  & Model-aided PMU
  & 200-bus WECC-like
  & Formal MLE stability certificate \\
Dasgupta 2015 \cite{dasgupta_pmu-based_2015}
  & Transient (RA)
  & Model-free PMU
  & IEEE 162-bus
  & Coherency detection; binary only \\
Wei 2018 \cite{wei_model-free_2018}
  & Transient (RA)
  & Recursive PMU MLE
  & IEEE 39-bus
  & Slope-trend decision; noise sensitive \\
Dasgupta 2012 \cite{dasgupta_pmu-based_2012,dasgupta_pmu-based_2015}
  & Voltage (STVS)
  & Model-free PMU
  & WSCC 9-bus
  & Foundational STVS-LLE framework \\
Ge 2018 \cite{ge_improved_2018}
  & Voltage (STVS)
  & Phase-corrected PMU
  & IEEE 39-bus
  & Resolves oscillation contamination \\
Gubitz 2018 \cite{gubitz_online_2018}
  & Voltage (STVS)
  & Streaming ALE/TALE
  & IEEE 39-bus
  & Severity quantification; BESS trigger \\
Khaitan 2017 \cite{khaitan_thrust_2017}
  & Multi-mode (joint)
  & PMU SSI
  & Up to 13029-bus
  & Largest-scale LLE validation \\
Zhang 2024 \cite{zhang_koopman_2024}
  & Voltage (STVS)
  & Koopman-DMD + LLE
  & Nordic32
  & 0.5\,s detection; IBR pending \\
Peng 2025 \cite{peng_measuring_2025}
  & Voltage (STVS)
  & GVD / VTE-based LLE
  & IBR test systems
  & IBR-compatible; energy-physical basis\\
Verdejo 2015 \cite{verdejo_application_2015}
  & Stochastic stability
  & Almost-sure MLE
  & Up to 68-machine
  & Stochastic stability margin \\
\bottomrule
\end{tabular}
\end{table*}

%

%
%

%
\section{Chaos Detection, Bifurcation Analysis, and Control}
\label{sec:chaos}

This section reviews the literature that uses Lyapunov
exponents (LEs) to detect, characterize, and suppress
chaotic oscillations in power systems.
The section is organized as follows.
Section \ref{subsec:routes_chaos} covers routes to chaos
and classical bifurcation analysis in power system models.
Section \ref{subsec:chaos_smib} reviews chaos in
single-machine and reduced-order models, the largest class
of studies.
Section \ref{subsec:chaos_components} covers
component-level chaos in generators, converters, and
renewable energy devices.
Section \ref{subsec:chaos_networks} addresses chaos
in multi-machine networks and microgrids, including the
newly identified class of hidden-attractor-driven
sub-synchronous oscillations.
Section \ref{subsec:chaos_ferrores} covers ferroresonance
and sub-synchronous resonance.
Section \ref{subsec:chaos_control} reviews the diverse
body of chaos suppression methods, organized by controller
type.
Section \ref{subsec:chaos_sync} addresses chaos
synchronization as a control paradigm.
Section \ref{subsec:chaos_fractional} covers the growing
sub-literature on fractional-order chaos models and
controllers.
Section \ref{subsec:chaos_gaps} identifies open problems.

\subsection{Routes to Chaos and Bifurcation Analysis}
\label{subsec:routes_chaos}

Chaos in power systems does not arise abruptly; it follows
structured pathways from stable equilibria through a
sequence of bifurcations. Understanding these pathways is
a prerequisite for designing targeted control interventions.

\subsubsection{Hopf Bifurcation and Period-Doubling
Cascade}

The primary route to chaos documented in the power
systems literature is the \emph{Feigenbaum period-doubling
cascade}: as a bifurcation parameter (mechanical input
power $P_m$, load reactive demand $Q$, 
or damping coefficient $D$) is varied, the system
undergoes a sequence of bifurcations
$T \to 2T \to 4T \to \cdots$, converging to a chaotic
attractor at a finite parameter value
\cite{srivastava_chaotic_1993, li_chaotic_2009,
al-hussein_chaos_2021}. In many models this cascade is
preceded by a Hopf bifurcation-the emergence of a limit
cycle from a stable equilibrium-and followed by
\emph{boundary crises} in which the chaotic attractor
collides with an unstable manifold and is destroyed,
leaving the trajectory to collapse or diverge
\cite{nayfeh_bifurcations_1995,
srivastava_elimination_1998}.

The foundational demonstration in a realistic DAE model
is due to Li and Ca\ {n}izares \cite{li_chaotic_2009},
who showed that the IEEE 14-bus system with standard
transient-stability DAE equations-without any
simplifying assumptions-exhibits a full sequence of
Hopf bifurcation, period-doubling, and chaotic attractors
under a single-contingency scenario. Critically, they
also showed that adding a power system stabilizer (PSS)
eliminates the chaotic regime, establishing a direct
link between chaos detection by LLE and the practical
effectiveness of conventional damping controllers.

The 1995 study by Nayfeh \textit{et al.}\cite{nayfeh_bifurcations_1995}
provided the most complete bifurcation taxonomy for a
reduced model with dynamic loads: saddle-node, subcritical
Hopf, supercritical Hopf, cyclic-fold, period-doubling,
limit points of periodic orbits, and
chaos were all identified. Example chaotic Lyapunov
spectra reported are $(0.32,\, 0,\, -4.75,\, -80.50)$
and $(0.38,\, 0,\, -22.44,\, -104.8)$, confirming
high-dimensional chaotic attractors near voltage-collapse
conditions. The paper also introduced a nonlinear feedback
control $u = K w^3$ that converts a subcritical Hopf into
a supercritical Hopf, widening the stable operating
region-an early application of bifurcation engineering
for power systems.

The survey by Yibei \textit{et al.}\cite{yibei_research_2011}
systematized these results across the literature,
identifying three routes to chaos in power systems
(period-doubling, quasiperiodic, and intermittency),
and five bifurcation types (saddle-node, Hopf,
period-doubling, cyclic-fold, and torus) alongside their
LLE signatures. The survey emphasized that LEs are the
primary quantitative diagnostic for distinguishing these
regimes.

\subsubsection{Stochastic and Parameter-Uncertainty-Induced
Chaos}

A conceptually important extension of deterministic
bifurcation analysis is the observation that random
parameter variations can \emph{induce} chaos even in
systems that are safely periodic under their nominal
parameters. Qin \textit{et al.}\cite{qin_random_2014}
studied the forced SMIB model with randomized damping:
\begin{equation}
\ddot{\theta} + (\bar{\lambda} + \rho\xi(t))\dot{\theta}
+ \eta\sin\theta = k\sin(\omega t),
\label{eq:stoch_smib}
\end{equation}
where $\xi(t)$ is zero-mean noise of intensity $\rho$.
Using a stochastic Melnikov method, a necessary condition
for noise-induced chaos is derived:
%
\begin{equation}
\rho^2 \frac{32\pi \sqrt{\eta}}{\eta} S_0 
+ \left( 2k \frac{\pi}{\sqrt{\eta}} \, \text{sech}\left(\frac{\omega \pi}{2\sqrt{\eta}}\right) \right)^2 
> \left( \frac{8\lambda}{\sqrt{\eta}} \right)^2
\label{eq:melnikov}
\end{equation}
where $\theta_0'(t)$ is the heteroclinic orbit of the
unperturbed system and $\mathcal{M}_d$ is the
deterministic Melnikov function. As $\rho$ increases
beyond a threshold, the LLE transitions from negative to
positive, and safe-basin erosion accelerates. This result
has direct implications for grid security assessment:
measurement uncertainty and ambient disturbances can push
a nominally stable system into chaotic operation without
any discrete fault event.

\subsection{Chaos in Single-Machine and Reduced-Order
Models}
\label{subsec:chaos_smib}

The single-machine infinite-bus (SMIB) swing equation
remains the most widely used testbed for chaos analysis,
owing to its mathematical tractability and its role as
the simplest nonlinear oscillator in power engineering.

\subsubsection{Classical Forced SMIB Analysis}

The foundational forced SMIB model:
\begin{align}
\dot{x}_1 &= x_2, \notag \\
\dot{x}_2 &= -c\,x_2 - b\sin x_1 + f\sin(x_3), \notag \\
\dot{x}_3 &= \omega,
\label{eq:forced_smib}
\end{align}
with forcing amplitude $f$ and frequency $\omega$ as
bifurcation parameters, is studied in multiple papers.
Chen \textit{et al.}\cite{chen_dynamic_2005} performed
a comprehensive study including: local stability of
equilibrium points via a direct Lyapunov function
$V = \frac{1}{2}x_2^2 + \frac{b}{2}(1 - \cos x_1)$,
bifurcation in $f$, phase portraits, Poincar\'{e} maps,
LLE computation, chaos suppression by feedback, and
\emph{chaotification} (intentionally generating chaos
from a stable operating point using time-delay feedback).
The Lyapunov spectrum at $f = 2.41$ is reported as
$(\lambda_1,\lambda_2,\lambda_3) = (0.0543,\, 0,\,
-0.5543)$, confirming chaos; the sum equals the negative
damping coefficient, consistent with Liouville's theorem.

Nangrani and Bhat \cite{nangrani_fractional_2018}
studied the IEEE 14-bus system under transient turbine
torque ripple, modeled as a damped sinusoidal disturbance
superimposed on the mechanical torque. A Duffing-like
approximation of the rotor dynamics confirmed that
lower-frequency ripple ($\approx$1.5--1.6 rad/s) produces
strange attractors, and that a fractional order proportional integral (FOPI) controller
can mitigate the resulting instability by injecting a
corrective torque through state feedback.

Nangrani \textit{et al.}\cite{nangrani_chaotic_2015}
showed that turbine torque ripple alone, without any
electrical forcing, can drive the SMIB rotor into
chaotic oscillation, producing phase trajectories that
transition from limit cycles to strange attractors at
specific ripple frequencies. This underscores that chaos
can originate from the mechanical side of the
generator-a pathway often neglected in electrical-domain
models.

Lo and Zhu \cite{lo_chaotic_1998} introduced the
concept of \emph{Arnold diffusion} to power system
analysis: in a multi-machine Hamiltonian model, random-like
chaotic drift can carry the system state into any region
of the phase space consistent with energy conservation,
unlike ordinary chaos, which is confined to a specific
attractor region. The LLE was used to confirm chaotic
motion, and the phase-space diffusion metrics distinguished
ordinary chaos from Arnold diffusion.

Srivastava \textit{et al.}\cite{srivastava_chaotic_1993}
applied the complete attractor-classification scheme
of Table \ref{tab:attractor_types} to a disturbed Three-Bus test system
with OLTC, SVS, and dynamic load dynamics,
reporting a chaotic spectrum $(+0.331,\, 0,\, -2.03,\,
-33.6)$ and connecting chaotic dynamics to a
Hopf-bifurcation-to-voltage-collapse route.

A cautionary result is provided by
Lai \textit{et al.}\cite{lai_illustrating_1993},
who showed that complex nonlinear transients near
voltage-collapse boundaries are \emph{not always chaotic}:
for their test case the computed Lyapunov spectrum was
entirely nonpositive, and trajectories eventually
converged. This is a methodological warning: positive LLE
must be verified with consistent Poincar\'{e} section and
power spectrum analysis before chaos is claimed.

Gupta and Singh \cite{noauthor_nonlinear_nodate}
studied a three-machine infinite-bus system, tracing
a period-doubling route to chaos in rotor-angle
dynamics as internal parameters vary, and using LLE
alongside bifurcation diagrams to characterize the
transition. Gupta \textit{et al.}\cite{gupta_chaos_2022}
extended this to a small-scale grid with combined
wind variability, periodic load, and Gaussian noise,
demonstrating multistability and safe-basin erosion
through simultaneous use of LLE, Lyapunov dimension,
and attractor-coexistence analysis.

\subsection{Chaos in Components: Generators, Converters,
and Renewable Devices}
\label{subsec:chaos_components}

\subsubsection{Synchronous Generator with Excitation}

The extension of chaos analysis to generator-excitation
dynamics requires a higher-order model. Min
\textit{et al.}\cite{min_fractional-order_2017} and
\cite{min_fractional-order_2018} developed fractional-order
SMIB models that include excitation-system states,
yielding 4D nonlinear systems. In the 2017 paper, chaos
onset occurs when the fractional order exceeds
$q \approx 3.992$ (product of four per-state orders);
the LLE becomes positive at this threshold, confirming
that fractional memory effects expand the chaotic
parameter region relative to the integer-order model.

Li \textit{et al.}\cite{li_controlling_2019}
studied chaos in a six-dimensional power system model
combining rotor, excitation, and turbine-governor dynamics.
The uncontrolled model is confirmed chaotic with
$\lambda_1 = 0.4359$ (positive LLE confirmed by phase
portrait, broadband power spectrum, and Poincar\'{e} map).
A key modelling insight is that incorporating
turbine-governor dynamics significantly changes the
chaotic parameter range compared with the simpler 2D and 4D
models that dominate the earlier literature.

Shariatmadar \textit{et al.}\cite{shariatmadar_lyapunov_2013}
studied the effect of internal stator winding faults on
machine chaos using an extended discrete Poincar\'{e}
map. The winding-function-based inductance model under a
turn-to-turn fault breaks the periodicity of the
machine's limit cycle, inducing bifurcation and chaotic
motion whose LLE exceeds zero. This is an early paper
connecting component-level fault physics directly to
chaos diagnosis, foreshadowing the protection
applications discussed in Section \ref{sec:protection}.

Wang and Song \cite{wang_analysis_2019} studied
nonlinear oscillations in the rotor-mechanical system
(electromagnetic-mechanical coupling) under parametric
excitation, identifying quasi-periodic and chaotic
regimes in the bifurcation diagram with the excitation
amplitude as the control parameter.

\subsubsection{Doubly-Fed Induction Generator (DFIG)}

The DFIG has received increasing attention as a source
of chaotic oscillations in renewable-dominated systems.
Cao and Song \cite{cao_chaotic_2025} constructed a
nine-dimensional nonlinear DFIG model with states
$\mathbf{x} = [\omega_w, \omega_r, \theta_s, x_1,
\ldots, x_4, i_{rd}, i_{rq}]^T$, computed the full
Lyapunov spectrum, and mapped the chaos-parameter
boundary in the $(k_{p2}, k_{i2})$ controller-gain plane.
Chaos occurs when any exponent becomes positive, and
the chaos boundary is a direct design constraint for
converter tuning.

Chen \textit{et al.}\cite{chen_criterion_2023}
derived an analytical Hopf bifurcation criterion for a
4-state DFIG model using the Routh-Hurwitz stability
condition applied to the characteristic polynomial:
\begin{equation}
\lambda^4 + a_3\lambda^3 + a_2\lambda^2
+ a_1\lambda + a_0 = 0,
\label{eq:dfig_char}
\end{equation}
identifying the parameter condition under which a
conjugate pair of eigenvalues crosses the imaginary axis.
Beyond the Hopf point, the system follows the
period-doubling route to chaos as confirmed by positive
LLE. This provides an explicit analytical instability
boundary rather than a purely numerical one.

Chen
\textit{et al.}\cite{chen_supercritical_2025}
(fractional-order DFIG) identified supercritical
pitchfork bifurcations as a second mechanism for chaos
onset in converter-connected renewable generators.
Using the Central Manifold Theorem to reduce the
3D fractional DFIG to a 1D normal form near the
critical equilibrium, it was shown analytically that
a supercritical pitchfork bifurcation occurs at
$u_d = \gamma - 1$ (normalized rotor d-axis voltage).
Decreasing the fractional order $q$ below a threshold
eliminates chaos entirely, suggesting that fractional-order
control is a potential anti-chaos mechanism.

Fang \textit{et al.}\cite{fang_analysis_2025} extended
the analysis to an induction motor under periodic forcing,
identifying forced nonlinear oscillation, quasi-periodicity,
and chaos transitions, with LLE used to delineate the
boundaries.

\subsubsection{Converter-Interfaced Small Wind Turbines}

Orlando \textit{et al.}\cite{orlando_new_2012}
identified a new class of chaotic phenomena: voltage-sag
events can trigger chaotic oscillations in back-to-back
converter small wind turbine systems through interaction
between the sag-induced transient and the converter
control dynamics. LLE confirmed positive values during
these chaotic post-sag trajectories, and bifurcation
maps in the (sag magnitude, sag duration) plane
delineated the chaotic zone. This result is relevant to
grid-code LVRT requirements, since converter control
parameters designed for normal operation may inadvertently
produce chaotic dynamics under fault conditions.

\subsection{Chaos in Multi-Machine Networks and Microgrids}
\label{subsec:chaos_networks}

\subsubsection{Multi-Machine Systems and Distributed
Generation}

Srivastava \textit{et al.}\cite{srivastava_elimination_1998}
performed the first comprehensive chaotic-bifurcation
survey on multi-bus systems (a 3-bus and a 9-bus network),
identifying torus bifurcation and limit points of periodic
orbits for the first time in power system models.
For the 9-bus system under increasing loading, multiple bifurcation phenomena are observed. A saddle-node bifurcation occurs at $l = 3.598273$. Subcritical Hopf bifurcations are identified at $l = 3.577746$ and $l = 3.588495$, while torus bifurcations occur at $l = 3.249934$ and $l = 3.037992$. Period-doubling bifurcations are observed at $l = 3.135110$, $3.235849$, $3.042431$, and $3.558070$. Additionally, limit points of periodic orbits are detected at $l = 3.070598$, $3.268290$, and $3.022379$.

Gupta \textit{et al.}\cite{gupta_analysis_2018}
built a 4-bus distributed generation model combining a
synchronous machine, a wind turbine generator, a dynamic
load, and an SVC device, and computed the full Lyapunov
spectrum as the inverter current $i_0$ varies, reporting
$(L_1,L_2,L_3,L_4) = (6.485, < 0, < 0, \ll 0)$
at the chaotic operating point. The Kaplan-Yorke dimension
was also computed from the spectrum. This is one of the
few papers in the corpus that uses the full LE spectrum-
rather than only the LLE-as a characterization tool.

Kozlov and Trosko \cite{kozlov_chaotic_2016} showed
that chaotic electromagnetic and electromechanical dynamics
can emerge from classical Park-Gorev bilinear equations
with control lags and excitation saturation, reporting
two positive exponents $\lambda_1 = 0.3547$,
$\lambda_2 = 0.3597$ (hyperchaos) in the parameter
range $D \in [0.5,\, 4.0]$. The structural analogy
between power system bilinear equations and the Lorenz
system is explicitly drawn, placing power system chaos
within the broader theory of coupled nonlinear oscillators.

Sun \textit{et al.}\cite{sun_observation_2011}
demonstrated chaos in a smart-grid model with a wind
turbine as a distributed generator, confirmed by one
positive LLE and a negative sum of all exponents.
Rysev \textit{et al.}\cite{rysev_identifying_2017}
proposed a real-time chaotic mode detection algorithm
from WAMS data (rotor angle and frequency), estimating
the LLE %
and reporting $\lambda_1 \approx +0.23$ for a simulated
3-generator chaotic swing. This method bridges the
gap between the theoretical chaos-in-power-systems
literature and the practical WAMS-based monitoring
discussed in Section \ref{sec:stability}.

\subsubsection{Hidden Attractors and Sub-Synchronous
Oscillations in Microgrids}

A qualitatively new class of chaotic phenomena has been
identified in islanded microgrids composed of
grid-forming (GFM) converters.
Yang \textit{et al.}\cite{yang_sub-synchronous_2023,
yang_sub-synchronous_2025} discovered that sustained
sub-synchronous oscillations in islanded GFM microgrids
are \emph{not} caused by conventional Hopf-born
self-excited oscillations, but by \emph{hidden
attractors}-periodic, quasi-periodic, or chaotic
attractors that coexist with a stable equilibrium point
and have basins of attraction that can only be reached
by large disturbances (not small perturbations around
the equilibrium).

The Lyapunov spectrum is used to classify the hidden
attractors:
\begin{itemize}
  \item $(+,\, 0,\, -, -, -)$: hidden chaotic attractor,
        $\lambda_1 = 0.2191$ for droop gain
        $m_{p5} = 0.01$\,p.u.
  \item $(+,\, +,\, 0,\, -, -)$: hidden hyperchaotic
        attractor, $\lambda_1 = 0.223$,
        $\lambda_2 > 0$ for $m_{p5} = 0.0135$\,p.u.
\end{itemize}
Fractal basin boundaries between the stable equilibrium
and the hidden attractors produce \emph{final-state
sensitivity}: infinitesimally close initial conditions
following a disturbance may result in either synchronism
recovery or persistent sub-synchronous oscillation,
making deterministic prediction of post-disturbance
behavior impossible.

Three mitigation strategies are proposed: attractor
elimination by parameter tuning that removes the hidden
attractor without destabilizing the equilibrium;
disturbance magnitude limitation that keeps the post-fault
trajectory within the basin of the equilibrium; and
tripping of converter units whose droop parameters create
the hidden attractor. The Lyapunov dimension $D_L$ is
used to quantify attractor complexity and guide parameter
selection.

This work is the most conceptually advanced chaos result
in the corpus: it shifts the field from the classical
question is the system chaotic? toward the question
does the system have multiple coexisting attractors,
and what are their basins? The practical implication
is that IBR microgrid security cannot be assessed by
small-signal methods or LLE of the equilibrium alone.

\subsection{Ferroresonance and Sub-Synchronous Resonance}
\label{subsec:chaos_ferrores}

\subsubsection{Chaotic Ferroresonance}

Ferroresonance-nonlinear resonance in circuits
containing transformer magnetization inductance and
system capacitance-is a classical source of chaotic
overvoltages in power systems. Hui
\textit{et al.}\cite{hui_elimination_2009} modeled
the transformer magnetization with an 11th-order
two-term polynomial and the core loss with a 3rd-order
voltage polynomial, yielding a ferroresonant circuit
with richer nonlinear behavior than the linear-core-loss
model. The LLE comparison:
\begin{itemize}
  \item Linear core loss: $\lambda_1 = +0.0552$
        (chaotic ferroresonance),
  \item Nonlinear core loss: $\lambda_1 = -0.1673$
        (stable/periodic behavior),
\end{itemize}
demonstrated that the modeling of nonlinear core loss
is not a refinement but a fundamental structural change
that can eliminate predicted chaos. The paper also showed
that adding neutral resistance suppresses chaotic
ferroresonance in the neutrally-grounded configuration,
providing a practical mitigation measure.

\subsubsection{Sub-Synchronous Resonance Detection}

Cho \textit{et al.}\cite{cho_non-linear_2018} proposed
a measurement-based sub-synchronous resonance (SSR)
detection index derived from phase-space reconstruction
of power system time series. The method maps measured
signals to an amplitude--Floquet-multiplier plane
and applies an MLE-based divergence estimator to the
resulting 2D trajectory. Noh \textit{et al.}\cite{noh_statcom_2020}
extended this by embedding the phase-space geometric
index directly into a STATCOM current-regulator gain
controller: the controller adjusts the STATCOM gain
online based on the measured trajectory radius, providing
self-adaptive SSR damping without requiring explicit
system-model knowledge.

\subsection{Chaos Suppression: Controller Design
Methods}
\label{subsec:chaos_control}

The second major strand of this literature is the design
of controllers to suppress detected chaos. Across the
corpus, five controller families are used, and each paper
applies LEs in two distinct roles: as a \emph{diagnostic}
to confirm chaos in the uncontrolled system, and-in
more rigorous contributions-as a \emph{design guide}
or \emph{performance metric} for the controlled system.

\subsubsection{FACTS-Based Bifurcation Engineering}

Srivastava \textit{et al.}\cite{srivastava_elimination_1998}
established the paradigm of using FACTS devices to
eliminate dynamic bifurcation. By feeding CSC, PAR,
or SVC with generator speed deviation $\omega$ or
accelerating power $P_a$ (selected by participation
analysis), the Hopf bifurcation boundary can be shifted
beyond the maximum load, effectively eliminating the
chaotic window. This approach uses LEs only
diagnostically-to confirm that the FACTS-controlled
system is no longer chaotic-but the control design
itself is guided by continuation/bifurcation analysis.

\subsubsection{Sliding Mode Control}

Sliding mode controllers (SMC) are the most widely
applied chaos-suppression technique in the corpus.
Kumar and Singh \cite{kumar_chaos_2018} applied
a first-order SMC to a 4D SMIB model with excitation
regulator dynamics, constructing a switching surface
$s_i$ for each state and proving reachability using
the Lyapunov function $V_i = \frac{1}{2}s_i^2$ 
%
%
which guarantees $s_i \to 0$ in finite time. The LLE
$\lambda_1 = 0.1227$ confirmed chaos before control;
phase-plane simulations confirmed convergence to the
target equilibrium after SMC activation.

Li \textit{et al.}\cite{li_controlling_2019}
designed a \emph{fixed-time} sliding mode controller
for the 6D model, using a non-singular sliding surface
based on hyperbolic functions to avoid singularity and
reduce chattering. The controller converts
chaos-suppression into stabilization of a reduced
second-order subsystem in $(x_1, x_2)$, exploiting the
cascade structure of the 6D model to show that
single-input control propagates stability to all six states.

Rajagopal \textit{et al.}\cite{rajagopal_bifurcation_2019}
combined sliding mode with adaptive parameter estimation
for a fractional-order 4D model with uncertain damping
$D$ and mechanical input $P_m$. The adaptive law updates
parameter estimates online 
%
%
while the fractional-order Lyapunov inequality
$\frac{1}{2}D^q(\delta^2) \leq \delta\, D^q\delta$
is used to prove finite-time convergence of the
synchronization error and parameter estimates simultaneously.

\subsubsection{Synergetic Control}

Al-Hussein \textit{et al.}\cite{al-hussein_chaos_2021}
applied synergetic control theory-a smooth invariant-manifold
method that avoids the chattering of SMC-to suppress
chaos in the 4D voltage-stability benchmark. The system
was augmented with a STATCOM for reactive power and
an energy storage device for real-power damping, yielding
a 6D controlled system. 
%
The LE spectrum on the route to chaos (showing the
transition to positive LLE as $P_m$ enters the interval
$[1.1006,\, 1.102]$) was used to confirm the chaotic
operating regime before controller design, and the same
spectrum was used to verify chaos elimination after
controller deployment.

\subsubsection{ANFIS and Fuzzy Intelligent Controllers}

Ginarsa \textit{et al.}\cite{ginarsa_controlling_2009}
trained an ANFIS-based composite controller offline
using data from a conventional composite controller,
then deployed the trained ANFIS as a lightweight
nonlinear map for online chaos suppression. The control
input is the rotor-speed deviation $\Delta\omega$ and
its derivative; LEs are referenced qualitatively to
motivate the chaos hypothesis.
ANFIS training objective.


Luo \textit{et al.}\cite{luo_dynamic_2023}
proposed the most sophisticated intelligent controller
in the corpus for fractional-order PMSG synchronization:
a hierarchical type-2 fuzzy neural network (HT2FNN)
combined with fixed-time convergence, prescribed
performance functions, and a cost-minimization objective.
The controller achieves chaos suppression (via
synchronization to a stable drive system) and parameter
identification simultaneously, with closed-loop proof
using fractional Lyapunov stability inequalities, whereas Lyapunov exponents are used for dynamic diagnosis.

\subsubsection{Continuous Feedback Synchronization Control}
Noh \textit{et al.}\cite{noh_toward_2019} embedded
an LLE-based oscillation detector directly into a
STATCOM droop-gain controller: when the phase-space
reconstruction of measured power-system frequency or
rotor-angle data produces a positive MLE, the STATCOM
current-regulator gain is reduced via a nonlinear droop
function, increasing effective damping. This closed-loop
architecture is unique in the corpus: the LLE is not only
a diagnostic but a real-time trigger for proportional
corrective action.

\subsubsection{Optimal Control with Chaotic Forecast}

Dou and Liu \cite{dou_optimal_2004} integrated
chaos-based short-term prediction with optimal control
for unit load management.
The LLE is used to compute the
prediction horizon $T_{\max} = 1/\lambda_1$ (reported
$\lambda_1 = 0.0053$, giving $T_{\max} \approx 188$\,s),
which determines the effective forecast window and hence
the control horizon. This is the only paper in the corpus
that directly integrates the LE predictability horizon
into a feedback control design.

A related architecture for microgrid power-quality
control is proposed by Li \textit{et al.}\cite{li_optimal_2018},
who embedded an improved LLE-based chaotic predictor
for voltage/frequency evolution inside a situation-aware
converter controller. The predictor provides ultra-short-term
PCC forecasts that feed a multi-objective mode-selection
decision layer, with smooth strategy-switching between
STATCOM, MPPT, and APF operating modes.

\subsection{Chaos Synchronization as a Control Paradigm}
\label{subsec:chaos_sync}

In the chaos synchronization paradigm, a chaotic
\emph{response} system is coupled to a drive system
and the goal is to achieve $\|\mathbf{x}_r(t) -
\mathbf{x}_d(t)\| \to 0$ as $t \to \infty$, which
can serve either as a chaos suppression mechanism (if
the drive is a stable periodic orbit) or as a
coordination tool for interconnected chaotic systems.

Shahverdiev \textit{et al.}\cite{shahverdiev_chaos_2008}
studied synchronization of two identical forced SMIB
systems coupled via a term $K(x_2^r - x_2^d)$
(speed-deviation coupling). The \emph{conditional Lyapunov
exponent} (CLE) of the error dynamics-the LE of the
linearized error system with the drive trajectory as
background-is used as the synchronization criterion:
\begin{itemize}
  \item $\mathrm{CLE} < 0$: synchronized ($K > K_\mathrm{min}
        \approx 20$ for ODE case),
  \item $\mathrm{CLE} \geq 0$: desynchronized.
\end{itemize}
The delay-differential equation (DDE) variant with
$\sin(Rx_1(t-\tau))$ forcing requires stronger coupling
($K > 46$) due to the added complexity from the delay.
This establishes a quantitative design rule: the minimum
coupling gain needed for synchronization is the value
that drives the CLE to zero.

Ai \textit{et al.}\cite{ai_finite_2023} proposed
finite-time adaptive synchronization for a fractional-order
4D model with uncertain parameters, using synchronization
to a stable drive system as the suppression mechanism.
The uncontrolled fractional system is shown to be chaotic by phase trajectories, Poincaré section, broadband power spectrum, and a positive maximum Lyapunov exponent. At $Q1 = 11.3767$ and $\alpha = 0.999$, the paper reports one positive exponent, which is used as the main quantitative evidence of chaos.
\subsection{Fractional-Order Chaos: Models and Control}
\label{subsec:chaos_fractional}

Fractional-order (FO) dynamical systems have attracted
growing attention in the chaos-in-power-systems
literature because they can model long-memory effects
and heritage-dependent dynamics that integer-order
equations cannot capture.

\subsubsection{Fractional-Order Modeling}

The general commensurate fractional-order power system
is written using the Caputo derivative $D^q$:
\begin{align}
D^q x_1 &= x_2, \notag \\
D^q x_2 &= -\lambda x_2 + \rho - \sin(x_1)
           - \sigma\cos(\eta\tau)\sin(x_1)
           + \mu\cos(\gamma\tau),
\label{eq:fo_smib}
\end{align}
where $q \in (0,1]$ is the fractional order
\cite{min_fractional-order_2018}. The minimum order for
chaos onset is typically reported around $q \approx 0.97$--$0.9999$
for the SMIB model and $q \approx 0.4229$ for two-machine
models \cite{yan_dynamics_2021}. Below this threshold,
memory effects dominate and the system remains non-chaotic.
This provides an additional design parameter not available
in integer-order models.

\subsubsection{Fractional-Order Controllers}

Nangrani and Bhat \cite{nangrani_numerical_2017,
nangrani_fractional_2018} proposed optimized FOPI and
FOPID state-feedback controllers for a 4D benchmark
model, using the LLE \emph{as the performance index}:
the optimal fractional orders $\lambda$ and $\mu$ are
chosen as the values that minimize the LLE of the
controlled system. This inverts the usual role of the
LLE (from diagnostic to design objective), though the
optimization is numerical rather than analytical.

Rajagopal \textit{et al.}\cite{rajagopal_bifurcation_2019}
proved rigorous finite-time adaptive synchronization for
the 4D fractional-order model with uncertain parameters,
using a fractional Lyapunov composite function.

Luo \textit{et al.}\cite{luo_dynamic_2023}
and Ai \textit{et al.}\cite{ai_finite_2023}
pushed the frontier to coupled PMSG and 4D models,
respectively, combining FO modeling with HT2FNN
approximation and fixed-time convergence theory.

\subsection{LE-Based Cyber-Security Risk Assessment}
\label{subsec:chaos_cyber}

A notable lateral application is the use of conditional
Lyapunov exponents (CLEs) for cyber-physical security
monitoring. Yan \textit{et al.}\cite{yan_risk_2015}
proposed a framework that links cyber attack graphs
(modeled as state machines over SCADA vulnerabilities)
to the nonlinear dynamic stability of the grid, with CLEs
estimated from PMU data as the real-time stability risk
indicator. When CLEs increase above a threshold,
intrusion response actions are triggered. This is the
only paper in the corpus that connects LE theory directly
to cyber-physical system security.

\subsection{Open Problems in Chaos and Control}
\label{subsec:chaos_gaps}

Despite the substantial literature reviewed above,
several critical gaps remain.

\textbf{(1) IBR model realism.}
The vast majority of chaos studies use simplified
integer-order or fractional-order ODE models of
synchronous generators. The dynamics of GFM and GFL
converters under large disturbances are governed by
control-loop switching and saturation, not swing-equation
nonlinearities. The hidden-attractor work
\cite{yang_sub-synchronous_2023, yang_sub-synchronous_2025}
is the only systematic study that begins from a
converter-native model. More work is needed to determine
whether the control synthesis methods of
Sections \ref{subsec:chaos_control}--\ref{subsec:chaos_fractional}
are applicable to IBR-dominated networks.

\textbf{(2) Scalability from SMIB to multi-machine
networks.}
Most chaos-control papers are validated on SMIB or
3-bus benchmark models. Extension to realistic large-scale
multi-machine systems requires addressing: (i) the
exponential growth of phase-space dimension with
system size; (ii) the interaction between multiple
simultaneously chaotic machines; and (iii) the need
for distributed rather than centralized controllers.

\textbf{(3) From chaos suppression to chaos prevention.}
All reviewed chaos-control methods operate
\emph{reactively}: they suppress chaos after it has
been detected. A proactive paradigm-using LLE-based
stability margins to detect bifurcation approach and
adjust operating parameters before chaos onset-is
largely absent. The LLE-based PMU monitoring tools of
Section \ref{sec:stability} provide the detection
capability; the connection to preemptive control actions
is not yet made.

\textbf{(4) Experimental and field validation.}
Without exception, the chaos-control results in this
literature are validated by simulation on toy models.
No paper provides hardware experimental or field
validation of chaos suppression using an LLE-based
controller. Real-time implementation faces challenges
in LLE computation latency, noise robustness, and
closed-loop stability under digital control delays.

\textbf{(5) Fractional-order model interpretability.}
The physical interpretation of the fractional order
$q$ remains debated \cite{yan_dynamics_2021}. The
fractional swing equation describes systems with
long-memory damping, which may model electromagnetic
transients in complex cable networks but is not
physically motivated for standard synchronous machines.
Chaotic behavior attributable to the choice of
$q < 1$ should be clearly distinguished from behavior
arising from the physical nonlinearity of the system.

Table \ref{tab:chaos_summary} provides a structured
summary of the key papers reviewed in this section.

\begin{table*}[!t]
\renewcommand{\arraystretch}{1.3}
\caption{Summary of Key Papers in Chaos Detection,
Bifurcation Analysis, and Control}
\label{tab:chaos_summary}
\centering
\small
\begin{tabular}{p{3.0cm}p{2.4cm}p{2.2cm}p{1.6cm}p{4.4cm}}
\toprule
\textbf{Reference} &
\textbf{System model} &
\textbf{LE role} &
\textbf{Controller} &
\textbf{Key contribution} \\
\midrule
Srivastava 1993 \cite{srivastava_chaotic_1993}
  & 3-Bus + OLTC + SVS
  & Spectrum classification
  & None
  & Full LE spectrum maps chaos routes \\
Nayfeh 1995 \cite{nayfeh_bifurcations_1995}
  & 4D reduced w/ dynamic load
  & Chaos verification
  & Nonlinear feedback $Kw^3$
  & Bifurcation taxonomy; Hopf control \\
Srivastava 1998 \cite{srivastava_elimination_1998}
  & 3-bus, 9-bus
  & Chaos diagnosis
  & FACTS
  & First multi-bus FACTS chaos elimination \\
Li 2009 \cite{li_chaotic_2009}
  & IEEE 14-bus DAE
  & LLE sign
  & PSS
  & Chaos in realistic DAE; PSS cure \\
Qin 2014 \cite{qin_random_2014}
  & Stochastic SMIB
  & LLE transition
  & None
  & Random parameters induce chaos \\
Chen 2005 \cite{chen_dynamic_2005}
  & SMIB 3D
  & Spectrum 
  & Feedback / time-delay
  & Chaos + chaotification in same model \\
Li 2019 \cite{li_controlling_2019}
  & 6D PS model
  & LLE diagnosis
  & Fixed-time SMC
  & Single input suppresses 6D chaos \\
Al-Hussein 2021 \cite{al-hussein_chaos_2021}
  & 4D benchmark + STATCOM
  & Spectrum
  & Synergetic controller
  & Smooth chaos suppression + proof \\
Min 2017/2018 \cite{min_fractional-order_2017}
  & Fractional 4D SMIB
  & Chaos boundary mapping
  & Synchronization
  & FO model; min order for chaos \\
Rajagopal 2019 \cite{rajagopal_bifurcation_2019}
  & Fractional 4D uncertain
  & LLE diagnosis
  & FO adaptive SMC
  & Adaptive chaos suppression + ID \\
Cao 2025 \cite{cao_chaotic_2025}
  & 9D DFIG model
  & Full LE spectrum
  & None
  & Chaos param. map for converter gain \\
Chen 2023 \cite{chen_criterion_2023}
  & 4D DFIG
  & LLE + Hopf criterion
  & None
  & Analytical Hopf criterion for DFIG \\
Yang 2023 \cite{yang_sub-synchronous_2023}
  & GFM islanded microgrid
  & Spectrum; $D_L$
  & Attractor elimination
  & Hidden attractor sub-sync. oscillation \\
Noh 2020 \cite{noh_statcom_2020}
  & SSR + STATCOM
  & MLE trigger
  & STATCOM gain control
  & LLE embedded in feedback loop \\
\bottomrule
\end{tabular}
\end{table*}

%

%
%

%
\section{Forecasting, Prediction, and Predictability Analysis}
\label{sec:forecasting}

The application of Lyapunov exponents to power system forecasting rests on a single foundational observation: if a time series arises from a low-dimensional chaotic attractor, the largest Lyapunov exponent (LLE) bounds the prediction horizon, and the reconstructed phase space provides the geometric structure needed to build a nonlinear predictor that outperforms linear models in the short term. This domain encompasses 42 publications in our corpus, spanning from pioneer daily-peak-load studies in the mid-1990s \cite{choi_daily_1996, mori_short-term_1996, mizukami_forecasting_1996} to recent deep-learning-LE hybrids and interval prediction frameworks \cite{safari_novel_2018, ma_intelligent_2025, gu_dynamic-inner_2024}.

The section is organized as follows. Section \ref{subsec:forecast_pipeline} describes the standard chaotic forecasting pipeline common to all papers. Section \ref{subsec:predict_horizon} covers the fundamental predictability horizon result and its power engineering implications. Section \ref{subsec:load_forecast} reviews LLE-based load forecasting. Section \ref{subsec:wind_forecast} covers wind power and wind speed forecasting. Section \ref{subsec:pv_forecast} addresses photovoltaic power forecasting. Section \ref{subsec:price_forecast} covers electricity price forecasting. Section \ref{subsec:decomp_hybrid} discusses decomposition-and-hybrid architectures in which LEs guide component-wise model selection. Section \ref{subsec:forecast_gaps} identifies open problems.

\subsection{The Standard Chaotic Forecasting Pipeline}
\label{subsec:forecast_pipeline}

The overwhelming majority of the 42 reviewed papers follow a common five-stage pipeline that transforms a scalar power system time series into a short-term point forecast.

\textbf{Stage 1 - Preprocessing.} The raw time series $\{x(k)\}$ is detrended, stationarized, and denoised. Preprocessing methods used in the reviewed literature include Box-Cox transformation with differencing \cite{mohamed_variability_2007}, wavelet-threshold denoising \cite{ma_intelligent_2025, yufei_photovoltaic_2016}, and EEMD endpoint extension via LLE-based prediction \cite{bingjiang_hybrid_2018}.

\textbf{Stage 2 - Embedding parameter selection.} The time delay $\tau$ is chosen by the first minimum of the average mutual information (AMI) or the first zero of the autocorrelation function. The embedding dimension $m$ is chosen by false nearest neighbors (FNN) \cite{aghbalou_long_2018, bazine_new_2018} or by the C-C method \cite{si-min_application_2009, wang_power_2020, yapeng_probe_2017}. A less common but significant approach is choosing $m$ at the dimension where the LLE stabilizes: Mori and Urano \cite{mori_short-term_1996} and Michanos \textit{et al.}\cite{michanos_short-term_2003} both select the embedding dimension with the \emph{minimum} largest Lyapunov exponent, arguing that the weakest-chaotic embedding provides the statistically most tractable local dynamics.

\textbf{Stage 3 - Chaos verification.} After reconstructing the phase space $\mathbf{S}(k) = [x(k),\, x(k{+}\tau),\, \ldots,\, x(k{+}(m{-}1)\tau)]^T$, the LLE is computed to confirm that the series is chaotic ($\lambda_1 > 0$). The Wolf algorithm is the most widely used estimator across this literature \cite{wang_load_2006, niu_support_2008, niu_new_2009, wang_optimization_2008, dong_prediction_2009}. The Rosenstein small-data method is also used for shorter records \cite{chen_identification_2009, yufei_photovoltaic_2016, yapeng_probe_2017}. Reported LLE values across the corpus are consistently small and positive (typically $\lambda_1 \in (0.01,\, 0.5)$), reflecting weak chaos with a meaningful but finite
prediction horizon.

\textbf{Stage 4 - Prediction in reconstructed space.} Given the current reconstructed state $\mathbf{S}(k)$, the forecast $\hat{x}(k{+}h)$ is computed by one of several methods: local linear approximation on the $q$ nearest neighbors of $\mathbf{S}(k)$ in the attractor (local-region method); a weighted variant in which neighbors are downweighted by their distance to $\mathbf{S}(k)$ (adding-weight methods); or a machine learning regressor (SVM, ANN, LSSVM, RBF) trained on the full set of reconstructed state vectors.

The \emph{adding-weight one-rank local-region method} (also called the weighted first-order local method) is the most widely used pure-chaos predictor in this corpus. The one-step forecast is:
\begin{equation}
\hat{x}(k{+}1) = \sum_{i=1}^{q} w_i\, x\bigl(n_i + 1\bigr), \qquad w_i = \frac{e^{-\alpha\,\|\mathbf{S}(k) - \mathbf{S}(n_i)\|}}   {\sum_{j=1}^{q} e^{-\alpha\,\|\mathbf{S}(k) - \mathbf{S}(n_j)\|}},
\label{eq:weighted_local}
\end{equation}
where $n_1,\ldots,n_q$ are the indices of the $q$ nearest attractor neighbors of $\mathbf{S}(k)$ and $\alpha$ is a distance-weighting parameter \cite{si-min_application_2009,yufei_photovoltaic_2016, yapeng_probe_2017}.

\textbf{Stage 5 - Multi-step and recursive forecasting.} For $h$-step-ahead forecasts, predictions are made iteratively: each $\hat{x}(k+i)$ is appended to the series and used as the start of the next reconstruction. Accumulated error from iterative single-step prediction is a widely acknowledged limitation; several papers propose direct multi-step prediction formulations \cite{si-min_application_2009, safari_novel_2018} to mitigate this problem.

\subsection{The Predictability Horizon: A Model-Independent Upper Bound}
\label{subsec:predict_horizon}

A result that distinguishes the LE-based forecasting literature from conventional time-series work is the \emph{Lorenz predictability horizon}: the maximum time ahead over which any predictor can remain within a tolerable error, derived directly from $\lambda_1$.

Given an initial measurement uncertainty $\delta_0$ and an acceptable maximum prediction error $\Delta$, the predictability horizon is (from Section \ref{subsec:predictability}):
\begin{equation}
T_{\max} = \frac{1}{\lambda_1} \ln\!\left(\frac{\Delta}{\delta_0}\right).
\label{eq:tmax_sec4}
\end{equation}
The most rigorous application of \eqref{eq:tmax_sec4} in the power-forecasting literature is the study by Mohamed \textit{et al.}\cite{mohamed_variability_2007} of the Tunisian daily peak-load series.

\subsubsection{Predictability Horizon of Peak Load: The Tunisian Study}

Mohamed \textit{et al.}\cite{mohamed_variability_2007} do not propose a forecasting model. Instead, they ask the more fundamental question: \emph{how far ahead can the daily peak-load series be predicted, and how does this limit change over time?} This meta-forecasting perspective is unique in the corpus.

The daily peak-load data from the Tunisian electricity company (STEG) are stationarized via a Box-Cox transform with $L = 0.53$ and double differencing ($\nabla^1 \nabla^7$). The AMI criterion selects $\tau = 4$, and FNN selects $m = 5$. The LLE is estimated by Kantz's method, yielding $\lambda_1 \approx 0.032$ in the period 1996--2004, corresponding to $T_{\max} \approx 3$ days for $\Delta/\delta_0 = e$.

More significantly, the authors track $\lambda_1$ over sliding windows and find that it increased substantially after 2000, coinciding with rising air-conditioning penetration. This demonstrates that the predictability of load demand \emph{deteriorates} as the load mix becomes more weather-sensitive and nonlinearly coupled. The practical implication is stark: any load forecasting model, regardless of its sophistication, faces a shorter reliable horizon as cooling load penetration grows.

This is the only paper in the corpus that uses LEs purely as a \emph{meta-forecasting} tool-not to build a predictor, but to set a rigorous, physically derived upper bound on what any predictor can achieve. It should be a reference point for all papers that claim improved forecast accuracy on load or renewable time series.

\subsubsection{Predictability Horizon for Wind and PV}

Wang \textit{et al.}\cite{wang_wind_2025} analyze wind-power time series from a monitoring tower, concluding from the LLE that wind-power dynamics are weakly chaotic and exhibit a finite prediction horizon consistent with \eqref{eq:tmax_sec4}. The study reinforces the conceptual point that short-horizon accuracy claims in wind forecasting are physically bounded by the LLE of the underlying series.

Yufei \textit{et al.}\cite{yufei_photovoltaic_2016} report $\lambda_1 = 9.34 \times 10^{-5}$ for a rooftop PV output series, implying an extremely long predictability horizon for that dataset; however, this near-zero value raises questions about numerical robustness of the small-data estimator on a series whose variability is largely deterministic (driven by irradiance cycles) rather than truly chaotic.

\subsection{Short-Term Load Forecasting}
\label{subsec:load_forecast}

Load forecasting is the largest subcategory within the forecasting domain, with 18 papers spanning 1996--2026. The evolution of the literature follows a clear arc: from pure chaos-based local predictors in the 1990s, through chaos-informed model-structure selection for ANNs and SVMs in the 2000s, to multivariate chaos-market hybrid models by the 2020s.

\subsubsection{Pioneer Work: Chaos-Based ANN and Local Prediction (1990s)}

The earliest papers in this domain established the workflow that all subsequent papers follow. Choi \textit{et al.}\cite{choi_daily_1996} applied attractor reconstruction to KEPCO daily-peak-load data and used the reconstructed coordinates as ANN inputs for one-day-ahead forecasting, demonstrating that chaos-guided input design outperforms heuristic lag selection.

Mori and Urano \cite{mori_short-term_1996} introduced the distinctive criterion of choosing the embedding dimension at the \emph{minimum} largest Lyapunov exponent, arguing that the weakest-chaos setting produces the most regular local dynamics and therefore the most statistically tractable model. The Lyapunov spectrum was computed across dimensions $m = 1$ to $m = 7$; the minimum-LLE dimension ($m = 4$) was selected as ANN input size. This approach was independently confirmed by Michanos \textit{et al.}\cite{michanos_short-term_2003} on Greek power-system data, where both correlation dimension and minimum-LLE criteria independently yield $m = 3$.

Mizukami \textit{et al.}\cite{mizukami_forecasting_1996} proposed the most analytically motivated local predictor of this era: a Gram-Schmidt orthonormal-basis approximation in the reconstructed state space. Local geometry was approximated by orthonormal vectors spanning the neighborhood of the current embedded state, and the prediction was generated by evolving the current state along this locally fitted geometry. This approach anticipated later manifold-based prediction methods, though its multi-step accuracy deteriorated faster than ANN methods.

Wang \textit{et al.}\cite{wang_short-term_2009} proposed the Adding-weighted Largest Lyapunov Exponent (ALLE) forecasting method, which modifies the classical LLE-based trajectory-extrapolation formula by introducing a correction weight $w$ to reduce the prediction error from using only an average divergence rate. 

Comparison with the standard one-rank local-region method on New South Wales load data showed that ALLE achieves lower root-mean-square relative error for weekend loads.

\subsubsection{Chaos-Informed Model-Structure Selection for SVM and ANN (2000s)}

The primary contribution of the 2000s load-forecasting literature was not new chaos theory but a clearer methodology for using the LLE to determine the input dimensionality of machine learning regressors.

Niu \textit{et al.}\cite{niu_support_2008, niu_new_2009} established the workflow: embed the load series, estimate the LLE by Wolf's method as $m$ is varied, and select the embedding dimension where $\lambda_1$ stabilizes. Both papers report convergence at $m = 12$ for Inner Mongolia load data and demonstrate that this Lyapunov-chosen $m$ outperforms arbitrarily chosen dimensions in SVM regression accuracy.

Wang \textit{et al.}\cite{wang_optimization_2008} applied the same principle to optimize a BP neural network input dimension, finding stabilization at $m = 14$ for the same regional dataset.

Wang \textit{et al.}\cite{wang_load_2006} extended the basic chaos-SVM framework by adding an immunogenetic particle swarm incorporated intelligence algorithm for SVM hyperparameter tuning, combining Lyapunov-based embedding design with hybrid metaheuristic optimization. 

Niu \textit{et al.}\cite{niu_optimization_2010} further improved the SVM pipeline by adding gray-relation-analysis-based data mining, selecting only historically similar days as training data before SVM regression. The resulting DSVM model integrates three components: Lyapunov-based phase-space design, feature-based day selection, and SVM regression. 

Wu and Zhang \cite{wu_power_2005} introduced wavelet decomposition as a preprocessing step before chaos-informed LS-SVM forecasting, decomposing the daily peak-load series into trend, periodic, and random components and applying separate LS-SVM models to each wavelet level with chaos-based embedding.

Chen and Liu \cite{chen_identification_2009} contributed systematic chaos-identification study of Fujian province load data using the Rosenstein small-data method, reporting positive LLEs ($\approx 0.025$--$0.042$) for multiple sub-series and establishing the empirical basis for subsequent forecasting applications on that dataset.

\subsubsection{Multivariate Chaos and Market-Aware
Models (2010s--2020s)}

Wang \textit{et al.}\cite{wang_power_2020} extended the chaos-LSSVM framework to smart-grid conditions by incorporating real-time electricity price (RTP) as a second chaotic time series. Both load and price series were verified as chaotic ($\lambda_1^{\rm load} \approx 0.031$, $\lambda_1^{\rm price} \approx 0.0075$), and a multivariate phase space was reconstructed from both. The improved C-C method was used for joint delay and dimension selection. An adaptive PSO (APSO)-tuned LSSVM showed lower Mean absolute percentage error than chaos-LSSVM without price and RBF-NN on the same data.

Shen \textit{et al.}\cite{tianshi-shen_regulation_2016} embedded a standard LLE-based chaotic predictor into an operational demand-side regulation study: the predicted 24-hour load curve was used as the reference for scheduling electric-vehicle charging and cool-storage air-conditioning in the Shaanxi distribution network.

Yapeng \textit{et al.}\cite{yapeng_probe_2017} studied micro-grid short-term load forecasting, systematically analyzing how two design choices- sampling interval and historical data length-affect the accuracy of the weighted local prediction method. Sampling at 30-minute intervals with 7 days of history produced the lowest relative error, providing operational design guidance.

Niu \textit{et al.}\cite{niu_short-term_2015} applied phase-space reconstruction to a novel target variable: the \emph{sharp-degree} sequence derived from the load curve, an image-analysis-inspired metric that highlights inflection points. A chaotic radical basis  unction (RBF) neural network ($\lambda_1 = 0.3017$, $m = 6$) was trained on the sharp-degree series to predict load turning points-a practically useful output for dispatch planning.

Ma \textit{et al.}\cite{yang_research_2026} represents a recent direction in which chaotic mapping (specifically, Tent map initialization with provably high LLE) is used inside a whale optimization algorithm to initialize a large population of candidate Extreme Learning Machine (ELM) models, exploiting chaos theory at the optimization stage rather than the data analysis stage. This is methodologically distinct from the mainstream pipeline: the LLE here is a property of the \emph{optimizer}, not of the load time series.

\subsection{Wind Power and Wind Speed Forecasting}
\label{subsec:wind_forecast}

Wind forecasting is the second-largest forecasting subcategory, with 14 papers covering 2007--2025. The literature spans direct wind-power output forecasting from turbine measurements, wind-speed prediction as a physical proxy, Weibull distribution parameter prediction for long-term energy assessment, and interval/uncertainty prediction.

\subsubsection{Chaos-ANN Architecture (2007--2010)}

Dong \textit{et al.}\cite{lei_prediction_2007} was among the first to treat wind-farm power output as a scalar chaotic time series and reconstruct its phase space for direct ANN forecasting. The largest Lyapunov exponent was estimated using Wolf's method ($\lambda_1 = 0.1578$) and used to confirm the presence of chaos. The reconstructed lag vectors served as ANN inputs without meteorological inputs, demonstrating that short-horizon wind-power fluctuations exhibit exploitable attractor structure.

Dong \textit{et al.}\cite{dong_prediction_2009} and Wang \textit{et al.}\cite{wang_wind_2009} built on this approach by combining wavelet decomposition with chaos-informed ANN submodels. The wavelet transform decomposes the wind-power series into approximate and detail components; each component is analyzed separately for chaotic characteristics (positive LLE by Wolf's method), and a dedicated ANN submodel is trained in the reconstructed phase space of each component. Final predictions are recovered by summing component forecasts.

The support vector regression (SVR) variant of the same pipeline was introduced by Wang \textit{et al.}\cite{wang_short-term_2010}, who replaced ANN submodels with support vector regression on the reconstructed wind-speed series. The reported LLE for their wind-speed data is $\lambda_1 = 0.2429$, and SVR outperformed ANN on the same embedding in both one-step and multi-step prediction.

\subsubsection{Advanced Hybrid Architectures (2013--2025)}

Fan \textit{et al.}\cite{fan_wind_2013} proposed a \emph{chaotic operator network} (CON) for wind-speed prediction: instead of conventional hidden neurons, the hidden layer uses units governed by the chaotic map $x_{n+1} = \sin(\alpha x_n)$. The map's LLE is positive for most $\alpha > 1$ (confirmed by a Lyapunov diagram in the paper), and a genetic algorithm optimizes the operator parameters so that the network's internal dynamics follow the observed wind-speed dynamics. Compared with Autoregressive Integrated Moving Average (ARIMA) and standard BP networks, the CON produced lower RMSE on one Chinese wind-speed dataset.

Zhang \textit{et al.}\cite{bingjiang_hybrid_2018} proposed the most comprehensive wind-power hybrid in the corpus: a combination of endpoint extension via LLE prediction, EEMD decomposition, and Volterra Neural Network (VNN) modeling. The LLE is used in two separate roles: first to extend the series endpoints (reducing EEMD end effects), and second to characterize each EEMD component and guide VNN structure selection. The equivalence between a Volterra functional model and a three-layer feedforward ANN is used to derive the practical VNN implementation.

Mao \textit{et al.}\cite{mao_wind_2018} applied wavelet-packet decomposition to wind-power series, combined with a branched prediction strategy: \emph{chaotic} branches (positive LLE) are forecast by the weighted one-rank local-region method \eqref{eq:weighted_local}, while \emph{non-chaotic} branches (near-zero or negative LLE) are forecast by BP neural networks. This \emph{chaos-aware predictor routing} is one of the more principled hybrid designs in the corpus, since it applies the chaos predictor only where its assumptions hold.

Safari \textit{et al.}\cite{safari_novel_2018} proposed the most methodologically complete framework: multi-scale SSA-least squares-SVM (MSSSA-LSSVM). Wind-power series are first decomposed by Ensemble EMD (EEMD); then each component is classified as chaotic or non-chaotic by LLE. \emph{Only chaotic} components receive an additional singular spectrum analysis (SSA) treatment to suppress wild high-frequency fluctuations that degrade multi-step accuracy. Chaotic components use direct multi-step LSSVM prediction, while non-chaotic components use iterative multi-step prediction. Validation on a 3-year wind-farm dataset and NREL regional data showed the highest accuracy among all compared methods.

Aghbalou \textit{et al.}\cite{aghbalou_long_2018} shifted the target from short-term power to long-term Weibull distribution parameters of annual wind speed. The scale and shape parameters are treated as separate chaotic time series and their annual evolution is forecast using neural networks trained with Bayesian regularization on phase-space reconstructed sequences. This is a unique application of chaotic time-series analysis to distributional forecasting.

Song \textit{et al.}\cite{song_wind_2020, song_evaluating_2020} evaluated dendritic neuron models-architectures with nonlinear multiplicative computations at dendrite junctions-for wind-speed forecasting using chaotic phase-space inputs. The maximum LLE ranged from 0.05 to 0.077 across six Sotavento wind-speed datasets, confirming weak chaos. The evolutionary dendritic neuron model (EDNM) trained by L-SHADE differential evolution outperformed ELMAN, MLP, and SVM variants on 10-minute and 1-hour horizons.

Wang \textit{et al.}\cite{wang_analysis-forecast_2018} proposed an uncertainty forecasting system combining recurrence plot analysis, CEEMDAN decomposition, C-C phase-space reconstruction, and a multi-objective LSSVM interval predictor optimized by an improved water cycle algorithm. Lyapunov exponents are used as part of the chaos-characterization module alongside recurrence quantification metrics, providing a richer nonlinear diagnostic than LLE alone.

Ma \textit{et al.}\cite{ma_intelligent_2025} presented the most recent recent wind-speed interval prediction paper: a Nutcracker Optimization Algorithm (NOA) optimized wavelet-denoising plus QR-BiTCN-BiGRU architecture. The Wolf LLE ($\lambda_1 = 0.1862$) is computed after C-C-based reconstruction to verify chaoticity, and the phase-space coordinates are used as BiTCN-BiGRU inputs. The NOA optimizer tunes both the wavelet-threshold and the deep-learning parameters, achieving competitive interval coverage probability on benchmark wind-speed data.

Gu \textit{et al.}\cite{gu_dynamic-inner_2024} proposed the most recent point-prediction advance: a dynamic-inner manifold broad learning system (DiM-BLS) coupled with adaptive VMD. The LLE was computed to confirm chaos in both onshore ($\lambda_1 = 0.1582$) and offshore ($\lambda_1 = 0.1358$) datasets. The DiM-BLS architecture learns local dynamic manifold features from each VMD mode, providing physical interpretability beyond standard deep learning.

\subsection{Photovoltaic Power Forecasting}
\label{subsec:pv_forecast}

PV power forecasting using chaos theory is a smaller and more recent subcategory, with four papers in the corpus covering 2016--2018.

Yufei \textit{et al.}\cite{yufei_photovoltaic_2016} provided the earliest direct comparison of chaos-based and ANN-based PV forecasting on measured rooftop installation data. The C-C method selected $\tau$ and $m$ simultaneously; the small-data LLE ($\lambda_1 = 9.34 \times 10^{-5}$) is positive, supporting the chaos hypothesis. The adding-weight first-order local method \eqref{eq:weighted_local} produced a lower Mean absolute percentage error than a neural network for ultra-short-term (10-minute-ahead) prediction. The near-zero LLE value is noted but not critically examined; it suggests that PV output is dominated by near-deterministic diurnal cycles rather than strong chaos, which has implications for \eqref{eq:tmax_sec4}.

Bazine \textit{et al.}\cite{bazine_new_2018} applied the full three-stage chaos pipeline to daily PV production data from a Moroccan installation (chaos detection, phase space reconstruction, and forecasting), using both the LLE ($\lambda_1 = 0.2383$) and the 0--1 chaos test as complementary diagnostics before ANN and wavelet recurrent neural network (WRNN) prediction. The 0--1 test adds independent evidence for chaos beyond the LLE and reduces the risk of false-positive chaos identification. The WRNN with phase-space inputs outperformed both ANN without reconstruction and ANN with reconstruction on the
validation dataset.

\subsection{Electricity Price Forecasting}
\label{subsec:price_forecast}

Electricity price is the third main application series in this domain. Two papers apply chaos-based analysis to the New England spot price market, and their combined findings show that price data exhibit chaotic characteristics similar to those of load and wind data.

Yang and Duan \cite{yang_chaotic_2003} applied scalar phase-space reconstruction (delay $\tau = 8$, dimension $m = 9$) to hourly New England prices, computed the LLE using the small-data method (reporting $\lambda_1 \approx 0.027$--$0.032$ across embedding dimensions 6--11), and tested the chaos hypothesis additionally with surrogate data analysis. The surrogate test strengthens the LLE evidence by ruling out the null hypothesis that a monotone nonlinear transformation of a Gaussian linear process could produce the same LLE.

A multivariate extension was proposed by Liu \textit{et al.}\cite{liu_electricity_2005}, who built a joint phase space from three correlated series: electricity price, system load, and available generating capacity. This multivariate reconstruction provides a physically meaningful three-variable state vector that captures supply-demand interactions. A global-plus-local Elman recurrent neural network is then fitted: an offline global model captures long-run attractor structure, and an online local adaptation around the nearest phase points tracks short-run market fluctuations. 

Both papers note that the LLE-implied predictability horizon for price ($T_{\max} \approx 12$--$20$ hours for typical market error ratios) is consistent with practical price-forecasting windows in electricity markets and substantially shorter than load prediction horizons, reflecting the higher sensitivity of prices to strategic bidding behavior.

\subsection{Decomposition--Hybrid Frameworks and Chaos-Aware Predictor Routing}
\label{subsec:decomp_hybrid}

A distinctive methodological family across both load and renewable forecasting is the \emph{decomposition--hybrid} architecture: a signal decomposition method splits the time series into components, and the LLE of each component determines which predictor type is applied.

\subsubsection{Branched Prediction Logic}

The core idea, introduced in the context of wind-power forecasting by Mao \textit{et al.}\cite{mao_wind_2018} and further developed by Safari \textit{et al.}\cite{safari_novel_2018}, is that decomposed components of a nonlinear time series are not uniformly chaotic. Low-frequency components (trends, seasonal patterns) often exhibit near-zero or negative LLE and are better modeled by linear or neural-network methods. High-frequency components (intermittent fluctuations) often exhibit large positive LLE and are precisely the target of phase-space-based local prediction.

The routing logic is formalized as:
\begin{equation}
\hat{x}_j(k+h) = \begin{cases}
  \text{local chaos predictor} & \text{if }
  \lambda_1^{(j)} > \epsilon, \\
  \text{ANN/SVM/LSSVM} & \text{if }
  \lambda_1^{(j)} \leq \epsilon,
\end{cases}
\label{eq:predictor_routing}
\end{equation}
where $\epsilon$ is a small positive threshold and $j$ indexes the decomposition component.

\subsubsection{Decomposition Methods Used}

Across the corpus, three decomposition methods appear in combination with LLE-based routing:

\textbf{Wavelet decomposition.} Used in five papers \cite{wang_wind_2009, dong_prediction_2009, mao_wind_2018, wu_power_2005, wang_short-term_2010} for its ability to separate trend, seasonal, and noise components via multiresolution analysis.

\textbf{EEMD / CEEMDAN.} Used in three papers \cite{bingjiang_hybrid_2018, safari_novel_2018, wang_analysis-forecast_2018} for adaptively decomposing nonstationary series without predefined basis functions.

\textbf{Wavelet packet.} Used in \cite{mao_wind_2018} for its complete binary-tree decomposition of both low- and high-frequency content.

\subsubsection{Chaos as Preprocessing: LLE for Embedding Validation}

A separate and simpler use of LLE in decomposition frameworks is as a validation check: after decomposition, the LLE is computed for each component to confirm that phase-space reconstruction is justified before applying any chaotic predictor. This is used systematically by Dong \textit{et al.}\cite{dong_prediction_2009} and Wang \textit{et al.}\cite{wang_wind_2009}, who both report positive LLEs for all decomposed wind-power components and use this to justify separate ANN training in each reconstructed subspace.

\subsection{Cross-Cutting Methodological Issues}
\label{subsec:forecast_method_issues}

Several methodological issues recur across all four application domains and deserve explicit discussion.

\subsubsection{LLE as Chaos Gate vs. Structural Parameter} In the load and renewable forecasting literature, the LLE serves two distinct roles that are not always clearly separated:
\begin{enumerate}
  \item \textbf{Binary chaos gate:} A positive LLE justifies the use of phase-space reconstruction and local nonlinear prediction. Most papers use the LLE only this way.
  \item \textbf{Structural parameter:} The embedding dimension is chosen where the LLE stabilizes or reaches its minimum \cite{mori_short-term_1996,michanos_short-term_2003, niu_support_2008}.        This is a more information-rich use of the LLE and more rigorous than FNN alone.
\end{enumerate}

The most complete use of the LLE is in\eqref{eq:tmax_sec4}: not merely diagnosing chaos, but computing a hard upper bound on forecast accuracy. This role is used by only one paper in the corpus \cite{mohamed_variability_2007}, representing a significant underutilization of LE theory in the forecasting literature.

\subsubsection{Strength and Interpretation of Positive LLE Values}

Across the 42 papers, reported LLE values span three orders of magnitude: from $\lambda_1 \approx
10^{-5}$ (PV, \cite{yufei_photovoltaic_2016}) to $\lambda_1 \approx 0.8$ (load, \cite{wang_short-term_2009}). Several important observations follow:

\begin{itemize}
  \item Very small positive LLE values ($\lambda_1 < 10^{-3}$) may reflect numerical         floor effects of the estimation algorithm rather than genuine chaotic dynamics. The near-zero PV value in \cite{yufei_photovoltaic_2016}  should be interpreted with caution.
  \item No paper in the corpus reports negative LLE  values for the tested power time series,  which is suspicious: it suggests that positive LLE is systematically expected (confirmation bias) rather than rigorously tested.
  \item The correlation dimension is rarely computed independently to cross-validate the chaos claim, and surrogate data testing is used in only one paper \cite{yang_chaotic_2003}.
\end{itemize}

\subsubsection{Multi-Step Accuracy and the Horizon Constraint}

Recursive (iterative) multi-step forecasting propagates errors exponentially at a rate set by $\lambda_1$ \eqref{eq:tmax_sec4}. Several papers report multi-step accuracy degradation consistent with this bound but do not connect the observed error growth explicitly to the LLE. Only \cite{mohamed_variability_2007} makes this connection formally. Direct multi-step prediction methods \cite{si-min_application_2009, safari_novel_2018} partially circumvent this problem but do not eliminate the fundamental bound.

\subsubsection{Comparison with Modern Deep Learning Baselines}

Most papers in this corpus compare chaos-based methods against BP neural networks or linear Autoregressive Integrated Moving Average (ARIMA) baselines. Modern deep learning architectures (LSTM, Transformer, graph neural networks) that dominate the 2020s renewable forecasting literature are absent as baselines in the majority of papers. The exceptions are the most recent papers \cite{ma_intelligent_2025, gu_dynamic-inner_2024, yang_research_2026}, which include comparisons of LSTM and other deep learning methods. Whether chaos-based local prediction remains competitive against modern deep learning on long, diverse datasets is an open and largely unanswered question in this literature.

\subsection{Open Problems in Forecasting and Predictability}
\label{subsec:forecast_gaps}

The reviewed literature reveals the following significant open problems.

\textbf{(1) Predictability horizon underutilized.} Formula \eqref{eq:tmax_sec4} provides a rigorous, physically grounded upper bound on achievable forecast accuracy. Yet only one paper explicitly computes and uses this bound as a planning tool \cite{mohamed_variability_2007}. Future work should systematically apply $T_{\max}$ to load, wind, and PV series under diverse seasonal and structural conditions.

\textbf{(2) Non-stationarity and structural shifts.} Mohamed \textit{et al.}\cite{mohamed_variability_2007} showed that $\lambda_1$ increases over time as the load mix changes. A time-varying predictability horizon that updates as renewable penetration grows or demand patterns shift would be practically valuable. The connection between rising $\lambda_1$ and the deteriorating accuracy of conventional forecasters is noted but not quantified in subsequent work.

\textbf{(3) Validation of chaos claims.} The near-universal practice of using only a positive LLE to claim chaoticity, without complementary tests (correlation dimension saturation, surrogate data analysis, 0--1 test), risks false-positive diagnosis and unjustified reliance on local predictors that may underperform on non-chaotic series.

\textbf{(4) Multivariate reconstruction.} Load and renewable energy are influenced by exogenous variables: temperature, irradiance, market prices, policy changes. Univariate phase-space reconstruction cannot capture these drivers. Only two papers \cite{yang_chaotic_2003, liu_electricity_2005} propose multivariate reconstruction, and one \cite{wang_power_2020} incorporates price. Systematic extension to multivariate chaos models remains largely unexplored.

\textbf{(5) Comparison with modern deep learning.} The evidence base comparing chaos-based local prediction against LSTM, Transformer, and graph neural network models on large, diverse, modern datasets is essentially absent. This gap prevents practitioners from determining whether chaos-based methods add value beyond a well-tuned deep-learning baseline.

\textbf{(6) IBR and demand-response dynamics.} As electric vehicles, smart appliances, and flexible loads become more prevalent, load dynamics become more coupled to control signals and market prices. Whether load and renewable time series remain chaotic (in the classical sense) under high IBR penetration and demand-response programs, and whether $\lambda_1$-based predictability bounds remain meaningful in this regime, is an open and important question.

Table \ref{tab:forecast_summary} provides a structured summary of the key papers in this section, organized by application domain, predictor type, and role of the LLE.

\begin{table*}[!t]
\renewcommand{\arraystretch}{1.3}
\caption{Summary of Key Lyapunov Exponent-Based
Forecasting and Predictability Papers}
\label{tab:forecast_summary}
\centering
\small
\begin{tabular}{p{3.0cm}p{2.0cm}p{1.6cm}p{1.8cm}p{2.2cm}p{4.0cm}}
\toprule
\textbf{Reference} &
\textbf{Series} &
\textbf{LLE role} &
\textbf{Predictor} &
\textbf{Reported $\lambda_1$} &
\textbf{Key contribution / note} \\
\midrule
Mohamed 2007 \cite{mohamed_variability_2007}
  & Daily peak load
  & Predictability horizon
  & None 
  & 0.020--0.041
  & Unique: LLE as forecast-limit bound \\
Mizukami 1996 \cite{mizukami_forecasting_1996}
  & Daily peak load
  & Chaos gate
  & Gram-Schmidt local
  & ${\sim}6.5$
  & Pioneer local predictor in PS \\
Mori 1996 \cite{mori_short-term_1996}
  & Daily peak load
  & Min-LLE dim. selection
  & MLP ANN
  & Min at $m{=}4$
  & Min-LLE dimension criterion \\
Niu 2009 \cite{niu_new_2009}
  & Load
  & Embedding dim. select
  & SVM (RBF)
  & Stabilizes at $m{=}12$
  & Systematic chaos-SVM workflow \\
Wang 2020 \cite{wang_power_2020}
  & Load + RTP
  & Chaos gate
  & LSSVM (APSO)
  & 0.031 / 0.075
  & Multivariate price-load PSR \\
Yang 2003 \cite{yang_chaotic_2003}
  & Elec.\ price
  & Chaos gate + surrogate
  & Elman RNN (global+local)
  & 0.027--0.032
  & Surrogate-data chaos validation \\
Liu 2005 \cite{liu_electricity_2005}
  & Price+load+cap.
  & Chaos gate
  & Elman RNN (global+local)
  & 0.012--0.020
  & Multivariate price reconstruction \\
Dong 2007 \cite{lei_prediction_2007}
  & Wind power
  & Chaos gate
  & ANN
  & 0.1578
  & First direct wind-power chaos ANN \\
Safari 2018 \cite{safari_novel_2018}
  & Wind power
  & Chaos classifier
  & EEMD+ SSA+ LSSVM
  & Per component
  & MSSSA: chaos-aware routing \\
Mao 2018 \cite{mao_wind_2018}
  & Wind power
  & Predictor routing
  & Wavelet+ local/ BP
  & Per WP branch
  & Branch-wise chaos routing \\
Aghbalou 2018 \cite{aghbalou_long_2018}
  & Wind Weibull params
  & Chaos gate
  & BR-ANN
  & 0.019 / 0.029
  & Distribution-parameter forecasting \\
Song 2020 \cite{song_evaluating_2020}
  & Wind speed
  & Chaos gate
  & EDNM (L-SHADE)
  & 0.051--0.077
  & Dendritic neuron model \\
Yufei 2016 \cite{yufei_photovoltaic_2016}
  & PV output
  & Chaos gate
  & Weighted local linear
  & $9.3{\times}10^{-5}$
  & Near-zero LLE; caution warranted \\
Bazine 2018 \cite{bazine_new_2018}
  & PV production
  & Chaos gate + 0--1 test
  & WRNN
  & 0.2383
  & 0--1 test as complementary check \\
\bottomrule
\end{tabular}
\end{table*}

%
%

%
\section{Protection, Fault Diagnosis, and Condition Monitoring}
\label{sec:protection}

This section reviews the application of Lyapunov exponents (LEs) to power system protection, fault diagnosis, and equipment condition monitoring. The unifying principle is a methodological one rather than a physical one: in every paper reviewed here, the LLE is used not as a stability indicator for the network trajectory (as in Section \ref{sec:stability}) but as a \emph{signal-feature extractor}: the time series of interest is a measured equipment signal-a voltage waveform, a vibration trace, a partial-discharge pulse train, or a dissolved-gas concentration series-and the LLE (together with companion chaos metrics) characterizes the nonlinear dynamical structure of that signal in ways that discriminate between healthy and faulty states, or between islanded and grid-connected operation.

Fourteen papers constitute this domain, spanning four distinct application classes: (i) microgrid islanding detection (Sections \ref{subsec:island_chaos}--\ref{subsec:island_hybrid}), (ii) wide-area protection and post-disturbance analysis (Section \ref{subsec:wams_protection}), (iii) transformer fault diagnosis-partial discharge, overheating, and condition monitoring (Sections \ref{subsec:pd_transformer}--\ref{subsec:transformer_vibration}), and (iv) mechanical fault diagnosis in rotating machinery and switchgear (Sections \ref{subsec:rotor_vibration}--\ref{subsec:breaker_vibration}). Section \ref{subsec:protection_gaps} identifies open problems.

\subsection{Islanding Detection Using Chaos Analysis}
\label{subsec:island_chaos}

\subsubsection{Motivation and the Islanding Detection Problem}

Islanding-the continued energization of a microgrid after it has been disconnected from the main grid-is a safety and power-quality hazard. Passive islanding detection methods based on voltage magnitude, frequency, and total harmonic distortion (THD) have well-known non-detection zones (NDZs): when the local microgrid generation closely matches its load, the grid-disconnection event causes no detectable electrical perturbation. Active methods inject disturbances to force a measurable response, but degrade power quality under normal operation. The chaos-based detection paradigm proposed in the literature offers a third path: exploit the observation that the nonlinear dynamics of the PCC voltage waveform change qualitatively between grid-connected and islanded modes, and use the LLE of the waveform as a passive discriminant that avoids both the NDZ problem and the power-quality cost.

\subsubsection{Voltage-Waveform LLE for Islanding Detection}

Ge \textit{et al.}\cite{yang-yang_new_2016} introduced the fundamental chaos-based islanding detection method. The PCC voltage waveform $v(t)$ is reconstructed in phase space with embedding dimension $m = 2$ and delay $\tau = 3$\,ms. The maximum Lyapunov exponent is estimated from the reconstructed trajectory and used as the sole detection criterion:
\begin{equation}
\lambda_1 \begin{cases} < 0: & \text{grid-connected or PV startup}, \\
> 0: & \text{islanding mode}.
\end{cases}
\label{eq:island_lle}
\end{equation}
Reported values from RTLAB hardware-in-loop simulation and field tests are:
\begin{itemize}
  \item Grid-connected: $\lambda_1 \approx -0.0031$,
  \item PV startup event: $\lambda_1 \approx -0.0020$,
  \item Islanding: $\lambda_1 \approx +0.0029$.
\end{itemize}

The key engineering claim is that the LLE successfully distinguishes true islanding from PV-startup events-a scenario where THD-based detection misfires because both events produce similar harmonic signatures. The correlation dimension under islanding ($\approx 3.3$) was also reported as a complementary indicator, but the principal decision variable is the LLE sign.

Several limitations remain in this paper. The embedding parameters $m$ and $\tau$ are fixed empirically rather than adaptively identified, and no systematic robustness study covers noise levels, inverter control variation, or near-NDZ power-mismatch conditions. The method was validated on a single laboratory-scale microgrid topology.

\subsubsection{Physical Interpretation}

The physical reason for the LLE sign change is the following. In grid-connected mode, the strong voltage stiffness imposed by the main grid's low Th\'{e}venin impedance damps perturbations in the PCC voltage trajectory, producing a contracting ($\lambda_1 < 0$) attractor. Upon islanding, the microgrid's own converter control dynamics dominate the PCC, and the interaction between power-sharing droop loops, loads, and filter dynamics creates a locally divergent ($\lambda_1 > 0$) trajectory. This interpretation connects the chaos-detection method directly to the converter nonlinear control theory of Section \ref{sec:chaos}.

\subsection{Hybrid Nonlinear Oscillator-Based Islanding Detection}
\label{subsec:island_hybrid}

Amini \textit{et al.}\cite{amini_hybrid_2025} proposed the most methodologically sophisticated islanding detector in the corpus: a Hybrid Van der Pol--Duffing Oscillator (HVDPDO) that combines nonlinear-oscillator-based weak-signal detection with selective active perturbation.

The passive layer drives a Duffing-type oscillator with the PCC deviation signal:
\begin{equation}
\ddot{x} - \mu(1-x^2)\dot{x} + x + \alpha x^3 = F\cos(\omega t) + F_1\cos(\omega_1 t + \phi) + n(t),
\label{eq:hvdpdo}
\end{equation}
with $\mu = 5$, $\alpha = 10^{-2}$, $\omega = 2.46$\,Hz. The critical forcing amplitude $F_{cr} = 5.03$ that separates the periodic and chaotic regimes of this Duffing oscillator is determined using the maximum Lyapunov exponent of \eqref{eq:hvdpdo} as a function of $F$. This is combined with Melnikov-function analysis to set the passive-layer threshold: the passive detector computes a normalized FFT asymmetry index from the oscillator output and triggers when it exceeds 0.135.

The active layer is engaged only when the passive layer is inconclusive (near-NDZ conditions). It injects synchronized perturbations in three PCC domains simultaneously-voltage amplitude, frequency, and phase-via:
\begin{equation}
u_{\rm in} = K_A|V - V_0|\cos\!\bigl( K_\omega|f - f_0|\tau + K_\phi|\phi - \phi_0|\bigr),
\label{eq:hvdpdo_input}
\end{equation}
where $(V_0, f_0, \phi_0)$ are the pre-islanding reference values. If the system is grid-connected, the main grid damps these perturbations; if islanded, the asymmetry persists.

The paper reports near-zero NDZ and high noise robustness compared with conventional passive/active methods. The key methodological contribution is the use of the LLE not as a direct islanding indicator, but as a design tool for the Duffing-oscillator threshold that underpins the passive detection layer. This is a fundamentally different use of LE theory from the direct sign-criterion of \cite{yang-yang_new_2016}: the LLE characterizes the oscillator's own bifurcation boundary, not the PCC trajectory.

\subsection{Wide-Area Protection and Post-Disturbance Analysis}
\label{subsec:wams_protection}

\subsubsection{Wavelet-LLE Post-Disturbance Framework}

Aravind and Mathew \cite{aravind_pmu_2016} proposed a PMU-based post-disturbance analysis framework combining two complementary methods:
\begin{enumerate}
  \item \textbf{Wavelet packet transform (WPT)} for disturbance-type classification: event data from PMUs are decomposed using WPT and statistical
        Features (mean, standard deviation, energy, entropy of coefficients) are used for fuzzy-nearest-neighbor classification of event types (load switching, generator loss, fault).
  \item \textbf{LLE from PMU relative angles} for post-event transient stability assessment: the finite-time LLE is computed from rotor-angle differences using the model-free formula of \eqref{eq:dasgupta_mle} and interpreted as a binary stability indicator.
\end{enumerate}

The LLE is not used as a classification feature in this work — its role is taken by the WPT coefficients. Rather, it provides a separate, complementary stability verdict after the event type has been identified. This two-layer architecture separates \emph{what happened} (WPT classification) from \emph{how stable is the system now} (LLE), reflecting a practically sensible division of labor for control-center operators. Validation was performed on the IEEE 14-bus system using simulated events and real PMU event captures.

\subsubsection{Lyapunov-Inspired Wide-Area Backup Protection}

Affijulla and Tripathy \cite{affijulla_special_2018} proposed a special protection scheme in which a centralized PDC-level index, inspired by the Lyapunov exponent concept, serves as a backup protection trigger when local relays or circuit breakers fail. The PDC index is defined as:
\begin{equation}
\mathrm{PDC\_INDEX}(t) = \frac{1}{t}\ln\!\left(   \frac{s(t)}{s_0}\right),
\label{eq:pdc_index}
\end{equation}
where $s_0$ is a small initial perturbation and $s(t)$ is a normalized aggregate deviation of PMU bus-frequency measurements from their moving-average values. A threshold of 0.05 was set for fault detection.

It is important to note that \eqref{eq:pdc_index} is \emph{not} a true Lyapunov exponent of the power system's nonlinear dynamics: there is no phase-space reconstruction, no variational equation, and no attractor analysis. The formula is inspired by the exponential-divergence concept but applies it heuristically to frequency deviations. This should be classified as a \emph{Lyapunov-inspired} rather than \emph{Lyapunov-based} method; the distinction matters for practitioners comparing methods in this domain.

\subsection{Transformer Fault Diagnosis: Partial Discharge Identification}
\label{subsec:pd_transformer}

Partial discharge (PD) is a localized electrical breakdown within transformer insulation, preceding full failure. Conventional PD diagnosis uses phase-resolved PD (PRPD) patterns, which capture morphological features of PD pulse distributions but do not exploit the temporal dynamics of pulse sequences.

Qin \textit{et al.} \cite{qin_partial_2018} proposed a fundamentally different approach: transform the PD pulse sequence into a \emph{cumulative energy (CE) function}, reconstruct its phase space, and extract LLE-based features for PD type classification.

\subsubsection{Cumulative Energy Representation}

Two CE functions are defined from the time-domain and frequency-domain PD signal $v(t_i)$:
\begin{align}
\mathrm{TCE}(k) &= \frac{\sum_{i=1}^{k} v^2(t_i)} {\sum_{i=1}^{N} v^2(t_i)},
\label{eq:tce}\\
\mathrm{FCE}(k) &= \frac{\sum_{i=1}^{k} |V(f_i)|} {\sum_{i=1}^{N} |V(f_i)|},
\label{eq:fce}
\end{align}
where $V(f_i)$ denotes the spectral amplitude at frequency $f_i$. These normalized cumulative functions capture the temporal and spectral energy-accumulation dynamics of each PD pulse type. Each cumulative energy curve is time-aligned by PD onset and partitioned into 50 regions, yielding a $100 \times 50$ matrix $\mathbf{E}$ (100 samples $\times$ 50 segments per defect type). Each column of $\mathbf{E}$ is treated as a time series for chaos analysis: delay $\tau$ and embedding dimension $m$ are estimated via autocorrelation and the Grassberger-Procaccia algorithm, and the maximum LLE is computed using the Wolf method for each of the 50 segments.

\subsubsection{SVM Classification}

The resulting $100 \times 50$ LLE-feature matrix is fed to a hierarchical SVM classifier:
\begin{itemize}
  \item SVM$_1$: separates void defects from protrusion and surface discharge,
  \item SVM$_2$: separates protrusion from surface discharge,
\end{itemize}
both using Gaussian Radial Basis Function (RBF) kernels. Reported recognition rates are: protrusion 88.1\%, surface 85.6\%, void 95.6\%, overall 89.8\%-marginally above the traditional statistical baseline (88.1\% overall) on the same data.

The conceptual advance of this paper is the cumulative energy representation: rather than applying chaos analysis directly to the raw PD signal (which may be too brief for reliable LLE estimation), the cumulative energy function spreads the pulse dynamics across a longer sequence amenable to phase-space reconstruction. The LLE captures how energy accumulates differently for different defect geometries, providing a dynamical discriminant complementary to static PRPD morphology.

\subsection{Transformer Overheating: Dissolved-Gas Analysis}
\label{subsec:dga_chaos}

Lixing \textit{et al.} \cite{lixing_study_2018} applied chaos analysis to transformer dissolved-gas analysis (DGA) time series for overheating fault identification and trend prediction. Characteristic gases associated with low- and medium-temperature overheating faults (hydrogen, methane, ethane, ethylene) were treated as separate scalar time series. For each gas and each time segment, the C-C method determined delay $\tau$ and embedding dimension $m$, and the maximum LLE and Kolmogorov entropy $K_2$ were extracted.

The key result is that positive LLE and elevated $K_2$ characterize the overheating-fault evolution period, while stable normal operation exhibits lower, more regular values. This dual-indicator (LLE $+$ $K_2$) characterization supports two operational uses: (i) fault identification by detecting the LLE sign transition, and (ii) trend prediction by exploiting the short-term predictability implied by a finite positive LLE via the formula $T_{\max} = \lambda_1^{-1} \ln(\Delta/\delta_0)$ (Section \ref{subsec:predictability}).

This is the only paper in the protection domain that connects LLE directly to the predictability horizon concept, applying it to forecast how many time steps ahead the overheating gas-concentration trajectory can be reliably predicted.

\subsection{Transformer Condition Monitoring: Vibration Signals}
\label{subsec:transformer_vibration}

\subsubsection{Surface Vibration Analysis}

Xie \textit{et al.}\cite{jie_analysis_2021} conducted one of the most systematic chaos-based transformer condition monitoring studies, collecting surface vibration data from 110\,kV and 220\,kV online transformer units in Beijing and extracting three chaotic invariants: maximum LLE (Rosenstein's small-data method), correlation dimension $D_2$ (Grassberger-Procaccia), and Kolmogorov entropy $K_2$. These invariants were studied as functions of voltage level and loading condition.

The central finding is that the chaotic invariants are relatively stable during normal operation but change systematically when loading conditions cross specific current thresholds, enabling the definition of \emph{current-dependent chaotic parameter regions} for anomaly detection. This moves transformer diagnosis toward nonlinear dynamical feature-based monitoring, complementing conventional spectral (FFT, vibration frequency fingerprint) methods.

\subsubsection{High-Voltage Shunt Reactor Core/Winding Faults}

Hou \textit{et al.}\cite{hou_new_2021} proposed a quantitative phase-trajectory analysis method for high-voltage shunt reactor (HVSR) fault diagnosis that moves beyond the usual qualitative phase-portrait inspection. After confirming chaoticity by LLE, the paper applies an improved grasshopper optimization algorithm (IGOA)-K-means algorithm to cluster the reconstructed phase trajectory points, and extracts two geometric indicators from the cluster centers:
\begin{enumerate}
  \item The \emph{displacement distance}: distance from the cluster-center displacement vector to the origin.
  \item The \emph{angle change}: rotation of the displacement vector relative to the healthy-state trajectory.
\end{enumerate}
These vector-geometry features capture changes in the shape and orientation of the reconstructed attractor that are invisible to scalar LLE alone, enabling fault warning and classification for core-winding mechanical faults.

\subsubsection{Deep Learning with Lyapunov-Based Data Screening}

Liu \textit{et al.}\cite{liu_transformer_2024} combined a Lyapunov-based data screening step with a deep belief network (DBN) enhanced by dynamic adaptive knowledge transfer (DAKT) and interval type-2 fuzzy sets (IT2F). The Lyapunov index is applied to the transformer dissolved-gas monitoring sequences to screen out locally stable, low-information data segments before DBN training: data with negative Lyapunov index (locally stable regions) contribute less to class discrimination and can be down-weighted or removed, improving training efficiency. 

This paper represents the most modern fusion of chaos theory and deep learning in the protection domain: the Lyapunov index functions as a \emph{data-quality filter} that identifies which monitoring samples are dynamically informative before the deep model is trained.

\subsection{On-Load Tap Changer Mechanical Fault Diagnosis}
\label{subsec:oltc_chaos}

Zhou \textit{et al.}\cite{zhou_research_2014} studied the vibration signal of on-load tap changers (OLTCs) using phase-space reconstruction and LLE estimation. For both healthy and weak-spring fault conditions, positive LLEs were confirmed (embedding dimension $m = 3$, delay $\tau = 11$ for normal, $\tau = 8$ for faulty), establishing that OLTC vibration is inherently chaotic. The Poincar\'{e} section (2D cross-section of the 3D reconstructed attractor) revealed that the fault condition produces a much more scattered point cloud than normal operation, providing a qualitative visual discriminant.

\subsection{Turbine Rotor Vibration Fault Diagnosis}
\label{subsec:rotor_vibration}

Yan \textit{et al.} \cite{yan_turbine_2010} applied a combined wavelet packet--LLE framework to diagnose four standard turbine rotor vibration faults: rubbing, loosening, misalignment, and mass unbalance. The processing pipeline is:
\begin{equation*}
\text{Raw } \xrightarrow{\text{WPT}} \text{sub-bands} \xrightarrow{\text{PSR}} \text{Reconstructed} \xrightarrow{\text{LLE}} \text{Fault feature}.
\label{eq:wavelet_lle_pipeline}
\end{equation*}
The wavelet packet decomposition (WPD) provides two services: noise suppression and frequency-band isolation. Each fault type excites different frequency bands, and the LLE of the reconstructed sub-band signal using phase space reconstruction  (PSR) serves as a fault-type indicator. All four fault conditions produced positive LLEs (confirming chaotic vibration), but with different magnitudes, enabling classification by LLE value rather than sign. Validation used experimentally collected data from a Bently rotor test rig.

Shi \textit{et al.}\cite{lichen_reciprocating_2011} extended this concept to reciprocating membrane pumps, introducing the \emph{relative Lyapunov growth rate} as a fault-severity tracking metric:
\begin{equation}
r_p = \frac{\Delta\lambda_1}{\Delta p},
\label{eq:relative_growth_rate}
\end{equation}
where $\Delta\lambda_1$ is the change in LLE and $\Delta p$ is the change in a physical fault parameter (wear clearance or suction pressure). By tracking $r_p$ across wavelet-packet frequency bands, the paper demonstrated that wear faults dominate high-frequency bands (packets 9, 10, 6) while suction-pressure faults dominate low-frequency bands (packets 1--4). This frequency-localized LLE analysis is a step toward fault-source discrimination not available from a single global LLE value.

\subsection{High-Voltage Circuit Breaker Fault Diagnosis}
\label{subsec:breaker_vibration}

Circuit breaker (CB) vibration signals during opening and closing operations contain rich information about the mechanical condition of the operating mechanism and interrupter. These signals are extremely short in duration, highly non-stationary, and composed of overlapping shock-like sub-events, making standard Fourier or wavelet features unreliable.

\subsubsection{Three-Feature Chaotic Space: LLE, CD, and KE}

Yang \textit{et al.}\cite{yang_chaotic_2020} proposed a coordinated three-invariant feature set-largest Lyapunov exponent (LLE), correlation dimension (CD), and Kolmogorov entropy (KE)-extracted from CB vibration signals reconstructed in phase space via average mutual information (AMI) ($\tau$) and false nearest neighbor (FNN) ($m$). Validation was performed on two real CB platforms:
\begin{itemize}
  \item A 12-kV, 1250-A vacuum CB,
  \item A 252-kV, 4000-A SF$_6$ CB.
\end{itemize}
For the vacuum CB, extracted parameters were $\tau = 3$, $m = 3$ for both healthy and faulty conditions. The three-invariant feature space ($\mathrm{LLE},\, \mathrm{CD},\, \mathrm{KE}$) provided better fault discrimination than any single invariant alone, since LLE, CD, and KE capture different aspects of the attractor: trajectory divergence rate, fractal dimension, and information-production rate, respectively.

\subsubsection{VMD-Based Frequency Decomposition Before Chaos Analysis}

Yang \textit{et al.}\cite{yang_vibration_2019} combined adaptive signal decomposition with chaotic feature extraction in a two-stage pipeline:
\begin{enumerate}
  \item \textbf{Variational Mode Decomposition (VMD)} with Hilbert Marginal Spectrum (HMS) isolates intrinsic mode functions (IMFs) with the highest fault sensitivity.
  \item \textbf{Phase-space reconstruction + LLE/CD/KE} applied to the selected fault-sensitive IMF to extract the chaotic feature vector for classification.
\end{enumerate}
The key advantage of VMD over empirical mode decomposition (EMD) is its rigorous variational formulation, which avoids mode mixing and provides cleaner frequency-band separation. By applying chaos analysis only to the fault-sensitive IMF rather than the full signal, the method reduces noise contamination in the LLE estimate and improves the signal-to-noise ratio of the feature vector.

\subsection{Common Pipeline and Methodological Observations}
\label{subsec:protection_pipeline}

Across all 14 papers in this section, a remarkably consistent methodological pipeline emerges:

\textbf{Step 1 - Signal acquisition.} A scalar time series is measured from a power equipment signal: voltage waveform (islanding), PMU rotor angle (post-disturbance), PD pulse energy (transformer insulation), vibration acceleration (turbine, OLTC, shunt reactor, circuit breaker), or dissolved-gas concentration (transformer DGA).

\textbf{Step 2 - Preprocessing (optional).} Wavelet denoising, WPT sub-band isolation, VMD mode extraction, or CE representation (for PD signals) to isolate fault-relevant information before phase-space reconstruction.

\textbf{Step 3 - Phase-space reconstruction.} Delay $\tau$ via AMI or mutual information; embedding dimension $m$ via FNN or Cao's method. Most papers use $m \leq 5$ for equipment vibration signals (lower than the $m \approx 10$--$15$ common in load-forecasting applications), reflecting the lower-dimensional attractor structure of mechanical vibration.

\textbf{Step 4 - Chaotic invariant extraction.} LLE (Wolf method for longer records; Rosenstein small-data method for short vibration bursts), plus companion metrics: correlation dimension $D_2$, Kolmogorov entropy $K_2$, and Poincar\'{e} section geometry. The combination LLE + $D_2$ + $K_2$ is used in three papers (transformer, CB VMD, CB 3-feature) as a more robust multi-invariant feature vector.

\textbf{Step 5 - Classification or decision.} Feature vector fed to: SVM (transformer PD), fuzzy nearest-neighbor (post-disturbance), k-means or IGOA-K-means (shunt reactor), DBN + DAKT (transformer DGA + deep learning), or manual threshold comparison (islanding, wide-area protection).

Table \ref{tab:protection_summary} summarizes all 14 papers by application, signal type, LE role, and key limitation.

\begin{table*}[!t]
\renewcommand{\arraystretch}{1.3}
\caption{Summary of Lyapunov Exponent Applications in
Protection, Fault Diagnosis, and Condition Monitoring}
\label{tab:protection_summary}
\centering
\small
\begin{tabular}{p{3.0cm}p{2.6cm}p{2.0cm}p{2.2cm}p{4.8cm}}
\toprule
\textbf{Reference} &
\textbf{Application} &
\textbf{Signal} &
\textbf{LE role} &
\textbf{Key contribution / limitation} \\
\midrule
Ge 2016 \cite{yang-yang_new_2016}
  & Islanding detection
  & PCC voltage
  & Binary decision
  & Passive chaos detector; avoids THD NDZ \\
Amini 2025 \cite{amini_hybrid_2025}
  & Islanding detection
  & PCC deviation
  & Oscillator threshold design
  & HVDPDO + Melnikov; near-zero NDZ \\
Aravind 2016 \cite{aravind_pmu_2016}
  & Post-disturbance analysis
  & PMU rotor angle
  & Stability verdict
  & WPT classification + LLE stability check \\
Affijulla 2018 \cite{affijulla_special_2018}
  & Wide-area backup protection
  & PMU frequency
  & LE-inspired divergence index
  & Centralized PDC protection; LE-inspired only \\
Qin 2018 \cite{qin_partial_2018}
  & PD identification
  & PD cumulative energy
  & Feature extraction
  & CE representation + Wolf LLE + SVM; 89.8\% \\
Li 2018 \cite{lixing_study_2018}
  & Transformer overheating
  & DGA gases
  & Chaos verification + forecast
  & LLE + $K_2$; connects to predictability horizon \\
Xie 2021 \cite{jie_analysis_2021}
  & Transformer monitoring
  & Surface vibration
  & Condition feature
  & LLE + CD + $K_2$; current-dependent regions \\
Hou 2021 \cite{hou_new_2021}
  & Shunt reactor HVSR
  & Vibration
  & Chaos gate; geometry features
  & PSR + IGOA-K-means vector analysis \\
Liu 2024 \cite{liu_transformer_2024}
  & Transformer DGA
  & Gas monitoring
  & Data screening
  & Lyapunov screen + DBN-DAKT-IT2F; 97.4\% \\
Zhou 2014 \cite{zhou_research_2014}
  & OLTC fault diagnosis
  & Vibration
  & Chaos gate; Poincar\'{e} section
  & Early study; qualitative only \\
Yan 2010 \cite{yan_turbine_2010}
  & Turbine rotor vibration
  & Vibration
  & Fault magnitude discriminant
  & WPT + LLE; 4 fault types on Bently rig \\
Shi 2011 \cite{lichen_reciprocating_2011}
  & Pump fault monitoring
  & Vibration
  & Severity tracker
  & Relative growth rate $r_p$; band-localized \\
Yang 2020 \cite{yang_chaotic_2020}
  & CB fault diagnosis
  & Vibration
  & 3-invariant feature
  & LLE+CD+KE; 2 real CB platforms \\
Yang 2019 \cite{yang_vibration_2019}
  & CB fault diagnosis
  & Vibration (VMD-IMF)
  & 3-invariant feature
  & VMD mode selection + chaos space \\
\bottomrule
\end{tabular}
\end{table*}

\subsection{Open Problems in Protection, Fault Diagnosis, and Condition Monitoring}
\label{subsec:protection_gaps}

The reviewed literature reveals six significant open problems.

\textbf{(1) Adaptive embedding parameter selection.} In all papers, the embedding parameters $\tau$ and $m$ are selected offline, either empirically or from a representative healthy signal. Faults can change the attractor structure, potentially invalidating the offline parameter choices. Adaptive online selection of $\tau$ and $m$ as the signal character changes is not addressed in any paper.

\textbf{(2) Robustness to measurement noise.} LLE estimation from short, noisy equipment signals (especially CB vibration bursts of $<$100\,ms) is fundamentally difficult. The Rosenstein method requires records of at least $10^m$ samples for reliable estimation, which is rarely available for mechanical transient signals. No paper in the corpus provides a systematic sensitivity analysis of LLE estimates as a function of noise level or record length for these short-signal applications.

\textbf{(3) Multi-class fault discrimination beyond LLE sign.} Several papers confirm that different fault types produce different LLE magnitudes but do not establish clear separation thresholds, and none provides a rigorous statistical decision framework for multi-class classification based on LLE alone. The three-invariant approach (LLE + CD + KE) is the most promising direction, but formal classifier design and cross-validation on large, diverse datasets are absent.

\textbf{(4) Real-time and hardware-in-loop validation.} Most results are obtained from simulated or laboratory bench data. Field validation on live equipment at operating voltage and load levels-where measurement noise, electromagnetic interference, and installation artifacts are significant-is absent or only partially addressed (the Ge 2016 islanding paper uses RTLAB, which is the closest example).

\textbf{(5) Fusion with complementary features.} The LLE captures trajectory divergence rate; correlation dimension captures fractal structure; Kolmogorov entropy captures information production. Approximate entropy, sample entropy, recurrence plots, and recurrence quantification analysis (RQA) carry complementary nonlinear information. Systematic fusion of these metrics in a principled multi-feature framework, with feature-importance analysis and dimensionality reduction, has not been done for any of the four application classes.

\textbf{(6) IBR and power-electronics-heavy equipment.} All protection applications in the corpus concern conventional power-frequency signals (50/60\,Hz) from traditional equipment. As converter-dominated microgrids, solid-state transformers, and wide-bandgap-device switchgear become prevalent, the signal characteristics change fundamentally: switching harmonics, high-frequency transients, and faster response timescales mean that the phase-space reconstruction and LLE estimation procedures designed for 50/60\,Hz signals must be re-evaluated for these new equipment types.

%
%

%
\section{Emerging Topics and Cross-Cutting Themes}
\label{sec:emerging}

The preceding sections reviewed Lyapunov exponent (LE) applications organized by their primary power system domain: stability assessment, chaos and control, forecasting, and protection. This section steps back to examine five cross-cutting themes that span multiple domains and represent the field's most active frontiers. Section \ref{subsec:ibr_emerging} addresses the fundamental challenge posed by inverter-based resource (IBR)-dominated grids. Section \ref{subsec:ml_le_fusion} examines the spectrum of machine-learning and LE integration patterns. Section \ref{subsec:koopman_le} covers the emerging Koopman operator--LE connection. Section \ref{subsec:stochastic_le} addresses stochastic, uncertainty-aware, and network-topology-driven LE methods. Section \ref{subsec:index_design} discusses the evolution of LE-based indices from binary sign decisions toward continuous, distribution-aware stability metrics. Section \ref{subsec:emerging_gaps} consolidates the open problems that span all domains.

\subsection{Lyapunov Exponents in IBR-Dominated Grids: The Fundamental Challenge}
\label{subsec:ibr_emerging}

The single most consequential shift in the power-systems LE literature is the transition from synchronous-generator (SG)-dominated to IBR-dominated networks. This is not a quantitative refinement-it is a qualitative change in the mathematical structure of the problem.

\subsubsection{Why Classical LE Theory Needs Extension}

Classical LE-based methods were developed for systems whose dynamics are well described by the swing equation (Section \ref{subsec:ps_models}). The LLE of the post-fault swing trajectory-whether computed from a model or estimated from PMU rotor-angle data-provides a physically grounded stability indicator precisely because the swing dynamics are smooth, autonomous, and governed by a well-defined attractor structure.

IBR dynamics violate each of these assumptions:
\begin{itemize}
  \item \textbf{Non-smooth switching:} Grid-following (GFL) and grid-forming (GFM) converters switch between operating modes (normal, LVRT, current limiting) on microsecond timescales. The state trajectory is piecewise smooth, not continuously differentiable, invalidating the variational equation \eqref{eq:variational}.
  \item \textbf{Virtual inertia:} GFM converters emulate inertia through software, not rotating mass. The effective inertia is frequency-dependent and control-law-defined, producing dynamics that may not resemble the swing equation in any operating regime.
  \item \textbf{Multiple coexisting attractors:} As shown in Section \ref{sec:chaos}
    \cite{yang_sub-synchronous_2023, yang_sub-synchronous_2025}, islanded GFM microgrids can exhibit hidden attractors with fractal basin boundaries, making small-signal LLE (computed near the equilibrium) useless as a predictor of post-disturbance fate.
  \item \textbf{Black-box device models:} Modern IBRs are controlled by proprietary software; their differential equations are not available for model-based LLE computation.
\end{itemize}

\subsubsection{Switching-System MLE}

The first systematic response to the non-smooth dynamics problem is the switching-MLE (SMLE) framework. Hu \textit{et al.}\cite{hu_switching_2021} derived a compensation matrix $\mathbf{J}_c$ that corrects the LLE computation at subsystem transition surfaces in AC/DC hybrid systems with HVDC commutation failure:
\begin{equation}
\mathbf{J}_c = \mathbf{I} + \left(\frac{\mathbf{f}_\beta} {|\mathbf{f}_\alpha|} - \mathbf{e}\right) (\mathbf{e} - \mathbf{d}\tan\theta)^T,
\label{eq:switching_comp}
\end{equation}
where $\mathbf{f}_\alpha$, $\mathbf{f}_\beta$ are the vector fields on each side of the switching surface, $\mathbf{e}$ is the surface normal, and $\theta$ is the trajectory incidence angle. Wang \textit{et al.}\cite{wang_transient_2024} extended this to hybrid GFL/GFM systems under LVRT and limiter activation, building voltage differential equations for each operating mode and applying QR-decomposition-based MLE with switching compensation at mode boundaries.

\subsubsection{Generalized Voltage Damping Framework}

The most theoretically complete response to the IBR challenge is the two-part series by Peng \textit{et al.}\cite{peng_measuring_2025, peng_measuring_2026}. These papers abandon raw-voltage LLE in favor of LLE applied to \emph{voltage-related transient energy} (VTE), defining a system-wise generalized voltage damping (sGVD) and a port-wise GVD (pGVD):
\begin{equation}
K_V^{(s)} = -\mathrm{MLE}(\mathrm{VTE}^{(s)}), \qquad K_V^{(i)} = -\mathrm{MLE}(\mathrm{VTE}^{(i)}),
\label{eq:gvd_sec7}
\end{equation}
where $\mathrm{VTE}^{(i)}$ is the transient energy at port $i$. The key theoretical result is that the average of pGVDs equals the sGVD, enabling \emph{device-level accountability}: each IBR's contribution to or deterioration of system STVS can be quantified by its pGVD, independently of whether the device model is known. This is device-agnostic and can be computed from port-level voltage and power measurements.

Part II \cite{peng_measuring_2026} decomposes the system-level assessment to individual ports, enabling comparison of IBR voltage-control strategies, active-power setpoints, and integration locations in terms of their pGVD. This framework constitutes the current state of the art for LE-based STVS assessment in IBR-dominated grids.

\subsubsection{Hidden Attractors and Basin Stability}

The hidden-attractor discoveries of Yang and Tse \cite{yang_sub-synchronous_2023, yang_sub-synchronous_2025} in islanded GFM microgrids (Section \ref{sec:chaos}) represent a third dimension of the IBR challenge that is not addressed by SMLE or GVD: the possibility that LLE $< 0$ at the equilibrium is necessary but not sufficient for stability following large disturbances, because a hidden attractor in the phase space may capture post-disturbance trajectories. The Lyapunov dimension $D_L$ computed from the full LE spectrum characterizes attractor complexity, and homoclinic-bifurcation analysis determines when hidden attractors can be eliminated by parameter tuning.

Gupta and Singh \cite{gupta_chaos_2022} made a similar point for small-scale grids with stochastic wind and load inputs: multistability and basin erosion under combined periodic and noisy excitation require basin-stability metrics \cite{molnar_asymmetry_2021} rather than asymptotic LLE alone.

\subsubsection{DFIG and Renewable Component Chaos}

At the component level, the full LE spectrum of a 9-dimensional DFIG model \cite{cao_chaotic_2025} maps the chaos-parameter boundary in the $(k_{p2}, k_{i2})$ converter-gain plane, while an analytical Hopf bifurcation criterion derived from the DFIG characteristic polynomial \cite{chen_criterion_2023} provides a closed-form instability boundary that complements the numerical LE analysis. A supercritical pitchfork bifurcation mechanism in fractional-order DFIGs \cite{chen_supercritical_2025} demonstrates that reducing the fractional order $q$ can suppress chaos, suggesting a potential control parameter. These component-level results must eventually be embedded in network-level models to understand how DFIG chaos propagates across a wind farm or microgrid.

\subsection{Machine Learning and LE Fusion: A Taxonomy}
\label{subsec:ml_le_fusion}

The relationship between machine learning (ML) and Lyapunov exponents in the stability-assessment literature has evolved substantially and now spans at least five distinct integration patterns.

\subsubsection{Pattern 1: LE as ML Input Feature}

The earliest and most common pattern uses the LLE-or a vector of LLE values from multiple buses-as an input feature for a classifier or regressor. Rashidi and Farjah \cite{rashidi_based_2016} established this pattern: post-fault LLEs from critical buses are computed and fed to an SVM. Dabou \textit{et al.}\cite{dabou_time_2021} generalized it to a 500-feature framework (HDSI), of which LLE-based features form one of five signal-processing engines, and showed empirically that frequency-domain features outperform LLE features in discriminative power for the studied scenarios.

\subsubsection{Pattern 2: LE as Training-Data Labeler}

A more recent pattern uses the LLE to \emph{generate labels} for ML training rather than as a live input feature. The online classifier receives raw PMU data; the LLE-based label was only needed during the offline training phase.

Pinzon \textit{et al.}\cite{pinzon_real-time_2019} used LLE to generate the stability/instability labels on which a multi-state classifier was trained. Babaali \textit{et al.}\cite{babaali_weighted_2023} labeled scenarios using a hybrid LLE + dynamic voltage stability index (DVSI) rule, where LLE captured oscillatory divergence and DVSI captured delayed voltage recovery. Their XOR combination creates three classes (stable, alert, unstable) for a weighted ensemble (Random Forest + LightGBM) classifier. Zhu \textit{et al.}\cite{zhu_temporal_2025} used LLE-based hysteresis thresholding to create \emph{confident} and \emph{ambiguous} sample labels for temporal contrastive learning pre-training, avoiding misleading supervision near the stability boundary.

\subsubsection{Pattern 3: ML Predicts LE Distribution}

A third, more sophisticated pattern trains ML models to predict the \emph{distribution} of the LLE trajectory rather than classifying stability directly. Liu \textit{et al.}\cite{jiacheng_power_2024} used a gated recurrent unit (GRU) to predict the probability distribution of the finite-time MLE from short post-fault PMU data, then applied diffusion kernel density estimation (DKDE) and neighborhood approximation estimation (NAE) for the final stability decision. This produces a probabilistic assessment that conveys not only whether the system is stable but \emph{how confident} the assessment is, directly addressing the fundamental weakness of binary LLE sign-thresholding.

\subsubsection{Pattern 4: LE Accelerated by ML Preprocessing}

A fourth pattern uses ML or data-driven methods to \emph{preprocess} the signal before LLE computation, with the goal of reducing data requirements or removing oscillation contamination. The Koopman-based approaches \cite{zhang_koopman_2024, gao_short-term_2025} (described in Section \ref{subsec:koopman_le}) and the EMD-based decomposition of \cite{almomani_novel_2025, zong_transient_2024} represent this pattern: the LLE is computed on a cleaned or trend-extracted signal rather than the raw measurement.

\subsubsection{Pattern 5: Physics-Informed ML with LE Structure}

The most ambitious pattern embeds LE-inspired physics constraints directly into the ML model architecture. Yao \textit{et al.}\cite{yao_physics-informed_2025} trained a physics-informed LSTM (Pi-LSTM) in which the loss function includes a penalty for predictions that violate the trajectory-divergence structure implied by the power system's nonlinear dynamics. This ensures that the model's outputs are consistent with LE theory even when trained on limited labeled data, improving generalization to unseen contingencies.

Table \ref{tab:ml_le_patterns} summarizes these five patterns with representative papers and key trade-offs.

\begin{table}[!t]
\renewcommand{\arraystretch}{1.3}
\caption{Taxonomy of Machine Learning and Lyapunov
Exponent Integration Patterns}
\label{tab:ml_le_patterns}
\centering
\small
\begin{tabular}{p{2.2cm}p{2cm}p{3.2cm}}
\toprule
\textbf{Pattern} & \textbf{Representative papers} & \textbf{Key trade-off} \\
\midrule
1. LE as ML feature &  \cite{rashidi_based_2016,  dabou_time_2021} &  Interpretable; delayed by LLE window \\
2. LE as data labeler &  \cite{pinzon_real-time_2019,  babaali_weighted_2023,  zhu_temporal_2025} &  Fast online; LE quality shapes label quality \\
3. ML predicts LLE distribution &  \cite{jiacheng_power_2024} &  Probabilistic output; training data intensive \\
4. ML preprocesses for LLE & \cite{zhang_koopman_2024,  gao_short-term_2025} &  Shorter window; preprocessing accuracy critical \\
5. Physics-informed ML & \cite{yao_physics-informed_2025} &  Best generalization; complex training \\
\bottomrule
\end{tabular}
\end{table}

\subsection{Koopman Operators and Their Connection to Lyapunov Exponents}
\label{subsec:koopman_le}

The Koopman operator framework has emerged as a powerful bridge between data-driven and physics-grounded stability analysis, and it connects naturally to LE theory.

\subsubsection{Koopman Operator Fundamentals}

The Koopman operator $\mathcal{K}$ is an infinite-dimensional linear operator on the space of observables $g:\mathcal{X} \to \mathbb{C}$, defined by $(\mathcal{K} g)(\mathbf{x}) = g(\mathbf{F}(\mathbf{x}))$ where $\mathbf{F}$ is the nonlinear flow map. While the original nonlinear dynamics are hard to analyze, the Koopman operator is linear; its eigenvalues $\lambda_j$ and eigenfunctions $\phi_j$ characterize the global dynamical behavior via spectral decomposition: $g(\mathbf{x}(t)) = \sum_j b_j\, \phi_j(\mathbf{x}_0)\, e^{\lambda_j t}$. The Koopman eigenvalues are directly related to the Lyapunov spectrum: for a fixed point or periodic orbit, the real parts of the Koopman eigenvalues approximate the finite-time Lyapunov exponents of the linearized flow.

\subsubsection{DMD-Based Koopman Preprocessing for LLE}

Zhang \textit{et al.}\cite{zhang_koopman_2024} exploited this connection by applying Dynamic Mode Decomposition (DMD) to a short post-fault voltage window to extract the dominant Koopman mode and eigenvalue. The voltage trend is reconstructed as:
\begin{equation}
\Delta x_j(t) = b_j\,\phi_j(\mathbf{x}_0)\, e^{\lambda_j t},
\label{eq:koopman_mode}
\end{equation}
and the LLE is then computed from this analytically smooth trend rather than from the raw oscillatory voltage, eliminating contamination from damped oscillatory modes. On the Nordic32 system, only 0.5\,s of post-fault data was sufficient to extract the dominant mode, compared with 1.24--2.66\,s required by conventional MLE. Total computation time was reported at 21\,ms, well within the sub-100\,ms target for emergency control.

Gao \textit{et al.} \cite{gao_short-term_2025} extended this with Hankel-matrix-enhanced EDMD (HEDMD), which provides a higher-dimensional Koopman approximation that is more robust to multi-mode oscillations than standard DMD. The predicted dominant-operator trajectory is then fed to a predictive MLE estimator.

\subsubsection{Koopman Eigenvalues as Stability Indices}

An alternative use appears in Khodaparast \textit{et al.} \cite{khodaparast_static_2022}: dynamic Koopman eigenvalues and eigenvectors, derived from linearized time-varying (LTV) system analysis, are connected to LE values through a mode-vector stability certificate. The LE serves as the Lyapunov stability certificate for the LTV trajectory, bridging static (power-flow Jacobian) and dynamic eigenvalue analysis in a unified stability framework.

\subsubsection{Open Directions}

The Koopman--LE connection suggests several important directions. First, for IBR systems where the swing-equation model is unavailable, Koopman-based linearization from PMU data can provide an approximation of the system's Jacobian, enabling LLE computation without a physical model. Second, the Koopman spectrum provides a richer characterization than the LLE alone-the imaginary parts of Koopman eigenvalues encode oscillation frequencies, enabling simultaneous stability assessment and oscillation mode identification. Third, operator-theoretic certificates based on Koopman semigroup properties may provide formal stability guarantees for data-driven LLE estimates that currently lack theoretical backing.

\subsection{Stochastic, Uncertainty-Aware, and Network-Topology-Driven Extensions}
\label{subsec:stochastic_le}

\subsubsection{Almost-Sure LLE for Stochastic Systems}

The almost-sure LE framework of Verdejo \textit{et al.}\cite{verdejo_stability_2012, verdejo_application_2015} (Section \ref{subsec:stochastic}) established the theoretical foundation for stochastic power-system stability analysis. The key theoretical tool is Oseledets' theorem applied to linear systems under bounded Markovian perturbations: the almost-sure LE $\lambda$ is unique (under a nondegeneracy condition), satisfies $\lambda < 0$ iff the system is almost-surely exponentially stable, and can be computed by trajectory-averaging, sphere projection, or unit-sphere simulation methods. This provides a rigorous stability margin $\rho^*$ (the maximum disturbance intensity before $\lambda \to 0$) that is absent from all deterministic LE-based methods.

The Kazma--Taha framework \cite{kazma_stability_2025, kazma_generalizable_2025} extends this to the nonlinear DAE setting: the state-deformation tensor and its $\log\det$ metric track how uncertain renewable injections propagate as perturbation growth rates through the full NL-ODE flow, with a provably near-optimal (99\%) greedy renewable allocation algorithm derived from the submodularity of the $\log\det$ objective.

\subsubsection{Network Topology and Damping Asymmetry}

An unexpected connection between LE-based stability and network topology was identified by Molnar \textit{et al.}\cite{molnar_asymmetry_2021}. Using the Jacobian eigenvalue (as a small-signal LLE proxy) of the swing-equation network model, they showed that the optimal stability configuration requires \emph{heterogeneous damping}-a finding they termed \emph{Converse Symmetry Breaking} (CSB). The optimization:
\begin{equation}
\min_{\boldsymbol{\beta}} \lambda_{\max} \bigl(\mathbf{J}(\boldsymbol{\beta})\bigr), \quad \mathbf{J} = \begin{bmatrix}   \mathbf{0} & \mathbf{I} \\
  -\mathbf{P} & -\mathbf{B}(\boldsymbol{\beta}) \end{bmatrix},
\label{eq:csb_opt}
\end{equation}
(where $\boldsymbol{\beta}$ is the vector of per-generator damping coefficients, $\mathbf{P}$ the synchronizing power matrix, and $\mathbf{B}$ the damping matrix) yields heterogeneous $\beta_i$ values that minimize the maximum real eigenvalue (real-part LLE) and increase the basin of attraction. Validated on the NPCC (48-gen) and German (69-gen) grids, this result has direct implications for IBR parameter tuning: GFM virtual damping coefficients should be individually optimized by network position rather than set uniformly.

\subsubsection{Stochastic and Noisy Excitation: Multistability and Basin Erosion}

Qin \textit{et al.} \cite{qin_random_2014} demonstrated that random parameter perturbations (via the stochastic Melnikov method) can induce chaos even in deterministically stable systems. Gupta and Singh \cite{gupta_chaos_2022} extended this to a small-scale grid under combined periodic wind and noisy load disturbances, showing that multistability and safe-basin erosion are the principal nonlinear stability risks in such systems. These results collectively argue that the future of LE-based stability analysis must move beyond asymptotic single-trajectory LLE toward: (i) almost-sure stability certificates for genuinely stochastic inputs, and (ii) basin-stability metrics \cite{molnar_asymmetry_2021} that quantify the \emph{volume} of the stable basin rather than only its local divergence rate.

\subsection{From Binary LLE Sign to Continuous, Distribution-Aware Stability Metrics}
\label{subsec:index_design}

A clear evolution is visible across the corpus in how the LLE is used as a stability decision metric. The literature has progressed through four generations of index design.

\textbf{Generation 1 (1994--2012): Binary sign decision.} The LLE is estimated from a finite window of PMU data and its sign determines stability: $\lambda_1 > 0 \Rightarrow$ unstable, $\lambda_1 < 0 \Rightarrow$ stable \cite{liu_detection_1994, dasgupta_pmu-based_2012}. This is fast and simple but gives no severity information and is vulnerable to oscillatory contamination and window effects.

\textbf{Generation 2 (2013--2018): Corrected sign and slope trend.} Modifications address the sign-contamination problem: phase-correction via damped-mode fitting \cite{ge_improved_2018}, modified denominator referencing \cite{thilakarathne_modified_2018}, and slope-trend monitoring \cite{wei_model-free_2018} improve reliability without abandoning the sign-decision framework.

\textbf{Generation 3 (2018--2023): Continuous severity quantification.} Several papers moved beyond binary classification to continuous severity scoring. Gubitz \textit{et al.} \cite{gubitz_online_2018} introduced M-TALE, a time-weighted LLE variant correlated with distance from the stability boundary. Almomani \textit{et al.} \cite{almomani_novel_2025} and R\"{u}eger \textit{et al.}\cite{rueger_lyapunov_2019} combined LLE with amplitude and frequency metrics into multi-criteria indices for graded disturbance ranking. Aghtar \textit{et al.}\cite{aghtar_lyapunov_2023} used the average Lyapunov growth rate $\lambda_a = \sum_i \lambda_i / n$ (from the full spectrum) as a continuous contingency-ranking severity metric.

\textbf{Generation 4 (2024--present): Distribution-aware and component-decomposed indices.} The most recent papers treat the LLE as a stochastic quantity whose distribution encodes stability confidence, not only a scalar sign. Almomani \textit{et al.}\cite{almomani_novel_2025} proposed novel LE-based indices that quantify both recovery and oscillation stability separately via KL divergence between observed and reference LE distributions, building on EMD decomposition to separate these components. Liu \textit{et al.}\cite{jiacheng_power_2024} trained a GRU to predict the LLE distribution itself, providing a probabilistic stability certificate. The GVD framework \cite{peng_measuring_2025, peng_measuring_2026} decomposes the system-level LLE into per-device contributions (pGVD), enabling \emph{device-level responsibility attribution}.

This evolution reflects a maturing field: from asking is the system stable? (Generation 1) to asking how stable is each device, how confident are we, and how far is the system from the boundary? (Generation 4).

\subsection{Comparative Index Analysis Across Stability Modes}
\label{subsec:index_comparison}

Several recent papers have directly compared LLE-based indices against other stability metrics across different stability modes and operating scenarios.

Li \textit{et al.}\cite{li_analysis_2023} compared six indices (LLE, Simplified Branch Transient Transmission Capability (sBTTC), Transient Stability Index (TSI), power-current criteria, RoCoF, and Critical Correlation Ratio (CCR) dominant-mode indicator) across rotor-angle, voltage, and frequency instability scenarios in renewable-rich systems. The study found that the LLE correctly identifies instability for both rotor-angle and voltage modes but is susceptible to misjudgment near the critical stability boundary and requires careful window selection. The sBTTC was most reliable for voltage stability; LLE was competitive for rotor-angle instability but slower than TSI.

Bano \textit{et al.}\cite{bano_comparative_2022} compared model-based MLE, model-free MLE with zero-crossing logic, and a voltage stability boundary criterion across 39-bus scenarios, finding that model-free MLE methods with augmented decision rules (Huang's zero-crossing plus $\Delta\omega$ sign) were significantly faster than model-based MLE with comparable accuracy.

These comparative studies are essential for practitioners and are underrepresented in the corpus: only a handful of papers provide systematic head-to-head comparisons of LLE against alternative real-time stability indices on the same dataset.

\subsection{Scalability, Real-Time Deployment, and Computational Complexity}
\label{subsec:scalability}

\subsubsection{Computational Cost of Model-Based LLE}

Model-based LE computation scales as $\mathcal{O}(n^2)$ in the number of state variables (from the variational equation $n \times n$ matrix integration), making it prohibitive for large systems without the efficiency gains of the discrete-QR method \cite{hayes_viable_2018}. Hayes and Milano demonstrated the approach on a 1,479-bus system; Khaitan \textit{et al.}\cite{khaitan_thrust_2017} validated the THRUST framework on a 13,029-bus model. These are encouraging scale demonstrations, but both represent offline batch computation rather than continuous real-time streaming.

\subsubsection{Computational Cost of Data-Driven LLE}

Data-driven LLE estimation scales more favorably: the Rosenstein method is $\mathcal{O}(N^2)$ in the record length $N$ (for the nearest-neighbor search), but the online recursive variant of Wei \textit{et al.}\cite{wei_model-free_2018} reduces this substantially by maintaining only a sliding window of recent phase-space points. The Koopman-DMD preprocessing of \cite{zhang_koopman_2024} achieves total computation in 21\,ms for a 0.5\,s window, which is within the requirement for sub-cycle emergency control.

The PSLE\_TOOL \cite{kamble_psle_tool_2015} provides a MATLAB reference implementation for offline LLE computation that serves as a benchmark for practitioners, though its computational profile is not characterized for large-scale deployment.

\subsubsection{Communication Infrastructure Requirements}

Real-time LLE-based monitoring requires synchronized PMU data streams at sufficient resolution. The framework of Safavizadeh \textit{et al.}\cite{safavizadeh_framework_2019} explicitly addressed communication delay, packet loss, and measurement noise in its WAMPAC architecture, demonstrating that the LLE-based STVS indicator remained reliable with practical communication impairments at 30\,samples/s PMU rate. No paper in the corpus provides a full worst-case analysis of how communication degradation propagates into LLE estimation error, which is an important gap for industrial deployment.

\subsection{Open Problems Spanning All Domains}
\label{subsec:emerging_gaps}

Synthesizing the domain-specific open problems from Sections \ref{sec:stability}--\ref{sec:protection} with the cross-cutting themes above, we identify seven overarching research priorities.

\textbf{(1) A unified LE framework for IBR-dominated grids.} The GVD/pGVD framework addresses STVS; the SMLE addresses commutation-failure transients; hidden-attractor theory addresses microgrid sub-synchronous oscillations. A single, mathematically consistent framework covering all three phenomena for fully IBR-dominated networks (zero synchronous inertia) has not yet been established.

\textbf{(2) Rigorous probabilistic stability margins.} The almost-sure LE theory provides a noise-intensity margin $\rho^*$. A corresponding margin for nonlinear (not linearized) systems under non-Gaussian, non-stationary renewable disturbances-connecting LLE to formal probabilistic stability guarantees-is absent.

\textbf{(3) Real-time large-scale deployment.} No paper demonstrates continuous, 24/7 real-time LLE-based stability monitoring on a real utility system. Demonstration projects analogous to those that validated PMU-based angle monitoring on the WECC and European grids are needed.


\textbf{(4) Cross-domain method transfer.} Methods developed for one domain are rarely tested in another. For example, the CE-based chaos feature of the PD identification paper \cite{qin_partial_2018} could potentially be applied to PMU voltage sequences for STVS; the wavelet-LLE pipeline of turbine fault diagnosis \cite{yan_turbine_2010} might improve LLE estimation for short post-fault PMU windows. Systematic cross-domain experimentation has not been done.

\textbf{(5) Explainability and operator trust.} LLE-based indices are physically interpretable in principle (trajectory divergence rate), but operators need not only an alarm but also an explanation of \emph{which device or bus} is driving the instability and \emph{what control action} is recommended. The pGVD decomposition \cite{peng_measuring_2026} is a step toward device-level explainability; extending this to explicit control recommendations is an open problem.

\textbf{(6) LLE under cyber-physical attacks.} Yan \textit{et al.}\cite{yan_risk_2015} connected conditional Lyapunov exponent (CLEs) to cyber-intrusion risk, but the interaction between adversarial measurement manipulation and LLE estimation-specifically, can an attacker inject PMU data to fool a LLE-based stability monitor?- has not been studied.

\textbf{(7) Standardized benchmarks.} Each paper in this corpus validates its method on a different test system, fault scenario, and LLE estimation procedure. A standardized benchmark (common test system, common contingency set, common LLE implementation, common accuracy metrics) analogous to the IEEE transient stability test suite would allow rigorous method comparison and tracking of field progress.

%
%

%
\section{Conclusions and Future Research Directions}
\label{sec:conclusion}

This paper has presented a comprehensive review of the application of Lyapunov exponents to power system problems, covering more than 165 publications over the period 1991--2026. Four application domains were examined in depth: stability assessment and real-time monitoring (Section \ref{sec:stability}), chaos detection and control (Section \ref{sec:chaos}), forecasting and predictability analysis (Section \ref{sec:forecasting}), and protection, fault diagnosis, and condition monitoring (Section \ref{sec:protection}). Section \ref{sec:emerging} synthesized cross-cutting themes spanning all domains. This final section draws the main conclusions (Section \ref{subsec:conclusions_main}), and distills the principal lessons learned (Section \ref{subsec:lessons}).

\subsection{Main Conclusions}
\label{subsec:conclusions_main}

\subsubsection{The Lyapunov Exponent Is a Uniquely Versatile Power System Metric}

The body of literature reviewed here demonstrates that the Lyapunov exponent occupies a distinctive position among power system analysis tools: it is simultaneously a \emph{stability certificate} ($\lambda_1 < 0$ implies asymptotic stability), a \emph{chaos diagnostic} ($\lambda_1 > 0$ implies sensitive dependence on initial conditions), a \emph{forecast limit} ($T_{\max} = \lambda_1^{-1} \ln(\Delta/\delta_0)$ bounds any predictor), and a \emph{signal feature} for fault discrimination in equipment monitoring. No other single metric in the power systems literature spans all four of these functions, which explains the breadth of its application and the sustained research interest it has attracted over three decades.

\subsubsection{Stability Assessment: From Offline Theory to Online Monitoring}

The trajectory of the stability-assessment literature is the clearest success story in the corpus. Starting from the theoretical demonstration that the LLE correctly identifies CCT and stability regions in post-fault swing-equation models \cite{wadduwage_power_2013, wadduwage_investigation_2013}, the field progressed through a systematic transition to PMU-based real-time estimation \cite{liu_detection_1994, dasgupta_real-time_2013, dasgupta_pmu-based_2015}, resolving the oscillation-contamination problem that threatened practical deployment \cite{ge_improved_2018}, extending to voltage stability \cite{dasgupta_pmu-based_2012, safavizadeh_framework_2019}, and reaching the most recent generation of energy-based, device-resolved indices for IBR-dominated grids \cite{peng_measuring_2025, peng_measuring_2026}. The scalable discrete-QR method \cite{hayes_viable_2018} and the submodular renewable-allocation framework \cite{kazma_stability_2025} represent the current frontier of model-based LE analysis.

This progression from offline academic tool to near-deployable online monitor-spanning three decades and culminating in computationally efficient algorithms validated on systems up to 13,029 buses \cite{khaitan_thrust_2017}-constitutes one of the most complete application-development arcs in modern power system research. The remaining step-continuous real-time deployment on a live utility system with full communication infrastructure-has not yet been demonstrated.

\subsubsection{Chaos and Control: Rich Theory, Limited Practical Reach}

The chaos detection and control literature is the oldest and most theoretically developed domain. Its central achievements are: the systematic characterization of routes to chaos in power system models from the 1990s onward \cite{srivastava_chaotic_1993, nayfeh_bifurcations_1995, li_chaotic_2009}; the demonstration that stochastic parameter variation can induce chaos independently of the nominal operating point \cite{qin_random_2014}; the development of a diverse toolkit of chaos suppression methods including FACTS-based bifurcation engineering \cite{srivastava_elimination_1998}, sliding-mode \cite{kumar_chaos_2018}, synergetic \cite{al-hussein_chaos_2021}, and fractional-order adaptive controllers \cite{rajagopal_bifurcation_2019}; and the recent discovery of hidden-attractor-driven sub-synchronous oscillations in GFM microgrids \cite{yang_sub-synchronous_2023, yang_sub-synchronous_2025}, which represents the most significant conceptual advance in this domain in a decade.

Despite this theoretical richness, the practical reach of chaos-control methods remains severely limited: virtually all results are validated on SMIB or three-bus toy models, no paper provides hardware or field validation, and the models bear little resemblance to the IBR-dominated systems now being built. The LLE plays a diagnostic rather than constructive role in most chaos-control papers: it confirms chaos in the uncontrolled system but is rarely used to \emph{design} or \emph{tune} the controller, with the notable exception of the LLE-as-performance-index approach of \cite{nangrani_numerical_2017, nangrani_fractional_2018} and the chaos-forecast-integrated optimal controller of \cite{dou_optimal_2004}.

\subsubsection{Forecasting: A Mature Workflow with Underexploited Theory}

The 42-paper forecasting domain is methodologically cohesive: a five-stage pipeline (preprocessing, embedding, LLE verification, local prediction, multi-step extrapolation) has been applied across load, wind, PV, and price time series with remarkable consistency since the mid-1990s.

Three observations stand out from the review. First, the predictability horizon formula $T_{\max} = \lambda_1^{-1}\ln(\Delta/\delta_0)$ \cite{mohamed_variability_2007} is the deepest theoretical result in this domain-a model-independent, empirically computable bound on the achievable accuracy of any forecaster-yet it is explicitly used in only one of the 42 papers. The remaining 41 use the LLE only as a binary chaos gate to justify phase-space reconstruction, leaving the most powerful implication of LE theory for forecasting almost entirely unexploited. Second, the minimum-LLE embedding dimension criterion of \cite{mori_short-term_1996, michanos_short-term_2003} (selecting the dimension that minimizes $\lambda_1$ rather than maximizing it or testing for positivity) is a practically useful and theoretically motivated contribution that has not been adopted as standard practice. Third, the decomposition-based chaos-aware predictor routing of \cite{safari_novel_2018, mao_wind_2018}-applying local chaos predictors only to components with positive LLE and conventional models to non-chaotic components-represents the most principled hybrid architecture in the corpus and warrants broader adoption.

\subsubsection{Protection and Fault Diagnosis: A Coherent but Narrow Domain}

The 14 protection and fault diagnosis papers share a consistent methodology (five-stage pipeline from signal acquisition through chaotic feature extraction to classification) applied across diverse equipment types: transformer (partial discharge, overheating, vibration), circuit breaker (vibration), turbine rotor (vibration), OLTC (vibration), shunt reactor (vibration), and microgrid (islanding).

Three methodological advances are noteworthy. First, the cumulative energy representation of \cite{qin_partial_2018} enables LLE estimation from PD pulse sequences that are individually too short for phase-space reconstruction. Second, the three-invariant feature space (LLE, $D_2$, $K_2$) of \cite{yang_chaotic_2020, yang_vibration_2019} provides richer fault discrimination than the LLE alone while preserving interpretability. Third, the use of the Lyapunov index as a \emph{data-quality filter} before deep learning in \cite{liu_transformer_2024} represents the most modern integration of chaos theory and deep neural networks in this domain.

The principal limitation is that no paper provides field validation under realistic noise levels, and the extension to IBR and power-electronics equipment (solid-state transformers, wide-bandgap switches) has not begun.

\subsection{Principal Lessons Learned}
\label{subsec:lessons}

Synthesizing the findings across all four domains, we identify six lessons that carry broad methodological significance.

\textbf{Lesson 1: The finite-time LLE is the practical object, not the asymptotic LE.} The mathematical LE requires $t \to \infty$ for convergence; practical power system applications use finite-time estimates over windows of 100\,ms to 500\,s. The behavior of these estimates-their bias, variance, convergence rate, and sensitivity to oscillation, noise, and window length-is the central practical challenge of the entire field. Understanding this distinction is the prerequisite for all applied LE work.

\textbf{Lesson 2: Model-based and data-driven methods are complements, not competitors.} Model-based LLE computation (GSR, discrete-QR) provides rigorous stability region mapping and theoretical guarantees but requires model fidelity and computational resources. Data-driven LLE estimation (Rosenstein, Kantz, online recursive) enables real-time deployment from PMU data but is sensitive to noise and embedding parameters. The most robust systems combine both: offline model-based analysis for baseline calibration, online data-driven monitoring for real-time decisions.

\textbf{Lesson 3: The LLE alone is not sufficient; the LE spectrum and companion metrics add essential information.} Most papers use only the largest LE. The full LE spectrum determines the Kaplan-Yorke dimension, the metric entropy, and the sum-to-zero constraint (Liouville's theorem). Companion metrics-correlation dimension $D_2$ and Kolmogorov entropy $K_2$-capture different aspects of attractor geometry that the LLE misses. The three-invariant feature space \cite{yang_chaotic_2020} and the full-spectrum contingency ranking \cite{aghtar_lyapunov_2023} demonstrate the practical value of going beyond the scalar LLE.

\textbf{Lesson 4: Chaos verification requires more than a positive LLE.} The cautionary result of \cite{lai_illustrating_1993}- that complex nonlinear transients near voltage collapse can be mistaken for chaos if only the LLE is computed-remains relevant throughout the corpus. Robust chaos confirmation requires consistent Poincar\'{e} sections, broadband power spectra, surrogate-data testing \cite{yang_chaotic_2003}, and correlation dimension saturation. The near-universal practice of using only a positive LLE as a chaos claim, without these complementary tests, is a methodological weakness that this review explicitly flags.

\textbf{Lesson 5: Physical interpretability is the key advantage of LE over pure data-driven methods-preserve it.} The LLE's physical meaning (exponential trajectory divergence rate) is what distinguishes it from black-box stability classifiers. When LEs are combined with ML, this advantage should be preserved by using the LE to anchor physical meaning (as a feature, labeler, or constraint) rather than simply absorbing it into a deep-learning black box. The physics-informed LSTM of \cite{yao_physics-informed_2025} and the pGVD device decomposition of \cite{peng_measuring_2026} exemplify how physical interpretability can be retained in modern ML-augmented frameworks.

\textbf{Lesson 6: Every domain is now facing the same IBR challenge.} IBR-dominated dynamics appear as an open problem in every section: STVS assessment (Section \ref{subsec:stability_gaps}), chaos detection (Section \ref{subsec:chaos_gaps}), forecasting (Section \ref{subsec:forecast_gaps}), and protection (Section \ref{subsec:protection_gaps}). The GVD/pGVD framework \cite{peng_measuring_2025, peng_measuring_2026}, the SMLE \cite{hu_switching_2021}, and the hidden-attractor analysis \cite{yang_sub-synchronous_2023} represent first steps, but the fundamental question of how to define and compute meaningful Lyapunov exponents for fully IBR-dominated, converter-native power systems without classical swing-equation dynamics remains the paramount open problem of the field.

\subsection{Closing Remarks}
\label{subsec:closing}

The field reviewed in this paper has achieved something remarkable: a single mathematical concept from 1960s dynamical systems theory-the Lyapunov exponent-has been developed into a practical monitoring tool, a diagnostic instrument, a forecasting limit, and a fault-classification feature, all within the same engineering discipline. The breadth of application is matched by the depth of the theoretical foundation: from Oseledets' ergodic theorem to Takens' embedding theorem to Koopman operator theory, each layer of mathematics has found a practical role.

Yet the field stands at a pivotal moment. The power grid is transitioning from the synchronous-generator dynamics for which all existing LE methods were developed toward a converter-dominated regime whose physics are fundamentally different. The GVD/pGVD framework, the SMLE, and the hidden-attractor analysis are the first credible responses to this challenge, but they cover only fragments of the problem. The next decade of LE-based power system research will be defined by whether the community can build a coherent, theoretically grounded, and practically validated framework for Lyapunov analysis in IBR-dominated grids-one that preserves the physical interpretability that makes LEs uniquely valuable while adapting to a system whose dynamics no longer resemble the swing equation.

The 169 research priorities, methodological lessons, and terminological recommendations documented in this review are offered as a foundation for that effort.

\bibliographystyle{IEEEtran}
\bibliography{%
  control_and_Stability,%
  Forecasting__Prediction___Machine_Learning_Applications,%
  Fundamental_Theory___chaose,%
  Protection___Fault_Diagonistic%
}

\begin{IEEEbiography}[{\includegraphics
[width=1in,height=1.25in,clip,
keepaspectratio]{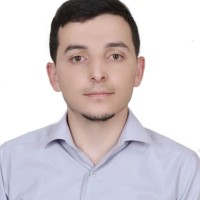}}]
{Mohammad Almomani}
is a Ph.D. candidate in the Department of Electrical and Computer Engineering at Iowa State University. His research focuses on nonlinear dynamics, power system stability, and data-driven stability assessment for modern power grids with increasing penetration of inverter-based resources. His interests include Lyapunov-exponent-based stability analysis, voltage stability, transient stability, oscillation detection, reactive power control, wide-area monitoring using PMU/WAMS data, transmission and distribution co-simulation, and transmission system operator–distribution system operator interaction. His recent work investigates how nonlinear dynamical-system tools, particularly Lyapunov exponents and phase-space reconstruction, can support stability monitoring, chaos characterization, renewable-energy forecasting, protection, and fault diagnosis in power systems. This research is motivated by the growing need for measurement-driven and model-free methods capable of assessing the dynamic behavior of low-inertia, renewable-rich grids. Almomani received the B.Sc. degree in electrical engineering from Yarmouk University in 2017 and the M.S. degree in electrical engineering from Mutah University in 2021.\end{IEEEbiography}

\begin{IEEEbiography}[{\includegraphics
[width=1in,height=1.25in,clip,
keepaspectratio]{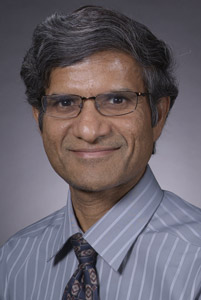}}]
{Venkataramana Ajjarapu }
(Fellow, IEEE) received the Ph.D. degree in electrical engineering from the University of Waterloo, Waterloo, ON, Canada, in 1986. He is currently a Thomas M Whitney Endowed Professor with the Department of Electrical and Computer Engineering, Iowa State University, Ames, IA, USA. His present research interests include reactive power planning, voltage stability
\end{IEEEbiography}

\end{document}